\documentclass[12pt]{amsart}
\usepackage[margin=1in]{geometry}
\usepackage[authoryear]{natbib}
\usepackage{graphicx}
\usepackage{amsmath}
\usepackage{amssymb}
\usepackage{amsthm}
\usepackage{epstopdf}

\newtheorem{assumption}{Assumption}
\newtheorem{condition}{Condition}

\newtheorem{theorem}{Theorem}
\newtheorem{corollary}{Corollary}
\newtheorem{proposition}{Proposition}
\newtheorem{lemma}{Lemma}
\newtheorem{example}{Example}
\usepackage{subfig,comment}
\usepackage[normalem]{ulem}
\newtheorem{prop}{{\bf\sc Proposition}}[section]

\title{Consistently Estimating Network Statistics Using Aggregated Relational Data}

\author[Breza et al]{Emily Breza}
\address[Breza]{Department of Economics, Harvard University, Cambridge, Massachusetts 02138, U.S.A.}
\author[]{Arun G. Chandrasekhar}

\address[Chandrasekhar]{Department of Economics, Stanford University, Stanford, California, 94305, U.S.A.}

\author[]{Shane Lubold, Tyler H. McCormick \and Mengjie Pan}
\thanks{E.B. and A.G.C were supported by the Alfred P. Sloan Foundation. T.H.Mc. is supported by the National Institute of Mental Health of the National Institutes of Health under Award Number DP2MH122405 and by the Eunice Kennedy Shriver National Institute of Child Health and Human Development research infrastructure grant, P2CHD042828, to the Center for Studies in Demography \& Ecology at the University of Washington. The content is solely the responsibility of the authors and does not necessarily represent the official views of the National Institutes of Health.}

\address[Lubold, McCormick, Pan]{Department of Statistics, University of Washington, Seattle, Washington 98195-4322, U.S.A.}

\begin{document}

\maketitle

\begin{abstract}
Collecting complete network data is expensive, time-consuming, and often infeasible.  Aggregated Relational Data (ARD), which capture information about a social network by asking a respondent questions of the form ``How many people with trait X do you know?'' provide a low-cost option when collecting complete network data is not possible. Rather than asking about connections between each pair of individuals directly, ARD collects the number of contacts the respondent knows with a given trait. Despite widespread use and a growing literature on ARD methodology, there is still no systematic understanding of when and why ARD should accurately recover features of the unobserved network.  This paper provides such a characterization by deriving conditions under which statistics about the unobserved network (or functions of these statistics like regression coefficients) can be consistently estimated using ARD. We do this by first providing consistent estimates of network model parameters for three commonly used probabilistic models: the beta-model with node-specific unobserved effects, the stochastic block model with unobserved community structure, and latent geometric space models with unobserved latent locations.  A key observation behind these results is that cross-group link probabilities for a collection of (possibly unobserved) groups identifies the model parameters, meaning ARD is sufficient for parameter estimation.  With these estimated parameters, it is possible to simulate graphs from the fitted distribution and analyze the distribution of network statistics. We can then characterize conditions under which the simulated networks based on ARD will allow for consistent estimation of the unobserved network statistics, such as eigenvector centrality or response functions by or of the unobserved network, such as regression coefficients.

 \end{abstract}
 
\maketitle

The empirical study of social networks has grown rapidly across a variety of disciplines, including but not limited to economics,  psychology, public health,  sociology, and statistics. The aim ranges from researchers trying to understand features of network structure across populations, to parameters in models of network formation, to how network features affect socio-economic behavior, to how interventions can affect the structure of the social network. 
Studying network structure and its relationship to other phenomena  can be demanding particularly in contexts where survey-based research methods are used: obtaining high quality network data from large populations can be expensive and often infeasible for cost, privacy, or logistical reasons.   The challenges associated with collecting complete network data mean that researchers must choose to either (i) reuse one of a handful of existing full graph datasets, likely not designed with their particular research goals in mind or (ii) postpone their research agenda while raising sufficient capital.

One recent approach to address these issues is known as Aggregated Relational Data (ARD), which solicit summaries of respondents' connections by asking for the number of people a respondent knows with a given trait.  ARD questions take the form ``How many people with trait $k$ are you linked to?'' and  can be integrated into standard probability-based survey sampling schemes because they do not directly solicit any connections in the graph \citep{breza2017using}. 
One major advantage of collecting ARD over more traditional network surveys is the reduced cost.  In the context of one large-scale randomized controlled trial studying the relationship between network structure and household finance in 60 villages, ~\cite{breza2017using} showed that ARD implementation is shorter (3 versus 8 months) and cheaper (\$34,000 versus \$189,000) compared to full network  enumeration 
and yet delivered the same economic conclusions that would have been obtained using the full network data. Because it is cheaper to collect, ARD also enables practitioners to collect panel data across multiple networks.

ARD was originally proposed to estimate the size of hard to reach populations, such as the number of HIV-positive men in the U.S. \citep{killworth1998estimation, scutelniciuc2012network, jing2014estimating}. Since then, the use of ARD has expanded significantly, particularly in the social sciences ~\cite{DiPrete:2011dn, feehan2017network,leung2013} where ARD enable researchers to estimate core features of respondent networks (e.g., a respondent's centrality or the extent of clustering).  In terms of methodology for analyzing ARD, \cite{mccormick2015latent} connects a model for ARD responses to network models of the fully observed graph. Specifically, ~\cite{mccormick2015latent} established a connection between ARD and the latent distance model, a common statistical approach for modeling fully observed network data.  The key result is that ARD are sufficient to identify parameters in a generative model for graphs, allowing inference about the distribution of graphs that plausibly correspond to the ARD. ~\cite{breza2017using} exploit this connection to generate a distribution over network statistics, such as the centrality of an individual or the average path length of the graph, and show examples where using statistics generated from ARD gives similar results to using statistics from the completely observed graph.  ARD has also been used to estimate common econometric models and outcomes, such as the linear-in-means model~\citep{boucher2020estimating}, choosing the optimal seeding for maximal information flow~\citep{sadler2022seeding}, and can be used to assess network model goodness-of-fit \citep{Lubold_GOF}.

Despite its increasingly widespread use, there is still little understanding of when or why ARD contains sufficient information to estimate model parameters or estimate network properties of the unobserved network.  We provide such a characterization in two steps. 
First, we show we can consistently estimate the parameters of a rich class of generative network models using only ARD. This fact relies on a simple but powerful observation that if the cross-type link probabilities allow us to identify the model parameters, then ARD is sufficient for consistent estimation. Critically, this insight allows us to sidestep maximizing the complicated log-likelihood directly and instead solve a system of equations based on the cross-trait linking probabilities.  We show that three common generative models fall into this class.

Next, we provide sufficient conditions to consistently estimate features of the underlying, observed network using ARD.  
The intuition is that, for sufficiently large graphs, some statistics of graphs converge to their expected value, where the expectation is taken over graphs from the same generative process.  In such cases, ARD suffices to recover the value of the graph statistics, so long as the statistics are not too sensitive to error introduced by using estimates for the generative network models parameters.
In such cases, the information in ARD is sufficient to consistently estimate generative model parameters as well as graph statistics of interest.

We investigate this  both theoretically and empirically in two settings.
 The first
  is when researchers can consistently estimate features of the underlying, unobserved network structure itself. Examples include  centrality or clustering measures for nodes. 
 This analysis studies the case of a single large network.
  The second 
   is when researchers can consistently estimate response functions of or by the network. That is, how do changes in network features correspond to changes in socio-economic outcomes or how might an intervention affect the structure of the network.  This analysis studies the case of many networks.
   
\section*{Aggregated relational data}
\label{sec:mle-proof}
We begin by defining ARD formally. Take an undirected, unweighted graph, ${g} = (V,E)$ with vertex  set $V$ and edge set $E$.  There are $n = |V|$ nodes, so we sometimes write $g_n$ to emphasize the graph size, and $g_{ij}={\bf 1}\{ij\in E\}$ denotes that $i$ and $j$ are connected in the graph. Suppose each node has one of $K$ traits, where $K$  is fixed and $K>3$.  Let $G_k$ denote the nodes with trait $k$, for $k = 1, \dotsc, K$, where $n_k = |G_k|$ is the number of nodes with trait $k$. We write $t^\star_i = k$ to denote that node $i$ has trait $k$. We suppose that the traits are binary and mutually exclusive, so every node has one of $K$ traits.  This is not a prohibitive assumption. To see this, imagine 
there are $L$ characteristics (e.g., rural/urban or college educated/not) and for simplicity assume that these are binary. Then it is clear that we can always construct $K$ traits, mutually exclusive, with $K = 2^L$. The extension to multi-valued characteristics is straightforward.  Additionally, traits constructed through intersecting characteristics (e.g., men with a given occupation below a particular age) also reduces the size of the target population, which can limit recall bias~\cite{mccormick2010many}.

To collect Aggregated Relational Data (ARD), the researcher asks $m$ randomly chosen nodes ``How many people with trait $k$ are you linked to?'' for each of these $K$ traits.  Linking is typically defined as knowing a potential connection (e.g., having interacted with the person in the past 2 years or recognizing the person if passing on the street).~\cite{feehan2016quantity} provide an extensive discussion and experimental evidence regarding the definition of linking. To simplify exposition, we will set $m = n$, meaning we have ARD from all nodes.  Our results also apply when $m<<n$, as is common in practice. In such cases, we would either need to impute parameters for nodes without ARD data (see~\cite{breza2017using}, for example) or make an assumption about node equivalence. For example, under a stochastic block model, all nodes in a given community have the same linking behavior.  So by collecting ARD from at least one node in each community, we can then estimate the parameters of all nodes.

Let $y_{ik}$ denote node $i$'s response to this question about trait $k$, with $y_{ik} = \sum_{j\in G_k} g_{ij}$. Critically, when collecting ARD, the researcher does not observe any edges, just how many edges are present between a node $i$ and people of a given trait. We use $\mathbf{\mathbf{y}}$ to denote the $m \times K$ matrix of ARD responses.
Since the $K$ traits are mutually exclusive, ARD responses count distinct alters across each trait group.
Otherwise, if trait group $A$ and trait group $B$ overlap, then a person in both groups would be counted twice, once in response to the ARD question about trait $A$ and once about $B$.  

To model the network, we consider a general graph model $\mathbb{P}(g_n | \theta^\star)$,  where edges form independently in the network,  conditional on the unknown parameter vector $\theta^\star$. We call such models conditional edge-independent graph models. The number of elements in the vector $\theta^\star$ can depend on the graph size $n$ (to accommodate node-level heterogeneity parameters, for example) but we omit this dependence to simplify the notation ($\theta^\star = \theta^\star_n$). %
In most settings, each component $\theta^\star_i$ is independently and identically drawn from $F$. In other cases, sometimes the distribution of $\theta^\star_i$ depends on the traits that node $i$ possesses, which we write as $\theta^\star_i | t^\star_i = k  \sim F_{k}$.  This conditional independence representation relies on exchangeability amongst nodes and, thus,  implies that the resulting asymptotic sequence of networks generated by these models are dense, meaning that the average degree for a given $n$ is a constant times $n$ \citep{orbanz2015bayesian}.

We define $p_{ij} = p_{ij}(\theta^\star)$ to be the probability that $i$ and $j$ connect, given the model parameters. The ARD response $y_{ik} = \sum_{j \in G_k}  g_{ij}$ is then is a sum of independent, but not identically distributed, Bernoulli$(p_{ij})$  random variables, which in various disciplines is known as either the Poisson's Binomial random variable or  the Poisson Binomial 
random variable \citep{Wang, Rosenman_Ecology, Rosenman_Political}. The probability mass function of $y_{ik}$ given $\theta$ in conditional edge-independent models is
\begin{equation*}
    f_{ik}(y | \theta^\star) := \sum_{A \subseteq \mathcal{A}_{y}} \prod_{j \in A} p_{ij}(\theta) \prod_{j \in A^c} \{1 - p_{ij}(\theta)\} \;,
\end{equation*}
where $\mathcal{A}_{y}$ is the set of subsets of $\{1, \dotsc, n_k\}$ that contain exactly $y$ elements.  In the case where $p_{ij} = p$, then this expression simplifies to the probability mass function of the Binomial$(n_k, p)$ random variable. An obvious first approach to modeling the ARD is to analyze the likelihood of the data $\mathcal{L}_n(\mathbf{y} \mid \theta) = \prod_{i = 1}^n \prod_{k = 1}^K f_{ik}(y_{ik} \mid \theta)$. Here, we have used the assumption of mutually exclusive traits, which allows us to write the likelihood of observing $({y}_{i1}, \dotsc, y_{iK})$ as $\prod_{k =1 }^K f_{ik}(y_{ik} | \theta)$. The conditional independence of edges, given $\theta$, allows us to write the joint distribution of ARD responses over all individuals as a product, which does not depend on whether traits are mutually exclusive. 

Proving consistency of $\hat \theta_n := \arg \max_{\theta} \mathcal{L}_n(\mathbf{y} \mid \theta)$ is challenging due to the complex nature of the log-likelihood, since each $\theta_i$ appears in $n$ terms of the likelihood.   
We explore a different approach to estimate $\theta^\star$.   Instead of looking at $f_{ik}(y_{ik} | \theta)$, where $\theta$ includes the parameters of node $i$ as well as those of all other nodes with trait $k$ (which are not observed with ARD), we look at the probability that node $i$ connects to an arbitrary node with trait $k$, $P_{ik}$, which is
\begin{equation*}
     P_{ik} := \mathbb{P}(g_{ij} = 1 | \theta^\star_i, j \in G_k) = \int_{\Theta_k} \mathbb{P}(g_{ij} = 1 | \theta^\star_i, \theta_j) dF_k(\theta_j) \;,
\end{equation*}
where again $\theta_j \sim F_{\theta, k}$ for nodes $i$ with trait $k$, and  $\Theta_k$ denotes the support of $F_{k}$.
 In the latent space model, $\Theta_k$ might be $p$-dimensional Euclidean, spherical, or hyperbolic space, and node locations are drawn according to a mixture model along the surface of the latent space \citep{hoffrh2002, tantrum, Lubold}. In the beta-model, $\Theta_k$ is a subset of the real line. 

To understand the utility of analyzing $P_{ik}$, rather than the full log likelihood $\mathcal{L}_n(\mathbf{y} \mid \theta)$, note that for any node $i$, 
\begin{equation}
\label{eq: WLLN_ARD}
    \frac{y_{ik}}{n_k} = \frac{1}{n_k} \sum_{j \in G_k} g_{ij} \overset{p}{\rightarrow} P_{ik} \;,
\end{equation} 
as $n_k \rightarrow \infty$, where $P_{ik}$ is again the probability that node $i$ connects with someone of trait $k$ and $n_k$ is the number of nodes with trait $k$. 
Here we have assumed that the weak law of large numbers applies to the average $y_{ik}/n_k$, as is the case for conditionally edge-independent graphs.  In the conclusion we discuss extensions for settings where edges could be correlated or where edge probability scales with the graph size. 

Supposing that (\ref{eq: WLLN_ARD}) holds, we can then equate the vector of normalized ARD responses with their respective edge probabilities $P_{ik}(\theta^\star)$, and use an estimating equation approach to estimate the model parameters. Supposing that this system has a unique solution in $\theta$ (or unique up to an isometry, as in the latent space model), this general approach allows us to derive estimators of model parameters and prove uniform convergence of these estimators in a host of rich and frequently used network models. When this system does have a unique solution, we say informally that such a model ``identifies'' the model parameters.

By equating observed ARD responses and the probability of connection between a node and nodes in a given trait group, we can invert that equation to solve for the parameters $\theta_i^\star$.
In the next three sections, we consider three common generative network models and derive consistent estimates of the parameters in each model using this intuition.

\section*{Beta-model} 
We first consider the generalized beta-model \cite{chatterjeeds2010, graham2014econometric}. The original version of this model states that an edge forms between nodes $i$ and $j$ with probability $\text{expit}(\nu_i^\star + \nu_j^\star)$ for some sequence of parameters $\nu_1^\star, \dotsc, \nu_n^\star$ that encode the popularity of nodes.  Here, $\text{expit}(x) = \exp(x)/(1+\exp(x))$.
The generalized beta-model includes a term that measures the effect of dyad-level covariates $X_{ij} \in \mathbb{R}^p$ on linking probability, so 
 \begin{equation*}
    \mathbb{P}(g_{ij} = 1 | \nu_i^\star, \nu_j^\star, \beta^\star) = \text{expit}(\nu_i^\star + \nu_j^\star + \beta^\star X_{ij}) \;.
\end{equation*}

\cite{chatterjeeds2010, graham2014econometric} propose estimates of the parameters using a fixed point procedure using the full network data. This procedure only requires the degree of a node. Suppose that we observe ARD about a collection of traits that are mutually exclusive and exhaustive.  Then the degree of node $i$ is $d_i = \sum_{k = 1}^K y_{ik}$. Let $\hat \nu_i$ and $\hat \beta$ denote the fixed-point estimates of $\nu_i$ and $\beta$ from \cite{chatterjeeds2010,graham2014econometric} computed using the ARD, which is by the preceding comments equivalent to the estimate computed from the full network data.

In the theorem below, we require that the support of the parameters in the beta-model be compact subsets of $\mathbb{R}$. This regularity condition was also imposed in \cite{graham2014econometric}. 

\begin{theorem}
\label{theorem: beta_consistency}
Suppose the support of each node effect $\nu_i^\star$ and of $\beta^\star$ are compact subsets of $\mathbb{R}$. Then, with probability $1 - O(1/n^2)$, 
\begin{equation*}
    \max_{1 \leq i \leq n} |\hat \nu_i - \nu^\star_i| \leq C \sqrt{\frac{\log(n)}{n}}
\end{equation*} 
for some constant $C$ that does not depend on $n$. In addition, $\hat \beta \overset{p}{\rightarrow} \beta$ as $n \rightarrow \infty$. 
\end{theorem}
Here, we have not made any assumption about the relationship between traits and the distribution of the node parameters. 

In cases where ARD is collected at the characteristic level and not at the trait level which creates a mutually exclusive partition, or when the mutually exclusive trades do not exhaust the space, then $\sum_{k =1}^K y_{ik}$ does not need to equal $d_i$, the degree of node $i$. In these cases, we can estimate degree of a node via other methods. One such example is the network scale-up method~\cite{killworth1998estimation},  which assumes that given a node's degree, ARD responses are modeled as $y_{ik} | d_i \sim \text{Binomial}(d_i, \frac{n_k}{n})$. This leads to the so-called ``ratio of sums'' estimator $\hat d_i = n \sum_{k = 1}^K y_{ik} / \sum_{k = 1}^K n_k$, where $n_k$ is the size of group $k$ and $n$ is the total size of the population \citep{killworth1998estimation, zheng2006many}. Typically, these ARD questions are based on characteristics with known group sizes, so that each $n_k$ is known. We can then plug in $\hat d_i$ in place of $d_i$ in the estimation procedures from \citep{chatterjeeds2010, graham2014econometric} to estimate the model parameters.

\section*{Stochastic block model}

We consider a generalized version of the stochastic block model (SBM), in which observable traits are dependent on, but potentially distinct from, latent community structure.   Edges are determined by community structure. This setting corresponds to a case where nodes belong to unobserved communities and a researcher observes traits that are (imperfectly) associated with community membership.  We show that ARD allows 
us to use links to observable groups to infer latent community membership.

We describe this model more formally. We first assign node communities $c^\star_i$ independently with probabilities $\pi_1, \dotsc, \pi_C$. Conditioned on these parameters, edges are generated independently with probabilities
\begin{equation*}
    \mathbb{P}(g_{ij} = 1 | c^\star_i = c, c^\star_j = c') =  P_{cc'} \;,
\end{equation*}
where $P$ is a $C \times C$ matrix of within- and cross-community edge probabilities. Here we suppose that the graph is undirected, so that $P$ is assumed to be symmetric. The intuition for this model is that the probability two nodes connect depends only on their latent group membership. 
In many cases of interest, the community structure is unknown \emph{a-priori} and unobservable but traits are observable. We let the $C \times K$ matrix $Q$ encode  the probability of having trait $k$, given that a node is in community $c$, so
\begin{equation}
\label{eq: prob_trait_SBM}
    \mathbb{P}(t^\star_i = k | c^\star_i = c) = Q_{ck} \;. 
\end{equation}
In the case of mutually-exclusive traits, each node is assigned exactly one of the $K$ traits with probabilities in (\ref{eq: prob_trait_SBM}). The intuition behind this model is that nodes with the same traits form edges in a similar way. 

We suppose that the ARD we have access to is about these $K$ traits and not about the unobserved community structure.  To estimate the parameters in the SBM, we begin by making the following assumption, 
which allows us to consistently cluster the ARD to estimate community structure in the unobserved graph. Specifically, we assume that no two communities have the same linking pattern to all other traits, which is clearly required for identification.

\begin{assumption}
\label{ass: distinct_clusters_CD}
The following condition holds: 
\begin{equation*}
    \min_{c, c'} ||Z_c - Z_{c'}|| > 0 \;,
\end{equation*}
where $Z_c := (\tilde P_{c1}, \dotsc, \tilde P_{cK})$ and $\tilde P_{ck} := \mathbb{P}(g_{ij} = 1 | c_i = c, t_j = k)$ is the probability that a node in community $c$ connects to a node with trait $k$: $\tilde P_{ck} = \left(\sum_{\ell = 1}^C Q_{\ell k} \pi_{\ell}\right)^{-1}\sum_{\ell' = 1}^C P_{c\ell'} Q_{\ell' k} \pi_{\ell'} .$

\end{assumption}

 To understand this assumption, let us consider a simple case when $C = K = 2$. Assumption \ref{ass: distinct_clusters_CD} then requires that
\begin{equation*}
    \begin{pmatrix}
    \tilde P_{11} \\
    \tilde P_{12}
    \end{pmatrix} \neq
       \begin{pmatrix}
    \tilde P_{21} \\
    \tilde P_{22}
    \end{pmatrix} \;.
\end{equation*}
We now analyze when these equalities do not hold. If the probability of belonging to community 1 and 2 are equal ($\pi_1 = \pi_2$), the first inequality is then equivalent to requiring that
\begin{equation*}
    (P_{11} - P_{21})Q_{11} + ( P_{12}- P_{22})Q_{21} \neq 0 \;.
\end{equation*}
If $P_{11} = P_{12} = P_{22} = P_{22}$, which corresponds to no community structure in the model, then Assumption \ref{ass: distinct_clusters_CD} is not satisfied for any $Q$ matrix. If $Q_{12} = Q_{21}$, which means that there is no relationship between traits and community membership, then Assumption \ref{ass: distinct_clusters_CD} is satisfied whenever $P_{11} - P_{21} \neq  P_{22} -  P_{21}$, which occurs in un-directed networks whenever $ P_{11} \neq P_{22}$ . In this case, even if there is no relationship between traits and network structure, Assumption \ref{ass: distinct_clusters_CD} is satisfied provided that communities behave differently in the network (i.e., there is meaningful community structure).

We now provide a classification algorithm to estimate the community membership of nodes. This procedure does not require us to know the number of communities. We initialize  $W=V$, the set of nodes in the sample, so $|W| = n$. Let $\tilde y_i = (y_{i1}/n_1, \dotsc, y_{iC}/n_C)$. While $W \neq \emptyset$, do the following:
\begin{enumerate}
\item Select a node $i$ randomly from $W$. Set $W = W \setminus \{i\}$. 
\item For any $j \in W$: If $||\tilde y_i- \tilde y_j||^2 \leq n^{-1}\log(n)$, assign node $j$ to be in the same community as $i$, and
second set $W = W \setminus \{j\}$.
\end{enumerate}

This procedure returns a consistent estimate of the community membership \emph{and} the number of communities. The distribution of ARD responses for people in a given community $c$ collapses to a point mass as the sample size grows, and so clustering in our problem is easier than clustering in general clustering problems, where the distribution of data does not need to change with the sample size. We therefore propose the algorithm above, over more standard clustering algorithms, because our clustering algorithm lends itself easily to concluding the uniform consistency in Theorem \ref{theorem: SBM_consistency} that we need later in Theorems \ref{thm: ARD_consistency} and \ref{thm: consistenct_OLS_cov}.

We prove in Theorem \ref{theorem: SBM_consistency} that this classification algorithm returns consistent community labels. 
Given the community memberships $\mathbf{\hat c}$, let $\hat C_c$ denote the set of nodes in our sample that are estimated to be in community $c$, under the membership vector $\mathbf{\hat c}$, with $|\hat C_c| =: m_c(n)$. We can estimate $P_{cc'}$ with
\begin{equation*}
    \hat P_{cc'} = \begin{cases}
    \frac{1}{m_c(n)} \sum_{i \in \hat C_c} \frac{y_{ic'}}{n_{c'}},\ \ \ \ c \neq c' \vspace{0.05in}\\ 
\frac{1}{m_c(n)} \sum_{i \in \hat C_c} \frac{y_{ic'}}{n_{c'} - 1}, \ \ \ c = c'
    \end{cases}
    \;.
\end{equation*}
where again $y_{ic}$ is the ARD response from node $i$ about trait $c$.  We can estimate $Q_{ck}$ with
\begin{equation*}
    \hat Q_{ck} = \frac{1}{m_c(n)} \sum_{i \in \hat C_c} \mathbf{1}\{t_i = k\}.
\end{equation*}
where $t_i$ is the observed trait of node $i$ and we can estimate $\pi$ with entries $\hat \pi_c = \frac{1}{n} \sum_{i = 1}^n \mathbf{1}\{\hat c_i = c\}$.

\begin{theorem}
\label{theorem: SBM_consistency}
Suppose Assumption \ref{ass: distinct_clusters_CD} holds. Then, up to a permutation on the community labels, the estimated community membership vector $\hat c$ satisfies
\begin{equation*}
    \max_{1 \leq i \leq n} \mathbf{1}\{\hat c_i \neq c^\star_i\} \overset{p}{\rightarrow} 0 \;,
\end{equation*}
as $n \rightarrow \infty$.
The estimated number of communities $\hat C$ as well as $\hat P$, $\hat  Q$, and $\hat \pi$ are all consistent as $n \rightarrow \infty$. 
\end{theorem}

\section*{Latent space model}

We consider a broad class of latent space models.
Broadly speaking, each node has a position in a latent (or unobserved) space, and the closer two nodes are in this space, the more likely they are to connect. Each node also has a gregariousness parameter, which controls the baseline edge probability for that node \cite{hoffrh2002, tantrum, Lubold, asta2014geometric, shalizi2017consistency}.

We formally define one variant of the latent space generative model, which we study in this work.  We draw the gregariousness parameter $\nu_i^\star$ from a distribution $F_\nu$ with compact support in $(a,0)$ for some $a<0$. We draw traits $t^\star_i \in \{1, \dotsc, K\}$ independently with probabilities $\pi_1, \dotsc, \pi_K$. Conditioned on traits, we also draw node positions $z_i^\star | t^\star_i = t \sim F_{t}$, where $F_t$ is some distribution over the latent surface $\mathcal{M}^p(\kappa)$. Here, $\mathcal{M}^p(\kappa)$ is a complete simply connected Riemannian manifold with constant curvature $\kappa$, which means by the Killing-Hopf theorem that it is either Euclidean, spherical, or hyperbolic space of dimension $p$ and curvature $\kappa$ \cite{Killing1891}. We suppose that $F_t$ is a symmetric distribution over $\mathcal{M}$ and is uniquely determined by its mean $\mu_t$ and variance $\sigma_t^2$. Some examples of this include the Gaussian distribution over $\mathbb{R}^p$ and the von-Mises Fisher distribution over the $p$-sphere.
In words, the node positions $z_i^\star$ is drawn from a mixture distribution on $\mathcal{M}^p(\kappa)$, with weights determined by $\pi_k = \mathbb{P}(t^\star_i = k)$.
Conditioned on these parameters, we draw edges independently with probability
\begin{equation}
\label{eq: LS_model}
    \mathbb{P}(g_{ij} = 1 | \nu_i^\star, \nu_j^\star, z_i^\star, z_j^\star) = \exp\{\nu_i^\star + \nu_j^\star - d(z^\star_i, z^\star_j) \} \;.
\end{equation}
Again, we suppose that we only have access to ARD about these $K$ traits. 
We write $\eta = (\mu_1, \dotsc, \mu_K, \sigma_1^2, \dotsc, \sigma_K^2)$ to refer to the "global" parameters.
To build our estimators $\hat \nu_i, \hat z_i,$ and $\hat \eta$, we proceed in two steps: (a) estimate the global parameters and  (b)  use them as a plug-in to estimate the node parameters. 
Our proof is based mainly on the following calculation. Consider the marginal probability of a connection between person $i$ and group $k$, $P_{ik}$.  The form of $P_{ik}$ comes from integrating across all individuals in group $k$ in~(\ref{eq: LS_model}), which is consistent with the information in ARD since no individual connections are observed.  Further, following~\citet{hoffrh2002} and~\citet{mccormick2015latent}, we can model $y_{ik} \mid  \nu^\star_i, z^\star_i \sim \text{Binomial}(n_k, P_{ik})$, where $P_{ik}$ is the probability that $i$ connects to a member of group $k$ (the explicit form is derived in the Supplementary Materials and is a function of $\nu_i^\star$ and $z_i^\star$) and $n_k$ is the number of nodes in group $k$. 

In step (a), we  derive the estimators for the global parameters, $\hat{\eta},$ in Section S.1 of the Supplementary Materials, but provide the intuition here. If we consider the probability of an arbitrary link between two members of the same group $k$, it does not depend on $\mu_k$ but only on the variance $\sigma^2_k$ and the expected shift in linking probability due to node effect $\nu_i$. Similarly, if we consider the probability of an arbitrary link across two groups $k$, $k'$ knowing the variance terms, then this provides information on centers $\mu_k, \mu_k'$. We can therefore equate the probability of connection between traits with the observed number of traits and solve for the parameter $\eta$. 
Given estimates of the global parameters $\eta$, we now estimate the node locations and fixed effects. Since $E(y_{ik})= P_{ik}/n_k$, we construct a system of equations by equating the ratio of the marginal probability of connection for person $i$ in group $k$ to that in group $k'$ ($P_{ik}/p_{ik'}$) to the ratio of sample averages ($y_{ik}n_{k'}/y_{ik'}n_k$), which does not depend on the fixed effect of node $i$. This allows us to estimate the locations of all nodes, up to a global isometry in the latent space. We then similarly estimate the node fixed effects, once we have estimated the node locations and global parameters, by equating $y_{ik}$ and $p_{ik}$. In summary, we construct Z-estimators of the global parameters, the node locations, and the node fixed effects by constructing 4 systems of equations, which allows us to consistently estimate all of the parameters in the latent space model. Equivalently, one can  interpret the moments based estimators for the location and fixed effects parameters as coming from maximizing a pseudo likelihood, which we describe in the Supplementary Materials.

We now state the assumptions for consistency of these estimators. $E_{kk'}[\exp\{-d(z, z')\}]$ denotes the expectation of $\exp\{-d(z, z')\}$, where $z \sim F(\mu^\star_k, \sigma^\star_k)$ is independent of $z' \sim F(\mu^\star_{k'}, \sigma^\star_{k'}).$

\begin{assumption}
\label{assumption: compact_param}
For each $k$, $\mu_k$ is in a compact subset of $\mathcal{M}^p(\kappa)$ and $\sigma_k$ is in a compact subset of $(0, \infty)$.
\end{assumption}

\begin{assumption}
\label{assumption: a_V}
The node effects $\nu_i^\star \overset{iid}{\sim} H$ satisfy   $E\{\exp(\nu^\star_i)\} < \infty.$
\end{assumption}

\begin{assumption}
\label{assumption: a_F}
The distribution $F$ is a symmetric distribution on $\mathcal{M}^{p}{(\kappa)}$ that is completely characterized by its mean and variance and satisfies the following two conditions. The function $z_i \mapsto E_k[\exp\{-d(z_i, z)\}]$ is Lipschitz for every $k \in \{1, \dotsc, K\}$ and $z_i \mapsto E_k[\exp\{-d(z_i, z)\}] / E_{k'}[\exp\{-d(z_i, z')\}]$ has a pseudo-inverse that is Lipschitz.
\end{assumption} 

 \begin{assumption}
 \label{assumption: lip_sigma}
Define the function $F_1: (z_i, \sigma_k, \sigma_{k'}) \mapsto E_k[\exp\{-d(z_i, z)\}] / E_{k'}[\exp\{-d(z_i, z')\}]$. The inverse function $F_1^{-1}$ is continuous in $\sigma$ and for every $k, k', \ell$, and $\ell'$, the following two functions are Lipschitz:
 \begin{equation*}
     \eta \mapsto \frac{E_{kk'}[\exp\{-d(z, z')\}]}{E_{\ell \ell'}[\exp\{-d(z, z')\}]}, \ \ \  \eta \mapsto  \frac{E_{kk'}[\{\exp(-d(z, z')\}]^2}{E_{\ell \ell'}[\{\exp(-d(z, z')\}]^2} \;.
 \end{equation*}
\label{assumption:sigmainvert}
 \end{assumption}

 Assumptions~\ref{assumption: a_F}-\ref{assumption: lip_sigma} ensure that the probabilities from~(\ref{eq: LS_model}) vary smoothly with changes in the distribution of points on $\mathcal{M}^p(\kappa)$. In the Supplementary Materials, we verify that common distributional choices (e.g., Gaussian in Euclidean space or von Mises Fisher on the hypersphere) satisfy these assumptions and discuss the pseudo-inverse defined in the assumptions above. For simplicity, we suppose that $n_k = n/K$ for each trait, so that traits are evenly divided amongst the nodes, and write $\tilde n = n/K$.

\begin{theorem}
\label{thm: LS_consistency}
Suppose Assumptions \ref{assumption: compact_param}, \ref{assumption: a_V}, \ref{assumption: a_F}, and \ref{assumption: lip_sigma} hold. The estimators $\hat z_i$ and $\hat
\nu_i$ computed from equating the ARD responses and the marginal probability of connections, as well as $\hat \eta$ (defined in the Supplementary Materials) are consistent for $z_i^\star, \nu_i^\star$, and $\eta^\star$ as $m, n \rightarrow \infty$, up to isometry on $\mathcal{M}^p(\kappa)$ and satisfy
\begin{align*}
    &\max_{1 \leq i \leq m(n)} d_{\mathcal{M}^p(\kappa)}(\hat z_i, z^\star_i) \leq \sqrt{\frac{3\log(\tilde n)}{2\tilde n}}, \\ &\max_{1 \leq i \leq m(n)} |\hat \nu_i - \nu^\star_i| \leq \sqrt{\frac{3\log(\tilde n)}{2\tilde n}},
\end{align*}
with probability $1 - O(m/\tilde n^3)$.
\end{theorem}

The proof of Theorem~\ref{thm: LS_consistency} and associated simulations are in the Supplementary Materials.

\section*{A taxonomy for estimating graph  statistics}
\label{sec:tax}

 We assume that data arise from one of three models considered in the previous work (beta-model, stochastic block model, or latent space model) and that ARD allows us to estimate the model parameters $\theta^\star_n$. We leverage Theorems \ref{theorem: beta_consistency}, \ref{theorem: SBM_consistency}, and \ref{thm: LS_consistency} and assume throughout the rest of this work that that the researcher has access to an estimator $\hat \theta_n(\mathbf{y})$ of $\theta^\star$. Here, $\theta^\star$ 
denotes the true parameters of one of the three models, and $\hat \theta_n(\mathbf{y})$ denotes the estimates of the model parameters from Theorems \ref{theorem: beta_consistency}, \ref{theorem: SBM_consistency}, and \ref{thm: LS_consistency}. We separate our discussion into two cases: (1) the researcher has a single large network with $n$ nodes; (2) the researcher has many independent networks. We recall for convenience that the user has access to an ARD survey from $m \leq n$ nodes.

\subsection*{Single large network}
\label{sec:singlebig}

Starting with the first case, assume the researcher is interested in estimating a network statistic, $S_{i}\left({ g}^\star_n\right)$ for node $i$ computed on the graph ${ g}^\star_n$. For simplicity we write this as a function of a single node, though it can easily be extended to functions of multiple nodes.  For the purposes of this argument, there is one actual realization of the graph, ${ g}^*_n$.  This is what we would have observed if we had collected information about all actual connections between members of the population, rather than collecting ARD. Importantly, the researcher collecting ARD cannot observe ${ g}^*_n$. This actual network realization does, however, come from a generative model with parameters that can, by  Theorems \ref{theorem: beta_consistency}, \ref{theorem: SBM_consistency}, and \ref{thm: LS_consistency}, be estimated from ARD.

In the following results, we characterize settings where network statics can be consistently estimated using only the $n\times K$ matrix of ARD, ${\mathbf{y}}$. For simplicity we set $m=n$, though our results hold when $m<n$ as well, though a researcher would need to sample a sufficiently large fraction of the graph to capture the structure of interest \citep{chandrasekharl2016}. Based on observing ARD, we compute ${E}\{S_{i}\left({ g}_n\right)\mid \hat \theta( \mathbf{y})\}$, where $\hat \theta({{\mathbf{y}}})$ is the estimator from Theorems \ref{theorem: beta_consistency}, \ref{theorem: SBM_consistency}, or \ref{thm: LS_consistency} using the ARD $\mathbf{y}$. We are interested in when $E\{S_i(g_n) \mid \hat \theta_n(\mathbf{y})\}$ is a good estimator of $E\{S_i(g_n) \mid \theta^\star_n\}$ and therefore of $S_i(g_n^\star)$.

There are two general conditions require to consistently estimate graph parameters from ARD.  First, the statistic of interest must be one that is relatively stable between draws from the graph generating process.  This condition is required since our estimators in the previous section concern parameters of the network formation model, but the goal is to estimate a statistic for a particular draw from this generating process, ${ g}_n^*$.  Second, we require that these estimates of generating model parameters are sufficiently precise and the form of the statistics is such we can control the variation in the estimated network statistic in the presence of small variance in the estimated model parameters.  We formalize these conditions in the following theorem. We use the notation $\theta_{j, n}^\star$ to refer to the $j$th entry of the vector of true parameter values $\theta^\star_n \in \mathbb{R}^n$. Finally, let the partial derivative with respect to the $i$th component be denoted  by $\partial_i E\{S_i(g_n) \mid \theta_n\}$.

\begin{theorem}
\label{thm: ARD_consistency}
Let $g_n^\star$ denote the graph of interest drawn from a conditional edge-independent graph models with parameters $\theta^\star_1, \dotsc, \theta_n^\star$, and let $\hat \theta_n$ denote estimates of these parameters. Suppose that 
\begin{enumerate}
\item $1/n \sum_{j} |\hat \theta_{j, n} - \theta_{j, n}^\star| \overset{p}{\rightarrow} 0$,
    \item $|E\{S_i(g_n) \mid \theta^\star_n\} - S_i(g_n^\star)| \overset{p}{\rightarrow} 0$, and
\item 
the function $\theta_n \mapsto E\{S_i(g_n) \mid \theta_n\}$ is differentiable and  
\label{eq: gradient_requirement}
    $$ \max_j \sup_{\theta_n} \partial_j E\{S_i(g_n) \mid \theta_n\} \leq C/n$$
for some finite constant $C> 0$.
\end{enumerate}
Then,  $|E\{S_i(g_n) \mid \hat \theta_n(\mathbf{y})\} - S_i(g^\star_n)| \overset{p}{\rightarrow} 0$ as $n \rightarrow \infty$. 
\end{theorem}

We provide a proof of Theorem \ref{thm: ARD_consistency} in the Supplementary Materials. The proof relies on a Taylor series approximation of the network statistic $E\{S_i(g_n) \mid \hat \theta_n(\mathbf{y})\}$. In particular, we require that the approximation term due to the estimation of $\theta_n^\star$ with $\hat \theta_n(\mathbf{y})$ disappear as $n \rightarrow \infty$. One sufficient condition for this to occur is given in Conditions 1-3 of Theorem \ref{thm: ARD_consistency}. 

Condition 1 of Theorem \ref{thm: ARD_consistency} requires that the average estimation error goes to zero in probability as the graph size grows. The estimators from Theorems \ref{theorem: beta_consistency}, \ref{theorem: SBM_consistency}, and \ref{thm: LS_consistency} satisfy Condition 1 of Theorem \ref{thm: ARD_consistency}, since the average estimation error is always upper-bounded by the maximum estimation error.  Thus, Theorem \ref{thm: ARD_consistency} implies that the researcher can use $E\{S_i(g_n) \mid \hat \theta_n(\mathbf{y})\}$ to estimate $S_i(g^\star_n)$, provided the network statistic $S_i(g^\star_n)$ satisfies Conditions 2 and 3.

Condition 2 of Theorem \ref{thm: ARD_consistency} equires that $|E\{S_i(g_n) \mid \theta^\star_n\} - S_i(g_n^\star)| \overset{p}{\rightarrow} 0$, which must be true regardless of the estimator used to estimate $\theta^\star_n$. Many network statistics are an average of terms, such as the clustering coefficient or the centrality coefficient, and so this condition holds for many statistics of interest. Condition 3 of Theorem \ref{thm: ARD_consistency} requires that changing the graph model parameters slightly does not change the value of $E\{S_i(g_n) \mid \theta_n\}$ too much. For many common network statistics, this condition is true, as we show in Corollary \ref{cor:MSE_density_diffcent}.

To clarify when the conditions of Theorem \ref{thm: ARD_consistency} hold and when they fail, we provide several pedagogical examples. Our first example is an obvious failure of the second condition.  Specifically, we show the statistic from a given realization does not converge to its expectation, then even after more nodes are observed, there is no increasing information, and  the mean-squared error of the estimate should not go to zero.
Let $p_{ij}(\theta^\star)$ denote the probability that nodes $i$ and $j$ connect.

\begin{corollary}\label{cor:MSE_link}Consider a sequence of distributions of conditional edge-independent graphs $\mathbb{P}(g_n | \theta^\star)$ on $n$ nodes, where $\theta^\star$ is known. Given an (unobserved) graph of interest, ${ g}_n^*$, and  $0<p_{ij}({\theta^\star})<1$, then the mean squared error for ${ E}\{S_{i}\left({ g}_n\right)\} = { E} \left(g_{ij} \right)$, the expectation of a draw from the distribution of any single link $g_{ij}$, is
\[
{ E}[\{ { E}  (g_{ij})-g^*_{ij}\}^{2}] =p_{ij}(\theta^\star)\{1 - p_{ij}(\theta^\star)\} \;.
\]
\end{corollary}

When a link exists, the mean squared error is $\{1-p_{ij}({\theta^\star})\}^2$ and when a link  does not, it is $p_{ij}({\theta^\star})^2$. In edge-independent models, node-level exchangeability ensures that $p_{ij}({\theta^\star})$ does not vanish with $n$, which means that the mean squared error cannot go to zero as $n \rightarrow \infty$. However, for graph models in which $p_{ij}$ tends to zero, then Condition 2 does hold. 

However, for many commonly used and non-trivial network statistics, the conditions of Theorem \ref{thm: ARD_consistency} do hold. By verifying the conditions of Theorem \ref{thm: ARD_consistency}, we have the following result.

\begin{corollary}
\label{cor:MSE_density_diffcent}
Suppose $g_n^\star$ is drawn from either the $\beta$-model, stochastic block model, or latent space model and $\hat \theta_n$ is computed from Theorems~\ref{theorem: beta_consistency},~\ref{theorem: SBM_consistency}, and~\ref{thm: LS_consistency}, respectively. For the following statistics $S_i({g}_n)$, we have that $|E\{S_i(g_n^\star) \mid \hat \theta_n(\mathbf{y})\} - S_i(g_n^\star)| \overset{p}{\rightarrow} 0$. 
\begin{enumerate}
\item Density (normalized degree): The density of node $i$ is $S_i({ g}_n) = \sum_j g_{ij}/n$. 
\item Diffusion centrality (nests eigenvector centrality and Katz-Bonacich
centrality): Define $S_i({ g}_n) = S_i({g}_n, q_n, T) =\sum_{j}\{\sum_{t=1}^{T}\left(q_{n}{ g}_n\right)^{t}\}_{ij}$ for some $q_n = C/n$ and any $T$. 

\item Clustering: Let  $N(i) = \{j: g_{ij}=1\}$ denote the neighbors of node $i$. The clustering coefficient is defined as
$    S_i({ g}_n) = { \sum_{j, k \in N(i)} g_{jk} }/(  { |N(i)| |N(i)-1|  } )$.
\end{enumerate}
\end{corollary}

Diffusion centrality is a more general form which nests eigenvector centrality when $q_{n}\geq 1/\lambda_1^n$,
and because the maximal eigenvalue is on the order of $n$, this meets our condition. Here $\lambda_1^n$ is the largest eigenvalue of the adjacency matrix of ${ g}_n$. It also nests Katz-Bonacich centrality. In each of these, $T\rightarrow\infty$. It also captures a number of other features of finite-sample diffusion processes that have been used particularly in economics \citep{banerjeegossip,banerjeecdj2013}. These notions each relate to the eigenvectors of the network---objects that are ex-ante not obviously captured by the ARD procedure but ex-post work since in this model statistics converge to their expectations.

\begin{figure*}
    \centering
    \includegraphics[scale = 0.25]{images/combined_fig1.pdf}
\caption{Scaled mean squared error of node-level and graph-level network features. These results corroborate the theoretical intuition we developed. Specifically, we show in Corollary~\ref{cor:MSE_link} that the mean squared error should be large for a single link and in Corollary~\ref{cor:MSE_density_diffcent} that the mean squared error should diminish for (normalized) degree and diffusion at the node level and clustering at the graph level.   
}
\label{fig:MSE}
\end{figure*}

These results give two practical extreme benchmarks. ARD should not perform well for estimating a realization of any given link in the network. In contrast, it should perform quite well for statistics such as density or eigenvector centrality. Other statistics may fall somewhere in the middle of this spectrum. For example, whether a notion of centrality such as betweenness - which relies on the specifics of the exact realized paths in the network - works well may depend on the specific statistic and network distribution. We explore these predictions empirically in Figure \ref{fig:MSE}.

\subsection*{Many independent networks}
\label{sec:manyindep}
Consider the setting where the researcher has $R$ networks each of size $n_r$, and the networks are over disjoint sets of nodes.  We use the terminology independent networks to refer to such a collection of networks For each network $r$ we observe ARD $n_r \times K$ matrix ${\mathbf{y}}_{r}$. We take $n_r = n$ for simplicity, but our results do not require this. Also, we drop the dependence on $n$ in the notation ${ g}_r$. 
Every network is generated from a network formation process with true parameter $\theta^\star_{r}$.  In this case of many networks, we consider how well the ARD procedure performs when the researcher wants to learn about network properties, aggregating across the $R$ graphs.  This is the case in a large literature \cite{caijs2013,beaman2016can, breza2016field}.

Let $S^*_{r}=S\left({g}^*_{r}\right)$ be a network statistic from the $R$ unobserved graphs generating the ARD.  For any given graph from the data generating process, define $S_{r}=S\left({g}_{r}\right)$. For notational simplicity, we consider network-level statistics, but the argument can easily be extended to node, pair, or subset-based statistics. We use the notation $\theta^\star_{i, n, r}$ to denote the $i$th entry of the vector of parameters $\theta^\star_{n, r} \in \mathbb{R}^n$ for network $r.$ We use similar notation for the estimator $\hat \theta_{i, n, r}$.

We consider two regression problems. In the first problem, 
the goal of the researcher is to estimate the model
\[
O_{r}=\alpha+\beta S^*_{r}+\epsilon_{r} \ \ \ r = 1, \dotsc, R \;,
\]
where $O_r$ is some socio-economic outcome of interest and 
and the parameter of interest is $\beta$. As before, $S^*_{r}$
is unobserved because ${g}^*_{r}$ is unobserved and the researcher
only has ARD, ${\mathbf{y}}_{ r}$. The researcher instead estimates the expectation of the statistic given using ARD, $\bar{S}_{r}= E\{S_{r} \mid \hat \theta_{n, r}\}$.  The regression becomes
\begin{equation}
\label{eq: mis-measured_cov}
O_{r}=\alpha+\beta\bar{S}_{r}+u_{r}.
\end{equation}
and $\hat \beta = \hat \beta_{n, R}$ is the ordinary least squares (OLS) estimator of $\beta$ from (\ref{eq: mis-measured_cov}). Critically, $\hat \beta$ depends on the size of each network $n$ and the number of networks $R$.

In the second regression model we consider, the network feature is an outcome that responds to an intervention, $T_r$:
\[
S^*_{r}=\alpha+\gamma T_{r}+\epsilon_{r}.
\]
We let $\hat \gamma_{n, R}$ denote the OLS estimator of $\gamma$ from the regression
\begin{equation}
\label{eq: mis-measured_outcome}
    \bar{S}_{r}=\alpha+\gamma T_{r}+\epsilon_{r}.
\end{equation}

\begin{theorem}
\label{thm: consistenct_OLS_cov}
Let $\hat \beta_{n, R}$ denote the OLS estimate from (\ref{eq: mis-measured_cov}) and let $\hat \gamma_{n, R}$ denote the OLS estimate from (\ref{eq: mis-measured_outcome}). Suppose that
\begin{enumerate}

\item the estimators of the parameters for the $r$th network, denoted by $\hat \theta_r(n)$, satisfy
\begin{equation*}
\max_{1 \leq r \leq R} \frac{1}{n}\sum_{i =1}^n |\hat \theta_{i, n, r} - \theta^\star_{i,n, r}| \overset{p}{\rightarrow} 0 \text{   as $n, R \rightarrow \infty$}
\end{equation*}
\item the functions $\theta_n \mapsto E\{S_{r} \mid \theta_{r, n}\}$ is differentiable for each network $r$ and each network size $n$. Suppose also that  
\begin{equation*}
\max_{1 \leq r \leq R} \max_{j} \sup_{\theta_{n, r}} \partial_j E\{S_{r} \mid \theta_{n, r}\} \leq \frac{C}{n} \;,
\end{equation*}
for some finite constant $C >0$. 
\end{enumerate}
If $E(\epsilon_r \mid S_r^\star) = 0$ and the design matrix has full rank, then $|\hat \beta_{n, R} - \beta| \overset{p}{\rightarrow} 0$ and $|\hat \gamma_{n, R} - \gamma| \overset{p}{\rightarrow} 0$ as $n, R \rightarrow \infty$. 
\end{theorem}

The following theorem shows that
the three conditions from Theorem \ref{thm: consistenct_OLS_cov} hold. 

\begin{theorem}
\label{thm: checking_cond_OLS_cov}
Suppose that each network $g^\star_{n, r}$ is known to be drawn from either the beta-model, stochastic block model, or latent space model and $\hat \theta_n$ is computed from Theorems~\ref{theorem: beta_consistency},~\ref{theorem: SBM_consistency}, and~\ref{thm: LS_consistency}, respectively. If $S_{r}^\star$ is the density, centrality, or clustering of a node in network $r$, as defined in Corollary \ref{cor:MSE_density_diffcent}, then Conditions 1 and 2 of Theorem \ref{thm: consistenct_OLS_cov} hold if $R n/\exp(n)\rightarrow 0$. 
\end{theorem}
In words, Theorem \ref{thm: checking_cond_OLS_cov} states that a researcher is able to run the regression in (\ref{eq: mis-measured_cov}) using the estimators in Theorems~\ref{theorem: beta_consistency},~\ref{theorem: SBM_consistency}, and~\ref{thm: LS_consistency} to consistently estimate $\beta$, the true effect of the network statistics on the observed socio-economic outcomes.

Take the most extreme example of a single link, where we know its presence cannot be identified in a single large network. Even if we were interested in a regression of
$y_{12,r}=\alpha+\beta g_{12,r}+\epsilon_{r}$,
where whether nodes 1 and 2 are linked affects some outcome variable of interest across all $R$ networks,
we can use $E\{g_{12,r} \mid \hat \theta(y_{r})\}$ in the regression to consistently estimate $\beta$.
Here, nodes 1 and 2 refer to arbitrarily labeled nodes and can be different across the $R$ networks. 
In contrast to the single network case, where the mean squared error of the estimate of $g_{12,r}$ does not tend to zero as $n$ grows, here simply having the conditional expectation is enough to estimate the slope of interest, $\beta$. Therefore, with many graphs, the ARD procedure works well under weaker conditions on the network statistics. However, despite the generality of Theorem \ref{thm: consistenct_OLS_cov},  Condition 2 of Theorem \ref{thm: consistenct_OLS_cov} still must hold. Some statistics are more sensitive to the input parameters and thus might not satisfy Condition 2. For example, the number of connected components has a higher mean squared error than the other statistics, which suggests that this statistic might lead to poor OLS estimators in (\ref{eq: mis-measured_cov}) and (\ref{eq: mis-measured_outcome}).
\section*{Simulation results}\label{sec:taxonomy_sims}

\subsection*{Single large graph}

We  explore the results for a single large graph through  simulation exercises. 
We first generate 250 graphs from the generating process in (\ref{eq: LS_model}) then randomly assign each node to one of $K$ traits.  Each network consists of 250 nodes, similar to the size of villages from  in~\cite{banerjeecdj2013}.  We then draw a sample of nodes from the graph and construct ARD using traits.  Our simulation does not reflect error in the ARD, which  may arise if, for example, a person is a member of a group but the respondent does not have this information (e.g.,~\cite{PK_etal03},~\cite{zheng2006many},~\cite{ezoe2012population}, or~\cite{feehan2016quantity}).  We then estimate graph statistics using the procedure outlined in~\cite{breza2017using}.

\begin{figure*}
    \centering
    \includegraphics[scale = 0.35]{images/combined_fig2.pdf}
\caption{Boxplots for the simulation experiments with multiple independent networks. In the left figure, we consider a regression where the node-level network statistics determine outcomes on one network. In the middle figure, we consider a regression where network-level statistics determines outcomes on multiple networks. In the right figure, we consider a regression where a treatment determines a network-level statistics. On the $x$-axis we provide the network statistics used and the $y$-axis represents the value of the regression coefficients estimators. The red line indicates the true value of the regression coefficients. These results corroborate the theoretical intuition developed in Theorems \ref{thm: ARD_consistency} and \ref{thm: consistenct_OLS_cov}.}
\label{fig:node}
\end{figure*}

 Figure \ref{fig:MSE} plots the mean squared errors of our estimation procedure across a range of common network statistics.  These mean sqaured errors reflect uncertainty in estimation of the model parameters and in the underlying network statistics. In order to make the mean squared errors comparable across statistics, we scale by ${1}/{{E}(S_i)^2}$. Subfigure (a) in Figure \ref{fig:MSE} focuses on node level statistics.  We compute ten node level statistics: (1) proximity (average of inverse of shortest paths); (2) average path length; (3) closeness centrality (the average inverse distance from $i$ over all other nodes); (4) 
 degree (the number of links); (5) diffusion centrality (as defined in \cite{banerjeecdj2013} -- an actor's ability to diffuse information through all possible paths); (6) eigenvector centrality (the $i$th entry of the eigenvector corresponding to the maximal eigenvalue of the adjacency matrix for node $i$); (7) the average distance from a randomly chosen seed (as in a diffusion experiment where the seed has a new technology or piece of information); (8) support (as defined in \cite{jacksonrt2012} -- whether linked nodes $ij$ have some $k$ as a link in common); (9) clustering (the share of a node's links that are themselves linked); (10) betweenness centrality (the share of shortest paths between all pairs $j$ and $k$ that pass through $i$); (11) whether link $ij$ exists.  The results from the simulation, ordered in terms of scaled mean squared error in the figure, are consistent with the theoretical results. Statistics such as density and centrality take values for each realization that are nearly their expectation, meaning that we can recover the statistics with low mean squared error.  For a single link this is not the case and, correspondingly, the simulations show higher error.

Subfigure (b) of Figure \ref{fig:MSE} focuses on graph-level statistics.  The graph level statistics are as follows: (1) share of nodes in the giant component; (2) average proximity (average of inverse of shortest paths); (3) average path length; (4) diameter; (5) the share of links across the two groups relative to within the two groups where the cut is taken from the sign of the Fiedler eigenvector (this reflects latent  homophily in the graph); (6) maximal eigenvalue; (7) clustering; (8) number of components. All network statistics, with the exception of the number of components one, have small scaled mean squared error. This reflects the intuition of Corollary \ref{cor:MSE_density_diffcent}. ARD recovers statistics that converge to their expectations, such as density, and might fail to recover statistics that do not.

We also evaluate our approach using observed, fully-elicited graphs. We use data from~\cite{banerjeecdj2013}, which consists of completely observed graphs from 75 villages in rural India. In each village, about one-third of respondents were asked ARD questions. ~\cite{breza2017using} compare statistics estimated with ARD (using estimated formation model parameters) from these graphs with the same statistics calculated using the complete graph.  We leverage these results and present a different aspect: how the mean squared error changes as the size of the graph grows.  We present results for individual-level statistics from these graphs and compute mean squared error across individuals.  Our results using graphs with real-world complexity and properties (e.g. density and community structure) confirm the results from our simulation experiments.   These results are presented in Figure S2 of the Supplementary Materials.

\subsection*{Many independent networks}
\label{sec:manyindep}
Multiple independent networks often arise in experiments, so we simulate a setting where we assign graph level treatment randomly to half of the graphs. Graphs in the control group have expected degree generated from a normal distribution with mean 15 and variance 25, while graphs in the treatment group are generated from a normal density with mean 25 and variance 25. Each graph has 250 nodes. All graphs have a minimum expected degree of 5 and a maximum expected degree of 35. Due to the association between density and treatment, we expect treatment effects on graph-level statistics, such as average path length and diameter. The average sparsity over all graphs is 20/250=0.08, which is a value similar to Karnataka data discussed in \cite{SavingsMonitors}.  
For individual measures, 50 actors are randomly selected in each network. For links measured between actors, 1000 pairs are randomly selected in each network. For network level measures, there is one measure per network, so the regression consists of $R$ data samples, where $R$ is the number of networks.

Figure \ref{fig:node} shows the simulation exercise with multiple independent networks. We use formation model parameters, $\theta^\star$, to get $\bar{S}_{ij,r}$ or $\bar{S}_{r}$ and include results using estimated model parameters in the Supplementary Materials (Figures S3, S4, and S5).
We present results with $R=200$ ($R=50, 100, 200$ are in the Supplementary Material). $\epsilon_r$ comes from a normal distribution with zero mean, and $\mathrm{var}(\epsilon_r)=\mathrm{var}(S^*_{ij,r})$ to maintain a 0.5 noise to signal ratio.

The first two panels in Figure \ref{fig:node} show the distribution of the estimate of $\beta$ in a  regression where the network statistic predicts an outcome of interest.  The middle line of each boxplot is the median $\hat{\beta}$, and the borders of boxes denote first and third quartiles. All boxplots have outliers removed.  The leftmost panel gives results for individual level measures while the center panel gives network level measures. Among the node level statistics we see that all estimated $\hat{\beta}$s are close to the simulation value of one.  The individual link measure, though empirically similar, is not centered around the true simulated value.  The downward bias is an example of attenuation bias or regression dissolution, since there is variability in the network statistic acting as the covariate.  The indicator of the presence/absence of a single link is the most variable of the network measures and, thus, bias persists for the link measure when it does not for the others.  For graph level measures, all estimated coefficients are centered around the generated values.

The rightmost panel in Figure \ref{fig:node} shows results for the case where the network statistic is the outcome and is predicted by another covariate, in this case treatment status.  
The percentage error is defined as $(\hat{\gamma}-\gamma)/\gamma$. 
Percent cut and diameter has large variation of percent errors than the other measures. This is due to the fact that the treatment effect, density differences between treatment and control, has a smaller effect on percent cut and diameter than on other measures. The average percent of variation explained by treatment in $S_r$ for percent cut and diameter is around 0.3, while it is around 0.5 for other measures.

\section*{Discussion}

Collecting full network data in large networks (e.g., a city) or across many networks (e.g., villages or schools) requires enumerating all egos and alters and therefore can be prohibitively expensive, logistically hard, or face privacy concerns. The use of ARD allows the researchers to overcome these problem by fitting frequently used and rich generative models, which can be then used to estimate socio-economic quantities and parameters of interest. This can include features of the network, but also responses in network structure to interventions as well as how socio-economic outcomes are affected by network structure. 

In this work, we first demonstrated that by using ARD we are able to
consistently estimate parameters in several families of frequently used generative network models, including ones where the number of parameters grows as the graph size grows. Second, we provided a taxonomy to describe when we may expect to estimate socio-economic features consistently using ARD. Together, our theoretical results and supportive simulations using empirical data, present new insights into settings where researchers can count on ARD to reliably estimate socio-economic quantities of interest. This makes the study of socio-economic networks much more accessible to a wide set of researchers; in our own setting using ARD delivers the same economic conclusions as the full network data deoes but at 80\% less cost \citep{breza2017using}.

There are several promising avenues for future work. First, the techniques studied here are likely more relevant for networks of the scale of villages or cities counties but certainly not necessarily things like large social media networks.  It is true that when the number of nodes is very large, one needs many more traits $K$ to exceed the number of latent communities $C$ (since presumably a large $C$ is needed to fit the network well). Note that geography can be included, to some degree, in a reasonably natural way. After all, one can imagine carving out a set of locations (as set if $L$ regions) and now a “type” $K$ is the sub-trait (e.g., caste) crossed with the location. So $K = T \times L$ and we would use $K > C$ in this way. This is not the only approach, but we leave a complete exposition of this strategy to future work. Second, we demonstrate consistent estimation for edge-independent network models.  Extending these results to a broader class of models, particularly those that are asymptotically sparse or which have correlated edges would extend the reach of our work and we believe much of the infrastructure we developed around the necessary properties of network statistics would still apply \citep{Peixoto_2022, wassermanp1996, snijders2002}.
Third, a natural question to ask is whether other data collection strategies might be more useful to deliver consistent estimates for quantities that fall outside of the taxonomy of statistics that are estimable with ARD.


 \section*{Supplementary Material}
 
\renewcommand{\theequation}{S.\arabic{equation}}
\renewcommand{\thesection}{S.\arabic{section}}
\renewcommand{\thefigure}{S.\arabic{figure}}

 We now outline the main parts of the supplementary materials. In Section \ref{sec: beta_SBM}, we provide proofs of Theorems \ref{theorem: beta_consistency} and \ref{theorem: SBM_consistency} in the main paper, which deal with consistency in the beta-model and the stochastic block model (SBM), respectively.  We then move to proving Theorem \ref{thm: LS_consistency}, which deals with consistency in the latent space model. First, Section \ref{sec: define_estimates} defines the estimates of the node locations and effects, and in Section \ref{sec: theorem_1_proof}, we prove Theorem \ref{thm: LS_consistency} in the main paper, which deals with the consistency of the estimates of the node locations and effects. The proof of Theorem \ref{thm: LS_consistency} relies on proving consistency of the estimates of the global parameters, which we do in Section \ref{sec: estimate_global}. Section \ref{sec: discussion_assumption} discusses the assumptions made in Theorem \ref{thm: LS_consistency} in the main paper and demonstrates that several conventional distributions used in the literature satisfies these assumptions. Section \ref{sec: proof_plugin} contains the proof of Theorem \ref{thm: ARD_consistency} in the main paper.  Section \ref{section: Other Proofs} provides proofs of the other theorems in the main paper.  Section \ref{sec: consist_OLS_many_network} contains the proof of Theorem \ref{thm: consistenct_OLS_cov} and Section \ref{sec: checking_conditions_Thm} contains the proof of Theorem \ref{thm: checking_cond_OLS_cov}. Sections \ref{sec:supsim} and \ref{sec: additional_sims} provide additional simulations.   Section \ref{sec: simulations} provides simulations to verify the consistency of the claims made in Theorem \ref{thm: LS_consistency}.  Section \ref{sec: add_lemma}  contains additional lemmas and results we use in the supplementary materials. 

In the proofs, we use $C$ to refer to constants or sequences of constants that can change from line to line, but critically these constants never depend on the graph size $n$ nor the number of nodes with trait $k$, $n_k$. 

\section{Consistency of Beta-Model and SBM Parameters (Theorems \ref{theorem: beta_consistency} and \ref{theorem: SBM_consistency})}
\label{sec: beta_SBM}

We begin with the beta-model. Before providing specifics, we first introduce the main ideas of the proof of Theorem \ref{theorem: beta_consistency}, which shows that the estimators, computed using just ARD, proposed in \cite{graham2014econometric} are consistent for the parameters of the beta-model. To do this, we first recall that \cite{graham2014econometric} proposes a fixed point estimator $\hat \nu_i$ that satisfies $\hat \nu_i(t+1) =  \phi(\hat \nu_i(t))$ for some known function $\phi$, which depends only on the degree sequence. They also propose a consistent estimator of the parameter $\beta$, which also only depends on the degree of the nodes. Since ARD allows us to recover the degree of nodes in the survey, we can then directly apply the results of \cite{graham2014econometric} to conclude Theorem \ref{theorem: beta_consistency}.  Before getting to the proof of Theorem \ref{theorem: beta_consistency},  we now re-state Theorem 3 of \cite{graham2014econometric}, which we use in our proof of Theorem \ref{theorem: beta_consistency}.
\begin{prop}[Theorem 3 of \cite{graham2014econometric}]
The fixed point estimator, as described in equations 17-18 of \cite{graham2014econometric}, satisfies
\begin{equation*}
    \max_{1 \leq i \leq \hat n} |\hat \nu_i - \nu_i^\star| \leq C \sqrt{\frac{\log(n)}{n}}
\end{equation*}
with probability $1 - O(1/n^2)$ for some constant $C > 0$. In addition, we have that $\hat \beta \overset{p}{\rightarrow} \beta$ as $n \rightarrow \infty$.
\end{prop}

\begin{proof}[Proof of Theorem \ref{theorem: beta_consistency}]
In the case of mutually exclusive and exhaustive traits, $d_i = \sum_{k = 1}^K y_{ik}$. Since the fixed point estimation procedure proposed in \citep{chatterjeeds2010, graham2014econometric}
depends only on the degree of each node, which we are able to estimate with ARD, we can then apply Theorem 3 of \citep{graham2014econometric} to conclude Theorem \ref{theorem: beta_consistency} of the main paper. Theorem 3 of \cite{graham2014econometric} requires several conditions (Conditions 1, 2, 3, and 5 of \cite{graham2014econometric}), which are all satisfied under the assumptions of Theorem \ref{theorem: beta_consistency} of the main paper. 

\end{proof}

\begin{figure}
\centering
\parbox{5cm}{
\includegraphics[width=6cm]{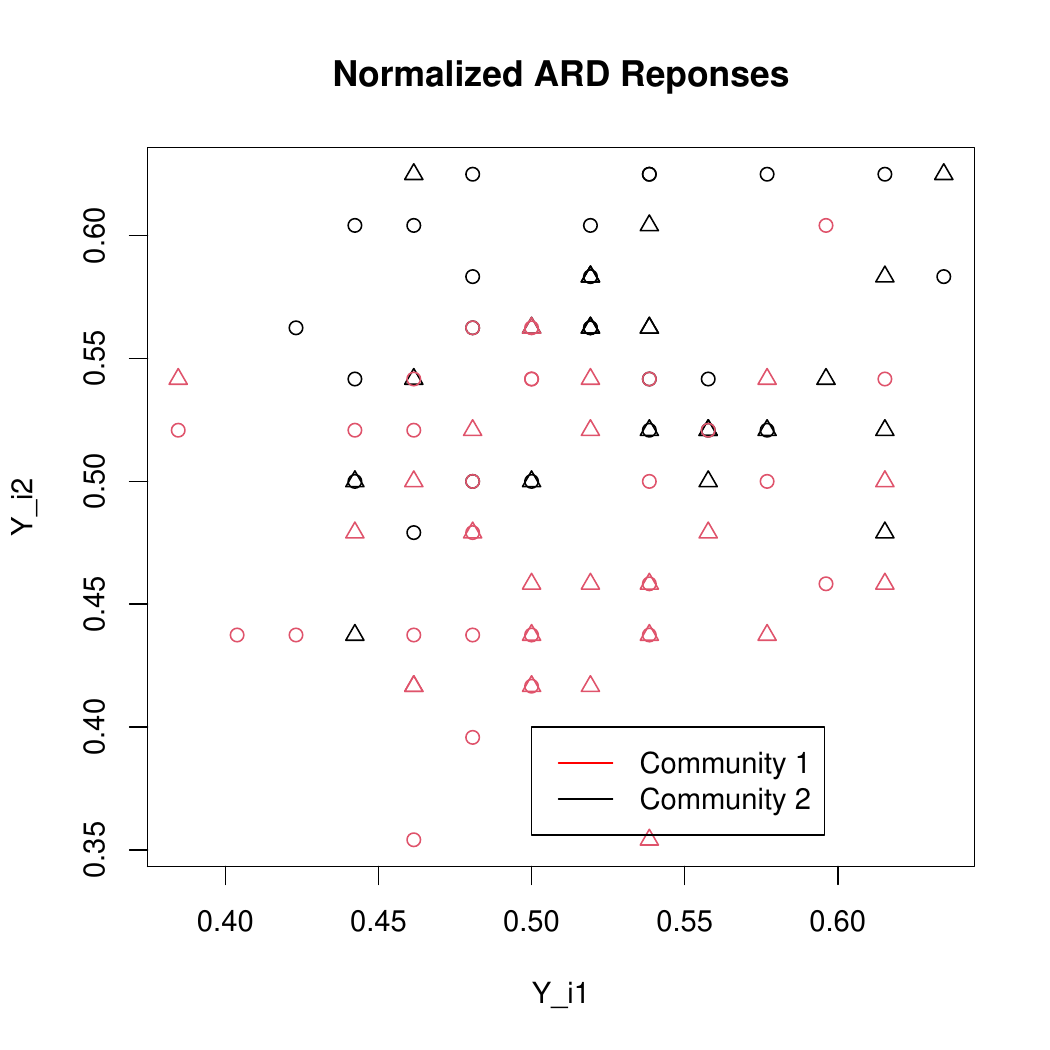}
}
\qquad
\begin{minipage}{5cm}
\includegraphics[width=6cm]{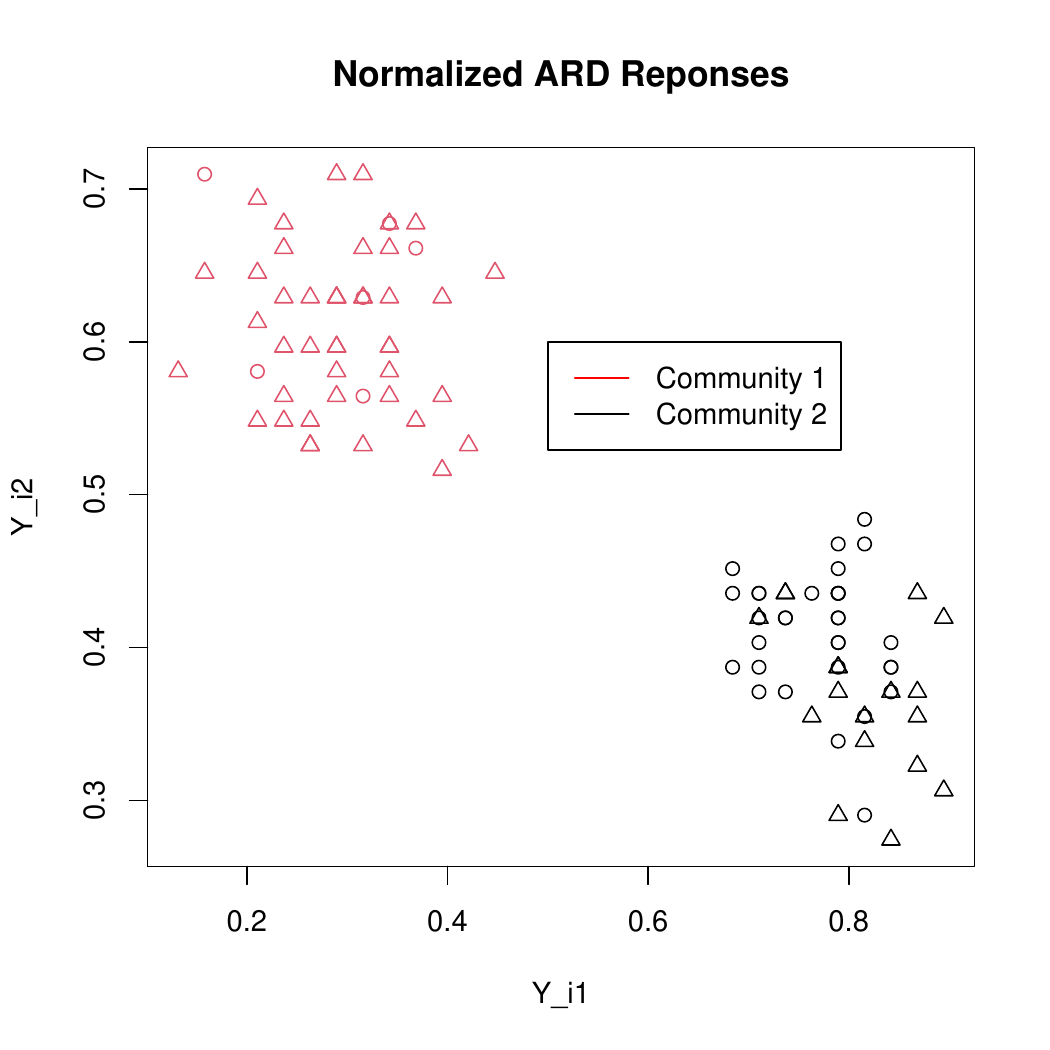}
\end{minipage}
\caption{Comparison of ARD responses in two different scenarios. On the left, we generate traits using the matrix $Q = \protect \begin{pmatrix}
    1/2 & 1/2 \\ 1/2 & 1/2
 \protect\end{pmatrix}. $ In this case, traits have no relationship with the community membership. In the left figure, we plot the normalized ARD responses, Here red indicates community 1, black indicates community 2, circles indicate trait 1, and triangles indicate trait 2. On the right, we repeat the simulation but using $Q = \protect \begin{pmatrix} 7/10 & 3/10 \\ 1/10 & 9/10 
 \protect \end{pmatrix}$. Here, there is a strong relationship between traits and community membership, and so K-means returns the correct clustering of the data.}
 \label{fig: clustering_ARD}
\end{figure}

We now give a brief overview of the proof of Theorem 2. The intuition is that the the ARD responses $\tilde y_i = (y_{i1}/n_1, \dotsc, y_{iC}/n_c)$ converge, by the weak law of large numbers, to $Z_i = (\tilde P_{i1}, \dotsc, \tilde P_{iC})$ at an exponentially fast rate in $n$. See Figure \ref{fig: clustering_ARD} for an illustration of this fact.  Therefore, two nodes in the same community will be classified together with probability going to 1, and since the by assumption the $Z_i$ are distinct, two nodes in different communities will eventually be classified into different communities. We want to emphasize again the differences between the problem we are studying here and classic clustering problems or community detection problems. Compared to classic clustering problems, in which the distribution of data does not change as the sample size grows, the data we are analyzing here, $y_{ik}/n_k$, is converging to its expectation at an exponentially fast rate. Therefore, as our sample size grows, it becomes easier to correctly cluster the ARD responses and therefore to correctly classify nodes into the right communities. Second, compared to more standard community detection problems, we do not observe the graph but instead observe ARD about the nodes \citep{bickelcl2011}. This ARD, because it is a sample average, converges exponentially fast to its mean, which allows us to perform fast community detection.

\begin{proof}[Proof of Theorem \ref{theorem: SBM_consistency}]
To begin, we pick a node randomly from $V$. Let $c_i$ denote its community membership. For any $j$, since $y_{jk}/n_k$ is a sum of (conditionally) independent random variables, by Hoeffding's inequality we have that $\mathbb{P}(|y_{jk}/n_k - p_{jk}| > \epsilon_n) \leq C\exp(-\epsilon_n^2 n)$ for some constant $C$. By recalling that $\tilde y_i = (y_{i1}/n_1, \dotsc, y_{iC}/n_C)$ is the normalized ARD response with mean $\tilde p_i = (\tilde P_{i1}, \dotsc, \tilde P_{iC})$, we can conclude by a union bound that
\begin{equation*}
 \mathbb{P}(\max_{j: c_j = c_i} ||\tilde y_{j} - \tilde p_{j}|| > \epsilon_n) \leq n C\exp(-\epsilon_n^2 n) \;.
\end{equation*}

By taking $\epsilon_n^2 = \log(n)/n$, we see that $\mathbb{P}(\max_{j: c_j = c_i} \mathbf{1}\{\hat c_j \neq \hat c_i\} > 0) \leq 1/n$.
In addition, since $\Delta > 0$, which gives us well-separated clusters, and $\epsilon_n \rightarrow 0$, we have that $\mathbb{P}(\max_{j: c_j \neq c_i} \mathbf{1}\{\hat c_j \neq \hat c_i\}) \rightarrow 1$ for any $j$ with $c_j \neq c_i$.  By definition of the classification algorithm, we can conclude that $\mathbb{P}(\max_{j: c_j = c_i} \mathbf{1}\{\hat c_j \neq \hat c_i\} > 0) \leq \mathbb{P}(\max_{j: c_j = c_i} ||\tilde y_{j} - \tilde p_{j}||) $.

Since the algorithm assigns nodes $j$ that are within $\epsilon_n$ away from $i$ into the same category, we see that the probability of any incorrect classification goes to zero for this community. The same argument applies to the second community, when looking at the set $V \setminus \hat C_i$. We then repeat this argument until all nodes are classified.

Given a consistent estimate of the community membership vector, it follows from the weak law of large numbers that $\hat Q, \hat P$, and $\hat \pi$ are consistent for $Q, \ P$ and $\pi$, where
\begin{equation*}
    \hat Q_{ck} = \frac{1}{m_c(n)} \sum_{i \in \hat C_c} \mathbf{1}\{t_i = k\} 
    \end{equation*}
    \begin{equation*}
    \hat P_{cc'} = \begin{cases}
    \frac{1}{m_c(n) m_{c'}(n)} \sum_{i \in \hat C_c} Y_{ic'}, \ \ \ c \neq c' \vspace{0.05in} 
    \\
       \frac{1}{m_c(n) (m_{c'}(n) - 1)} \sum_{i \in \hat C_c} Y_{ic'}, \ \ \ c = c' \;,
    \end{cases}
    \end{equation*}
    and $\hat \pi_c = \frac{1}{m_c(n)} \sum_{i =1}^n \mathbf{1}\{\hat c_i = c\}$, and $m_c(n)$ is the number of nodes that we estimate to be in community $c$ under the estimated community membership vector $\mathbf{\hat c}$.  
    
\end{proof}

\section{Consistency of Latent Space Model Parameters (Theorem \ref{thm: LS_consistency})}
\label{sec: define_estimates}
We now define the estimates of the node locations and the node effects. In the estimates provided below, we assume that we have estimates of the global parameters, which we denote by $\eta^\star = (\mu^\star_1, \dotsc, \mu^\star_K, \sigma^\star_1, \dotsc, \sigma^\star_K, E\{\exp(\nu^\star)\})$. In Section \ref{sec: estimate_global}, we provide estimates of $\eta^\star$ based on method-of-moment estimators.

Recall that the ARD data $y_{ik}$ satisfies $y_{ik} \mid  \nu^\star_i, z^\star_i, \eta^\star \sim \text{Binomial}(n_k, p_{ik})$ where $n_k$ is the size of group $k$ and $p_{ik}$, which we now define. With ARD data we do not observe any connections in the graph directly.  It is possible, though unlikely as long as the sample size is small compared to the population size when using simple random sampling, that we might observe an alter of one of the surveyed respondents.  That is, if person $i$ reports knowing 5 people named Michael, one of those people named Michael might also be in the survey.  Even in the unlikely event that this happens, we do not have access to this information through ARD since we do not observe any links.  When considering the Binomial representation, therefore, we are making a statement not about the connections between any two individuals (which we do not observe) but instead about marginal connections between a person and a population.  Respondent $i$ is almost certainly more likely to know some members of the group $k$ than others, but since ARD does not provide information on edges there is no way to specify that heterogeneity.  Instead, we focus on an aggregate summary of the relationship between respondent $i$ and members of group $k$ which does not differ between members of the group because ARD, unlike the complete graph, does not contain sufficient data to do so.  The power of our approach, however, is that, even under this limited information setting we still recover consistent estimates of model parameters.

Conditioned on node $i$'s effect $\nu^\star_i$ and latent space location $z^\star_i$, the probability node $i$ connects to an arbitrary node $j$ in group $k$, written as is $\mathbb{P}(g_{ij} = 1 \mid \nu_i^\star, z_i^\star, \eta^\star) := p_{ik}$,
\begin{equation}
\begin{aligned}
   p_{ik} &= \int_V \int_Z  \exp\{\nu^\star_i + \nu_j - d(z^\star_i, z_j)\} f_k(z_j) f_V(\nu_j) \ d\nu_j \ dz_j \label{eq: p_ik} \\
    &= \exp(\nu^\star_i) E\{\exp(\nu)\} \int_Z \exp\{-d(z_i^\star, z_j)\} f_k(z_j) \ dz_j \;.
\end{aligned}
\end{equation}
Here, we use the notation $\nu_i^\star$ to refer to a fixed but unknown parameter of interest, whereas $\nu_j$ represents a dummy variable that is integrated out.  Note here we have used the property that $\exp(a+b) = \exp(a)\exp(b)$. By assuming the link function is exponential, we can easily separate the terms in the expression for $\mathbb{P}(g_{ij} = 1\mid\nu_i, z_i,\eta)$. We believe we can extend these ideas to other link functions, as was done in \cite{Lubold}, but we leave that to future work.

We now motivate and then formally describe these method-of-moment estimators (or equivalently, Z-estimators). Since the ARD is Binomial, we can estimate $p_{ik}$ by equating $p_{ik}$ with $y_{ik}/n_k$. This then allows us to solve for the parameters $\nu_i$ and $z_i$ since $p_{ik}$ depends on these two parameters (and $\eta$, which we can consistently estimate). In total, we create two systems of equations (one for the node locations and one for the fixed effects). This section assumes that we know the true parameters $\eta^\star$, but in Section \ref{sec: estimate_global} we show how to estimate the parameters $\eta^\star$. 

We start with estimating the node locations.  To do this, we note that the ratio $y_{ik}/y_{ik'}$ converges in probability, by the weak law of large numbers, to the ratio 
\begin{equation*}
    {E_{\sigma_k}[\exp\{-d(z_i, z)\}]}/{E_{\sigma_{k'}}[\exp\{-d(z_i, z')\}]} \;,
\end{equation*}
which depends only on the variances of the distributions of node locations $\sigma_1, \dotsc, \sigma_K$ and the node location $z_i$, where we define the notation $E_{\sigma}[\exp\{-d(z, z_i)\}]$ to mean that the expectation is taken with respect to $\sigma$. Note that critically, in the ratio $p_{ik}/p_{ik'}$, the terms involving the node effects and $E\{\exp(\nu)\}$, which are all unknown at this point, cancel out. This is the reason we look at the ratio of two ARD responses. Here we also make the simplifying assumption that $n_k = n_{k'}$, although the results do not change significantly if we remove this assumption. This suggests that we should take our estimate of the node location, denoted by $\hat z_i$, to be the value of $z_i$ such that $y_{ik}/y_{ik'}$ is equal to the ratio ${E_{\sigma_k}[\exp\{-d(z_i, z)\}]}/{E_{\sigma_{k'}}[\exp\{-d(z_i, z')\}]}$. 

More formally, we define the function $G_1: \mathcal{M} \times (0, \infty)^2 \rightarrow \mathbb{R}$ by 
\begin{equation}
\label{eq: G_1}
    G_1(z_i; \sigma_k, \sigma_{k'}) = \frac{E_{\sigma_k}[\exp\{-d(z_i, z)\}]}{E_{\sigma_{k'}}[\exp\{-d(z_i, z')\}]} \;.
\end{equation}
We drop the dependence on $k$ and $k'$ for simplicity and just write $G_1$ without any mention of $k$ or $k'$. This function, when viewed as a function of $z_i$ for a fixed $\sigma_k, \sigma_{k'}$,  is not always invertible, but we can define a pseudo-inverse by $G_1^{-1}(x) = \{m \in \mathcal{M}: G_1(m) =  x\}$. In the following calculations, we will take the inverse to be chosen in a fixed way from this set. We discuss this condition further and give examples in Section~\ref{sec: discussion_assumption}. Our estimate of the node location, $\hat z_i$, solves $\log\{G_1(\hat z_i; \hat \sigma_k, \hat \sigma_{k'})\} = \log(y_{ik}/n_k) - \log(y_{ik'}/n_{k'})$ for two arbitrary and distinct entries $k, k'$. In practice, the user selects the values of $k$ and $k'.$ The user can estimate a location using each pair of indices $k \neq k'$. Taking an average (or the Fr{\'e}chet mean more generally) would improve the accuracy of the resulting estimate. Note that the log transformation simplifies the analysis of this estimator and allows us to use a proof technique that is similar to the one used to prove Theorem 1.3 in \cite{chatterjeeds2010} or Theorem 3 in \cite{ graham2014econometric}.

We now motivate our estimator of the the node effects. 
The idea is that ARD is a Binomial random variable and thus we can equate the probability of an edge between node $i$ and nodes in group $k$ (which depends on the node effect and the node location, which we have already estimated above) with the observed number of edges. We then solve for the node effect. To state this estimator more formally, define the function
\begin{equation*}
G_2(\nu_i, z_i) = E\{\exp(\nu)\} \exp(\nu_i) E[\exp\{-d(z_i, z)\}]  \;,
\end{equation*}
where here $z \sim F(\mu_k, \sigma_k^2)$. Since $y_{ik}/n_k$ converges in probability to $G_2(z^\star_i, \nu^\star_i)$, this motivates the following estimator
\begin{equation}
    \label{eq: def_hat_nu}
    \hat \nu_i = \log\left(\frac{y_{ik}}{n_k}\right) - \log(E[\exp\{-d(\hat z_i, z)\}]) - \log[\hat E\{\exp(\nu)\}] \;.
\end{equation}
where $z \sim F(\hat \mu_k, \hat \sigma_k)$ and the term $\log[\hat E\{\exp(\nu)\}]$ is the estimate of $\log[E\{\exp(\nu)\}]$ computed using $\hat \eta$. Again, as in the case of the node locations, the user can select the group index $k$ used in computing $\hat \nu_i.$ As in the case of the node location, we can compute $\hat \nu_i$ for all group indices $k$ and their average will be an improved estimate of $\nu_i^\star.$

In the next section, we prove Theorem \ref{thm: LS_consistency} in the main paper, which deals with showing that estimates of the node locations and node effects are consistent and satisfy a convergence rate of $\sqrt{3\log(\tilde n)/2 \tilde n}$ with probability at least $1 - O(m/\tilde n^3)$, where $\tilde n = n/K$ and $K$ is assumed to be fixed.   Our proof of Theorem \ref{thm: LS_consistency} is based on two separate lemmas: Lemma \ref{lemma: conc_ineq_z_i} proves the claimed convergence result for the node locations, and Lemma \ref{lemma: conc_ineq_nu_i} proves the claimed convergence result for the node effects.

  To begin with some notation, the estimates of the node locations and the node effects depend on the group parameters, which we denote by $\eta$. We let $\hat z_i(\eta)$ denote the estimate of $z_i^\star$ that is computed using the known and true $\eta$, and we let $\hat z_i(\hat \eta)$ denote the estimate based upon the plug-in estimate $\hat \eta$, which we define formally in Section \ref{sec: estimate_global}. 

\subsection{Proof of Theorem \ref{thm: LS_consistency}}
\label{sec: theorem_1_proof}

We now provide a proof of Theorem \ref{thm: LS_consistency} in the main text.  For clarity, we repeat the statement of the proof here along with the necessary assumptions.  The proof relies on consistent estimates of the global parameters.  For ease of exposition, we have moved the derivation of these estimates to the subsequent section.  We prove the result by constructing a series of Lemmas that, when combined, yield the desired result.  We begin by restating the necessary assumptions.  Additional discussion of the assumptions, including verification that they hold with distributional assumptions commonly used in practice is in Section~\ref{sec: discussion_assumption}.  Note that in the main part of the paper, the following four assumptions are labeled as Assumptions 2-5.

\begin{assumption}
\label{assumption: compact_param}
For each $k$, $\mu_k$ is in a compact subset of $\mathcal{M}^p(\kappa)$ and $\sigma_k$ is in a compact subset of $(0, \infty)$.
\end{assumption}

\begin{assumption}
\label{assumption: a_V}
The node effects $\nu_i^\star \overset{iid}{\sim} H$ satisfy   $E\{\exp(\nu^\star_i)\} < \infty.$
\end{assumption}

\begin{assumption}
\label{assumption: a_F}
The distribution $F$ is a symmetric distribution on $\mathcal{M}^{p}{(\kappa)}$ that is completely characterized by its mean and variance and satisfies the following two conditions. The function $z_i \mapsto E_k[\exp\{-d(z_i, z)\}]$ is Lipschitz for every $k \in \{1, \dotsc, K\}$ and $z_i \mapsto E_k[\exp\{-d(z_i, z)\}] / E_{k'}[\exp\{-d(z_i, z')\}]$ has a pseudo-inverse that is Lipschitz.
\end{assumption} 

 \begin{assumption}
 \label{assumption: lip_sigma}
Define $F_1: (z_i, \sigma_k, \sigma_{k'}) \mapsto E_k[\exp\{-d(z_i, z)\}] / E_{k'}[\exp\{-d(z_i, z')\}]$. The inverse function $F_1^{-1}$ is continuous in $\sigma$ and for every $k, k', \ell$, and $\ell'$, the following two functions are Lipschitz:
 \begin{equation*}
     \eta \mapsto \frac{E_{kk'}[\exp\{-d(z, z')\}]}{E_{\ell \ell'}[\exp\{-d(z, z')\}]}, \ \ \  \eta \mapsto  \frac{E_{kk'}[\{\exp(-d(z, z')\}]^2}{E_{\ell \ell'}[\{\exp(-d(z, z')\}]^2} \;.
 \end{equation*}
 \end{assumption}
 
Under the four assumptions above, we now restate Theorem \ref{thm: LS_consistency} in the main paper.

\begin{theorem}
Suppose Assumptions \ref{assumption: compact_param}, \ref{assumption: a_V}, \ref{assumption: a_F}, and \ref{assumption: lip_sigma} hold. The estimators $\hat z_i$ and $\hat
\nu_i$ and $\hat \eta$ are consistent for $z_i^\star, \nu_i^\star$, and $\eta^\star$ as $m, n \rightarrow \infty$, up to isometry on $\mathcal{M}^p(\kappa)$ and satisfy
\begin{align*}
    &\max_{1 \leq i \leq m(n)} d_{\mathcal{M}^p(\kappa)}(\hat z_i, z^\star_i) \leq \sqrt{\frac{3\log(\tilde n)}{2 \tilde n}}, \\
    &\max_{1 \leq i \leq m(n)} |\hat \nu_i - \nu^\star_i| \leq \sqrt{\frac{3\log(\tilde n)}{2 \tilde n}},
\end{align*}
with probability $1 - O(m/\tilde n^3)$.
\end{theorem}

\begin{proof}[Proof of Theorem \ref{thm: LS_consistency} in the main paper]
For readability, we split up the proof of Theorem \ref{thm: LS_consistency} in the main paper into several lemmas. Theorem \ref{thm: LS_consistency} claims a concentration inequality for the estimates of the node locations and node effects using the plug-in estimate $\hat \eta$ of the global parameters. We prove this result for the node locations (Lemma \ref{lemma: conc_ineq_z_i}) and for the node effects (Lemma \ref{lemma: conc_ineq_nu_i}) separately. These two lemmas require us to first prove the consistency (without a rate) on the estimates of node locations and effects, which we do in Lemma \ref{lemma: consistency_MLE}. The proofs of Lemmas \ref{lemma: conc_ineq_z_i} and \ref{lemma: conc_ineq_nu_i} are based on Lemmas \ref{lemma: z_i_error_known_eta} and \ref{lemma: conc_ineq_nu_i_known_sigma}, which prove the concentration inequalities using the true and unknown group parameter $\eta$. Combining the arguments in these lemmas proves the desired result.

\end{proof}

Our proof of Theorem \ref{thm: LS_consistency} starts with the following lemma, which states the estimates that maximize the pseudo-likelihood of the ARD are consistent as $m, n \rightarrow \infty$. We use this result later on to prove Theorem \ref{thm: LS_consistency}. We would like to emphasize that maximizing the pseudo likelihood, which we do in Section \ref{sec: add_lemma}, is equivalent to a method-of-moments estimator in this case. 

\begin{lemma}
\label{lemma: consistency_MLE}
Let the assumptions from Theorem \ref{thm: LS_consistency} of the main paper hold. Suppose that we have consistent estimates of the group parameters $\eta$, denoted by $\hat \eta$. Now suppose that $(\hat \nu_{1:m}, \hat z_{1:m})$ are the Z-estimators of the node effects and locations described in Section \ref{sec: define_estimates}. Then, $(\hat \nu_{1:m}, \hat z_{1:m})$ are consistent for $\nu^\star_{[1:m]}$ and $z^\star_{[1:m]}$ as $m, n \rightarrow \infty$, up to an isometry on $\mathcal{M}^p(\kappa).$

\end{lemma}

For readability, we have moved the proof of Lemma \ref{lemma: consistency_MLE} to Section \ref{sec: add_lemma}. The main idea of the proof follows the standard M-estimator consistency steps: showing a well-separated extremum and a uniform law of large numbers \citep{Vdv}.

\begin{lemma}
\label{lemma: conc_ineq_z_i}
With probability at least $1 - O(m/\tilde n^3)$, the following inequality holds up to isometry on $\mathcal{M}^p(\kappa)$.
\begin{equation*}
    \max_{1 \leq i \leq m(n)} d_{\mathcal{M}}(\hat z_i(\hat \eta), z^\star_i) \leq \sqrt{\frac{3\log(\tilde n)}{2 \tilde n}} \;.
\end{equation*}
\end{lemma}
\begin{proof}
By the triangle inequality,
\begin{equation}
\label{eq: triangle_z_i}
d_{\mathcal{M}}(\hat z_i(\hat \eta), z^\star_i) \leq     d_{\mathcal{M}}(\hat z_i(\hat \eta), \hat z_i(\eta)) + d_{\mathcal{M}}(\hat z_i(\eta), z_i^\star).
\end{equation}

We have two terms in the triangle inequality. We will only have to focus on the second one, because that will dominate the rate as we will soon show. We calculate that one below. The first one has an extremely fast rate as it tends to zero. This can be seen in a straightforward manner from using a Taylor expansion of the estimating equation in the usual way, because the estimating equation consists of an average taken over all pairs of groups and all pairs of potential links across every pair of group which gives order $O_P(1/\sqrt{K^2 m n})$, where again $m$ is the size of the ARD sample.  We will show later that this rate is much faster than the rate for the second term in the inequality, which means this term can be ignored when proving the rate of convergence on the term $d_{\mathcal{M}}(\hat z_i(\hat \eta), z^\star_i)$.

We now study the second term in the triangle inequality above. Now, using the definition of $\hat z_i(\eta)$ as $\hat z_i = G_1^{-1}(a;\hat \eta)$, we write 
\begin{equation*}
    d_{\mathcal{M}}(\hat z_i(\hat \eta), \hat z_i(\eta)) = d_{\mathcal{M}}(G_1^{-1}(a; \hat \eta), G_1^{-1}(a;  \hat \eta)) 
\end{equation*}
where $a = \log(y_{ik}/n_k) - \log(y_{ik'}/n_{k'}).$

Supposing that $G_1^{-1}(a; \sigma)$ is continuous in $\sigma$, which we assume in Theorem \ref{thm: LS_consistency} in the main paper,  we combine
Lemma \ref{lemma: consistencty_sigma} with the continuous mapping theorem to show that $d_{\mathcal{M}}(\hat z_i(\hat \eta), \hat z_i(\eta))$ converges to zero in probability. All we need to do now is show that the second term in (\ref{eq: triangle_z_i}) satisfies the claimed concentration inequality.  By Lemma \ref{lemma: z_i_error_known_eta}, which we state below, with probability at least $1 - O(1/n_k^3)$, 
\begin{equation*}
    d_{\mathcal{M}}(\hat z_i(\eta), z^\star_i)  \leq \sqrt{\frac{3\log(\tilde n)}{2\tilde n}} \;,
\end{equation*}
up to isometry on $\mathcal{M}$.  By a union bound,  and by recalling (\ref{eq: triangle_z_i}), we conclude that with probability at least $1 - O(m/\tilde n^3)$:
\begin{equation*}
   \max_{1 \leq i \leq m(n)} d_{\mathcal{M}}(\hat z_i(\hat \eta), z^\star_i) \leq \sqrt{\frac{3\log(\tilde n)}{2\tilde n}} 
\end{equation*}
up to isometry on $\mathcal{M}$. 
\end{proof}

The next lemma shows that the estimate of $\nu_i$, based on the plug-in estimate $\hat \eta$, satisfies a similar concentration inequality.  
 \begin{lemma}
\label{lemma: conc_ineq_nu_i}
The estimator $\hat \nu_i$ from (\ref{eq: def_hat_nu}) satisfies the following: With probability $1 - O(m/\tilde n^3)$, 
\begin{equation*}
    \max_{1 \leq i \leq m(n)} |\hat \nu_i(\hat \eta) - \nu^\star_i| \leq \sqrt{\frac{3\log(\tilde n)}{2\tilde n}} \;.
\end{equation*}
\end{lemma}
\begin{proof}
The proof follows the same argument that we used in the proof of Lemma \ref{lemma: conc_ineq_z_i}.  Since $\hat \eta$ is consistent for $\eta$, the second term in the definition of $\hat \nu_i$ can be ignored when proving the desired concentration inequality (again, this argument was used in the proof of Theorem 3 in \cite{graham2014econometric}). It therefore suffices to just argue that the term $\log(y_{ik}/n_k)$ satisfies the claimed concentration inequality. We can prove this inequality by Hoeffding's inequality. See Lemma \ref{lemma: conc_ineq_nu_i_known_sigma}, which proves this formally. Taking a union bound over all $i = 1, \dotsc, m(n)$ to proves the desired result.
\end{proof}

In the case where $d(z_i, z_j) = 0$ (only node effects determine connection propensity) and $m = n$ (meaning that we observe the entire graph and not just the ARD), then Theorem \ref{thm: LS_consistency} of the main paper simplifies to Theorem 3.3 of \cite{chatterjeeds2010}.

\begin{lemma}
\label{lemma: z_i_error_known_eta}
With probability at least $1 - O(m/\tilde n^3)$, the following inequality holds:
\begin{equation*}
    \max_{1 \leq i \leq m(n)} d_{\mathcal{M}}(\hat z_i(\eta), z^\star_i) \leq \sqrt{\frac{3\log(\tilde n)}{2\tilde n}} \;.
\end{equation*}
\end{lemma}
The proof is based on similar ideas found in \cite{chatterjeeds2010, graham2014econometric}. The intuition behind the proof is as follows. The estimator $\hat z_i(\eta)$ is based on the ARD $y_{ik}/n_k = 1/n_k \sum_{j \in G_k} g_{ij}$, which converges exponentially fast to $p_{ik}$ by Hoeffding's inequality.  This insight allows us to conclude the uniform control over the error in $\hat z_i(\eta).$ 
\begin{proof}
To begin, we recall that the estimator is $\hat z_i = G_1^{-1}(y_{ik}/n_k ; \eta)$. This function will not be invertible, but we can choose a representative from the set of $\{x: G_1(x; \eta) = y_{ik}/n_k\}$. Any choice will lead to the right answer, up to isometry. Note also that because of properties of $\mathcal{M}^p(\kappa)$, it is locally Euclidean. See \cite{Lubold} and its references for a more complete description of this point.  Since $\hat z_i(\hat \eta)$ converges to $z_i(\eta)$, up to isometry, we therefore only need to prove the argument for the Euclidean case (this follows from Lemma \ref{lemma: consistency_MLE}). The extension to the spherical and hyperbolic geometries follows since there is a neighborhood around $z_i$ in which the distances are approximately Euclidean distances, and thus the Euclidean arguments apply here too.

Since
\begin{equation*}
    a= \log(y_{ik} / n_k) - \log(y_{ik'}/n_{k'}) 
\end{equation*}
converges in probability, as $n \rightarrow \infty$, to $G_1(z_i)$, this motivates our estimate of $z_i$. We set $\hat z_i = G_1^{-1}(a).$ See Section \ref{sec: discussion_assumption} for a discussion on this inverse function. Since $G_1^{-1}\{\log(p_{ik}) - \log(p_{ik'})\} = z^\star_i$,
\begin{align*}
    \left\Vert \hat z_i(\eta) - z^\star_i \right\Vert &= \left\Vert G_1^{-1}(a) - G_1^{-1}\{\log(p_{ik}) - \log(p_{ik'})\} \right\Vert  \\
    &\leq C |\log(y_{ik}/n_k) - \log(y_{ik'}/n_{k'}) - \log(p_{ik}) - \log(p_{ik'})| \\
    &\leq \tilde C_n \{|y_{ik}/n_k - p_{ik}| +  |y_{ik'}/n_{k'} - p_{ik'}| \} \;.
\end{align*} 
for some sequence of constants $\tilde C_n$. We know that $\tilde C_n$ is on the order $n_k$ = $O(n)$ when $K$ is fixed (which we assume), since $x \mapsto \log(x)$ is Lipschitz on any interval $[a', b']$ with Lipschitz constant $1/a'$. In our case, with probability going to 1, $y_{ik} \geq 1$ and so $y_{ik}/n_k \geq 1/n_k$ and thus we can take $1/(1/n_k) = n_k$ to be the Lipschitz constant. We thus conclude that
\begin{equation}
\label{eq: prob_bound_zi}
    \mathbb{P}(\left\Vert \hat z_i(\eta) - z^\star_i\right \Vert > \epsilon) \leq \mathbb{P}
\left(\left|\frac{y_{ik}}{n_k} - p_{ik}\right| > \epsilon/\tilde C_n\right) + \mathbb{P}\left(\left|\frac{y_{ik'}}{n_{k'}} - p_{ik'}\right| > \epsilon/\tilde C_n\right) \;.
\end{equation}
We now show that both terms on the right hand side converge to zero exponentially fast. Since $y_{ik}$ is a sum of independent Bernoulli random variables, each with expectation $p_{ik}$, by Hoeffding's inequality \citep{Hoeffding1956}, 
\begin{equation*}
   \mathbb{P}\left(\left|\frac{y_{ik}}{n_k} - p_{ik}\right| > \epsilon/\tilde C_n\right) \leq 2 \exp\left(-2\frac{\epsilon^2 n_k}{\tilde C_n^2} \right) \;.
\end{equation*}
Set $\epsilon^2 = \frac{3}{2}n_k^{-1} \tilde C_n^2 \log(n_k) = O(\frac{3}{2} n_k^{-1} n_k^2 \log(n_k)$). Then, 
\begin{equation*}
   \mathbb{P}\left(\left|\frac{y_{ik}}{n_k} - p_{ik}\right| > \sqrt{\frac{3\log(\tilde n)}{2\tilde n}}\right) \leq 2 \exp\left\{-3 \log(n_k)\right\} = 2/n_k^3 \;.
\end{equation*}
Similarly, $ \mathbb{P}\left(\left|\frac{y_{ik'}}{n_{k'}} - p_{ik'}\right| > \sqrt{\frac{3\log(\tilde n)}{2\tilde n}}\right) \leq 2/n_k^3$. Putting this together, and recalling (\ref{eq: prob_bound_zi}), we see that 
\begin{equation*}
    \mathbb{P}\left(\left\Vert \hat z_i(\eta) - z^\star_i \right\Vert  > \sqrt{\frac{3\log(\tilde n)}{2\tilde n}}\right) \leq 4/n_k^3 \;.
\end{equation*}
By a union bound, with probability at least $1 - 4m /n_k^3$,
\begin{equation*}
    \max_{1 \leq i \leq m} \left\Vert \hat z_i(\eta) - z^\star_i\right\Vert  < \sqrt{\frac{3\log(\tilde n)}{2\tilde n}} \;.
\end{equation*}
\end{proof}

In the following lemma, we prove that the estimate $\hat \nu_i$ satisfies a similar type of concentration inequality. The proof is identical to the one given above, so we omit the details. 

\begin{lemma}
\label{lemma: conc_ineq_nu_i_known_sigma}
If each $z_i$ is known, and the global parameter $\eta$ is known,  the estimator $\hat \nu_i$ defined in (\ref{eq: def_hat_nu}) satisfies the following: With probability at least $1 - O(m/\tilde n^3)$, 
\begin{equation*}
    \max_{1 \leq i \leq m(n)} |\hat \nu_i(\eta) - \nu_i| \leq \sqrt{\frac{3\log(\tilde n)}{2\tilde n}} \;.
\end{equation*}
\end{lemma}

\subsection{Estimating Global Parameters in Latent Space Model}
\label{sec: estimate_global}
In this section, we provide estimates of the model parameters $\eta$.  Our discussion comes in three parts. We first show how to estimate the within-group variance terms. To estimate the within-group variances, we equate the ARD responses of people in a group $k$ to other nodes in the same group $k$ with the probability that an arbitrary edge exists between nodes in group $k$. Since this probability depends on only the within-group variance, as all nodes from a given group are distributed about the same group center, we can therefore estimate the group variance in this way.

To formally define our estimator, fix two groups $G_k$ and $G_{k'}$. The probability that an arbitrary node in group $k$ connects to other nodes in group $k$ is equal to, after integrating out all the parameters, $E\{\exp(\nu)\}^2 E_{kk}[\exp\{-d(z, z')\}]$, where $z, z'$ are independent and $z, z' \sim F(\mu^\star_k, \sigma^\star_k)$. Note critically that this does not upon the mean parameter $\mu^\star_k$.  

We let $m_k(n)$ be the number of nodes we sample that belong to group $k$. We define the quantity 
\begin{equation}
\label{eq: t_k}
    t_{kk'} = \frac{1}{m_k(n)} \sum_{i \in G_k} \frac{y_{ik'}}{n_{k'}} \;. 
\end{equation}
Then, for large $n$ (which implies that $|G_k| = n_k$ and $m_k(n)$ is large too), the ratio $t_{k}/t_{k'}$ converges in probability to 
\begin{equation}\label{eq: motiv_sigma_estimator}
     \frac{E\{\exp(\nu)\}^2 E_{kk}[\exp\{-d(z, z')\}]}{E\{\exp(\nu)\}^2 E_{k'k'}[\exp\{-d(z, z')\}]} = \frac{E_{kk}[\exp\{-d(z, z')\}]}{E_{k'k'}[\exp\{-d(z, z')\}]} \;.
\end{equation}
which depends again on just the unknown variance terms $\sigma_k^\star$ and $\sigma_{k'}^\star.$ In other words, by looking at the ratio $t_k / t_{k'}$, the term $E(\exp(\nu))^2$, which we have not yet estimated and do not know in practice, cancels. So this ratio depends only on the unknown variance vector $(\sigma^\star_1, \dotsc, \sigma_\star^2)$. Motivated by this description, we define an estimator $\hat \sigma^2(n) = \{\hat \sigma_1^2(n), \dotsc, \hat \sigma^2_K(n)\}$ as the root of the following system of equations
\begin{equation}
\label{eq: sigma_equations}
  \frac{ t_{kk}}{ t_{k'k'} } =  \frac{E_{kk}[\exp\{-d(z, z')\}]}{E_{k'k'}[\exp\{-d(z, z')\}]}\;.   
\end{equation}

If $K$ is large enough to ensure the above solution has a unique zero in the limit as $m, n \rightarrow \infty$,
this estimator is consistent for the true $(\sigma^\star_1, \dotsc, \sigma^\star_K)$.
\begin{lemma}
\label{lemma: consistencty_sigma}
The estimator $\hat \sigma^2(n) = \{\hat \sigma_1^2(n), \dotsc, \hat \sigma^2_K(n)\}$ that is the root of the system from (\ref{eq: sigma_equations}) is consistent as $n \rightarrow \infty$.
\end{lemma}
\begin{proof}
We first sketch an outline of our argument. We will define a sequence of random functions $\hat H_n$ such that $\lim_n E\{\hat H_n(\sigma^2)\} = 0$ only at the true $\sigma^\star$. This sequence of functions $\hat H_n$ is defined such that the estimator from the lemma minimizes this expression. Thus, to show consistency of the estimator, we can simply verify the two conditions from Theorem 5.7 of \cite{Vdv}, which for completeness we give in Section \ref{sec: add_lemma}. At a high level, Condition 1 requires that $H$ have a well-separated zero, and Condition 2 requires that $\hat H_n$ converge uniformly to $H$. Once we verify these two conditions, we can then conclude from Theorem 5.7 of \cite{Vdv} the desired consistency result.

By recalling the definition of $t_k$ in (\ref{eq: t_k}), we define the sequence of random functions $\hat H_n: (0, \infty)^K \rightarrow [0, \infty)$ by 
\begin{equation*}
    \hat H_n(\sigma^2) = \sum_{k = 1}^K \sum_{k' = 1}^K \left\{\frac{t_{kk}}{t_{k'k'} } - \frac{E_{kk}[\exp\{-d(z, z')\}]}{E_{k'k'}[\exp\{-d(z, z')\}]}\right\}^2 \;.   
\end{equation*}
We then define $H_n(\sigma^2) = E\{\hat H_n(\sigma^2)\}$ and $H(\sigma^2) = \lim_{n \rightarrow \infty} H_n(\sigma^2)$. By (\ref{eq: motiv_sigma_estimator}) and using the weak law of large numbers, combined with the continuous mapping theorem, it is clear that $H$ evaluated at the true $\sigma^2$ is zero. For sufficiently large $K$, this zero is unique, by using the same argument that we give in Lemma \ref{lemma: unique_max_log_lik} or by using Theorem 3 of \cite{breza2017using}. So Condition \ref{cond: well_sep} is satisfied. 

We now prove Condition \ref{cond: ULLN}. Recall that our goal is to show that 
\begin{equation*}
    \sup_{\sigma^2 \in S} |\hat H_n(\sigma^2) - H(\sigma^2)| \overset{p}{\rightarrow} 0 
\end{equation*}
It suffices to show that $\sup_{\sigma^2 \in S} |\hat H_n(\sigma^2) - H_n(\sigma^2)| = o_P(1)$, because $H_n$ converges uniformly to $H$ deterministically and hence also in probability. To show \emph{this} uniform law of large numbers, we will use Corollary 2.1 of \cite{Newey_1989}. For completeness, we provide this corollary in Section \ref{sec: add_lemma}. The pointwise convergence is automatically satisfied, by recalling (\ref{eq: motiv_sigma_estimator}). 
We now fix a $k, k'$ and expand inside the double sum in the expression for $\hat H_n$ as
\begin{equation*}
    \frac{t_{kk}}{t_{k'k'}} - 2  \frac{t_{kk}}{t_{k'k'}} \frac{E_{kk}[\exp\{-d(z, z')\}]}{E_{k'k'}[\exp\{-d(z, z')\}]} + \frac{E_{kk}[\exp\{-d(z, z')\}]^2}{E_{kk}[\exp\{-d(z, z')\}]^2} \;.
\end{equation*}
By comparing the terms inside the expression $| \hat H_n(\sigma^2) - \hat H_n(\tilde \sigma^2)|$, we see that there are just two terms to consider. To show the Lipschitz condition required to use Corollary 2.1 of \cite{Newey_1989}, let $\sigma, \tilde \sigma \in S \subseteq (0, \infty)^K$. To simplify the notation, we let $E_{kk}[\exp\{-d(z, z')\}]$ denote the expectation using the variance vector $\sigma$ and $\tilde E_{kk}[\exp\{-d(z, z')\}]$ to denote the expectation using the variance $\tilde \sigma$.

By assumption, the first term satisfies
\begin{equation*}
    2 \frac{t_{kk}}{t_{k'k'}} \left|\frac{E_{kk}[\exp\{-d(z, z')\}]}{E_{k'k'}[\exp\{-d(z, z')\}]} - \frac{\tilde E_{kk}[\exp\{-d(z, z')\}]}{\tilde E_{k'k'}[\exp\{-d(z, z')\}]}\right| \leq C \frac{t_{kk}}{t_{k'k'}} ||\sigma^2 - \tilde \sigma^2|| \;.
\end{equation*}
By assumption, the second term satisfies a similar Lipschitz condition:
\begin{equation*}
   \left|\frac{E_{kk}[\exp\{-d(z, z')\}]^2}{E_{k'k'}[\exp\{-d(z, z')\}]^2} - \frac{\tilde E_{kk}[\exp\{-d(z, z')\}]^2}{\tilde E_{k'k'}[\exp\{-d(z, z')\}]^2}\right| \leq C' ||\sigma^2 - \tilde \sigma^2||
\end{equation*}
Putting this all together, we see that
\begin{equation*}
    |\hat H_n(\sigma^2) - \hat H_n(\tilde \sigma^2)| \leq \sum_{k, k'} (C\frac{t_{kk}}{t_{k'k'}} + C')||\sigma^2 - \tilde \sigma^2|| \;. 
\end{equation*}
Since $\sum_{k, k'} E(C {t_{kk}}/{t_{k'k'}} + C') = O(1)$, we conclude by Corollary 2.1 of \cite{Newey_1989} that Condition 2 holds. By Theorem 5.7 of \cite{Vdv}, we conclude the consistency claim in the theorem. 

\subsubsection{Estimating Group Means}
\label{sec: group_means}
In this section, we show how to use the consistent estimates of the within-group variances $\sigma^\star_1, \dotsc, \sigma^\star_K$ to estimate the group mean parameters. Motivated by the same approach we used to prove consistency of $\sigma^\star_1, \dotsc, \sigma^\star_K$, consider now four group centers. The probability that nodes in the first two groups, say $k$ and $k'$ connect, divided by the probability that nodes in the last two groups, say $\ell$ and $\ell'$, connect is
\begin{equation*}
    \frac{E\{\exp(\nu)\}^2 E_{kk'}[\exp\{-d(z, z')\}]}{E\{\exp(\nu)\}^2 E_{\ell \ell'}[\exp\{-d(z, z')\}]} = \frac{E_{kk'}[\exp\{-d(z, z')\}]}{ E_{\ell \ell'}[\exp\{-d(z, z')\}]}  \;.
\end{equation*}
Having estimated the within-group variances terms, and noting that ${t_{kk'}}/{t_{\ell \ell'}}$ estimates the probability above, we can estimate the terms $\mu^\star_1, \dotsc, \mu^\star_K$ by solving the following system of equations: for every 4-tuple $(k, k', \ell, \ell')$ with distinct entries,
\begin{equation}
\label{eq: mu_system}
    \frac{t_{kk'}}{t_{\ell \ell'}} =  \frac{ E_{kk'}[\exp\{-d(z, z')\}]}{E_{\ell \ell'}[\exp\{-d(z, z')\}]} \;.
\end{equation}
The following lemma shows that this estimator is consistent as $n \rightarrow \infty$. 
\begin{lemma}
Let $\hat \mu_1(n), \dotsc, \hat \mu_K(n)$ be a root of the system in (\ref{eq: mu_system}). This estimator is consistent as $n \rightarrow \infty$, up to an isometry on $\mathcal{M}.$
\end{lemma}
\begin{proof}
The proof is nearly identical to the one given for Lemma \ref{lemma: consistencty_sigma}, so we only sketch the argument.  We define the sequence of random functions 
\begin{equation*}
    \hat H_n(\mu) = \sum_{k,k, \ell, \ell'}  \left\{\frac{t_{kk'}}{t_{\ell \ell'}} - \frac{  E_{kk'}[\exp\{-d(z, z')\}]}{ E_{\ell \ell'}[\exp\{-d(z, z')\}]} \right\}^2
\end{equation*}
We also define $H_n(\mu) = E\{\hat H_n(\mu)\}$ and $H(\mu) = \lim_{n \rightarrow} H_n$. At the true $\mu^\star$ parameter, $H(\mu^\star) = 0$ for sufficiently large $K$. For sufficiently large $K$, this is the only zero, up to an isometry on $\mathcal{M}$. (Again, by using the same argument that we give in Lemma \ref{lemma: unique_max_log_lik} or by using Theorem 3 of \cite{breza2017using}.) Thus, Condition 1 is satisfied. To show Condition 2, we use the same argument as we give in the proof of Lemma \ref{lemma: consistencty_sigma}. By assumption, we know that Condition 2 holds. Thus, by Theorem 5.7 of \cite{Vdv}, we can conclude the desired consistency result.
\end{proof}

\subsubsection{Estimating Node Effect Expectation}
\label{sec: node_effects}
In the previous two sections, we have shown how to obtain consistent estimates of the within-group variances and the group means. In this section, we show how to estimate the term $\tau = E[\{\exp(\nu)\}^2]$. The probability that any node in group $k$ connects with any node in group $k'$ is, after integrating out all parameters, 
\begin{equation}
\label{eq: prob_node_effect}
   E[\{\exp(\nu)\}^2] E_{kk'}[\exp\{-d(z, z')\}]\;,
\end{equation}
where $z \sim F(\mu^\star_k, \sigma^\star_k)$ and $z' \sim F(\mu^\star_{k'}, \sigma^\star_{k'})$.
By drawing $\hat z \sim F(\hat \mu_k, \hat \sigma_k)$ independently of $\hat z' \sim F(\hat \mu_{k'}, \hat \sigma_k)$, we can use $E_{kk'}[\exp\{-d(\hat z, \hat z')\}]$ to estimate the quantity $E_{kk'}[\exp\{-d(z, z')\}]$. Since 
\begin{equation*}
    t_{kk'} = \frac{1}{n_k} \sum_{i \in G_k} \frac{y_{ik'}}{ n_{k'}}
\end{equation*}
converges in probability to the expression in (\ref{eq: prob_node_effect}), we can estimate $E[\{\exp(\nu)\}^2]$ by
\begin{equation*}
    \hat \tau = \frac{t_{kk'}}{E_{kk'}[\exp\{-d(\hat z, \hat z')\}]}.
\end{equation*}
where $\hat z \sim F(\hat \mu_k, \hat \sigma_k)$ independently of $\hat z' \sim F(\hat \mu_{k'}, \hat \sigma_k)$. By the continuous mapping theorem and by recalling (\ref{eq: prob_node_effect}), we can consistently estimate $\tau$.
\end{proof}

\subsection{Discussion of Assumptions for Theorem \ref{thm: LS_consistency}}
\label{sec: discussion_assumption}
In this section we discuss two of the assumptions made in the main paper and discuss when these hold.

The $p$-dimensional normal distribution in $\mathbb{R}^p$ and the von-Mises Fisher distribution on the $p$-sphere are two models commonly used in the literature. We now argue that these two model satisfy this assumption. Recall that the term in question, in the case of a $p$-dimensional Gaussian distribution, is 
\begin{equation*}
    z_i \mapsto \int_{\mathbb{R}^p} \exp(-||z_i - z||) f(z) dz \;,  
\end{equation*}
where $f$ here is the pdf of the $p$-dimensional Gaussian distribution.  Note that $z \mapsto d(z_i, z)$ is Lipschitz, and $x \mapsto \exp(-x)$ is Lipschitz over $[0, \infty)$, and thus since $\exp(-x)$ is bounded by 1 on $(0, \infty)$, we conclude that $z_i \mapsto \exp\{-d(z_i, z)\}$ is Lipschitz. Because the integral of a Lipschitz function is again Lipschitz, we conclude that the assumption holds. 

We now look at the assumption that the inverse of the function $z_i \mapsto G_1(z)$ is invertible, 
where $G_1$ is defined in (\ref{eq: G_1}).  To begin the discussion, recall the simulation exercise in Figure \ref{fig: error_zi}. There are two group centers at $(2, 2)$ and $(-2, -2)$ in $\mathbb{R}^2$. The point we wish to estimate is at (0, 0), so the distance between each group center and this point is $2 \sqrt{2}$. There is a unique point in $\mathbb{R}^2$ that satisfies this constraint. However, consider the following two examples. 
\begin{example}
Consider two group centers at $(2, 2)$ and $(-2, -2)$ in $\mathbb{R}^2$. Suppose the point of interest $z_i$ is 2 unit away from the first point and 2 away from the second point. Then, the points $(2, -2)$ and $(-2, 2)$ will both solve the expression $F(z) = \log(p_{ik}) - \log(p_{ik'})$, where $p_{ik}$ depends on the distance between $z_i$ and the group centers. 
\end{example}
\begin{example}
Now let $\mathcal{M}^p(\kappa) = S^1(1)$, the circle with radius 1. Set two group centers at $(0, 1)$ and $(-1, 0)$ and suppose that the point of interest is $\pi/2$ away from the first group center and $3\pi/2$ away from the second group center. Then there are two points at $(0,1)$ and $(0, -1)$ that solve the expression $F(z) = \log(p_{ik}) - \log(p_{ik'})$, where $p_{ik}$ depends on the distance between $z_i$ and the group centers. 
\end{example}

The discussion above highlights the fact that the mapping $z \mapsto G_1(z)$ might not be invertible. We therefore suggest that the user select a representative element of the pseudo-inverse (hence our language in the main part of the paper).  

We now turn to discussing Assumption \ref{assumption: lip_sigma}.  We show that under mild distributional assumptions, the function $\sigma \mapsto \frac{E_{kk'}[\exp\{-d(z, z')\}]}{E_{\ell \ell'}[\exp\{-d(z, z')\}]}$ is Lipschitz. The discussion of the function $\mu \mapsto\frac{E_{kk'}[\exp\{-d(z, z')\}]}{E_{\ell \ell'}[\exp\{-d(z, z')\}]}$ is very similar. Suppose first that the function $\sigma_k \mapsto E[\exp\{-d(z_i, z)\}]$ is Lipschitz. Then, suppose that $g : (\sigma_k, \sigma_{k'}) \mapsto E(\exp\{-d(z_i, z)\}) /  E(\exp\{-d(z_i, z')\})$ is differentiable. It then has a gradient $\nabla g = (\nabla_k g, \nabla_{k'} g)$, where 
\begin{equation*}
    \nabla_k g =  \frac{\partial g}{\partial \sigma_k} = \frac{d}{d\sigma_k} E[\exp\{-d(z_i, z)\}]] / E[\exp\{-d(z_i, z')\}]]
\end{equation*}
Supposing that $E[\exp\{-d(z_i, z)\}]]$ is bounded away from zero, then this partial derivative is bounded because we assumed that the function $\sigma_k \mapsto E[\exp\{-d(z_i, z)\}]]$ is Lipschitz. The other partial derivative is given by 
\begin{equation*}
   \frac{\partial g}{\partial \sigma_{k'}} =  E[\exp\{-d(z_i, z)\}]] / \frac{d}{d\sigma_{k'}} E[\exp\{-d(z_i, z')\}]]
\end{equation*}
Supposing that the function $\sigma_{k'} \mapsto E[\exp\{-d(z_i, z')\}]]$ has a derivative that is bounded away from zero, we can thus conclude that $g$ is Lipschitz since each of its partial derivatives is bounded.

We now verify when the function $\sigma_k \mapsto E[\exp\{-d(z_i, z)\}]$ is Lipschitz. This function is given by 
\begin{equation*}
    \sigma_k \mapsto \int_{\mathcal{M}} \exp\{-d(z_i, z)\} f_k(\mu_k, \sigma_k) dz \;.
\end{equation*}
Supposing that $\sigma_k \mapsto f_k(\mu_k, \sigma_k)$ is Lipschitz, then we can use the Leibnitz rule (which allows us to pass the derivative inside the integral) to conclude that the function $\sigma_k \mapsto E[\exp\{-d(z_i, z)\}]$ is Lipschitz. By explicitly calculating the derivative of this expression in the cause of a Gaussian distribution, we see that $\sigma_k \mapsto f_k(\mu_k, \sigma_k)$ is Lipschitz. Since by assumption, each $\sigma_k$ is in a compact (and hence bounded subset of $(0, \infty)$), we can conclude that for each $z_i$, $\frac{\partial g}{\partial \sigma_k}$ is bounded. To show this, we need to show that $\frac{d}{d\sigma_{k'}} E[\exp\{-d(z_i, z')\}]]$ is bounded away from zero, which for a fixed $z_i$ is true because the $\sigma_k$ are by assumption in a compact subset of $(0, \infty). $ A similar argument applies to the function $\eta \mapsto \frac{E_{kk'}[\exp\{-d(z, z')\}]^2}{E_{\ell \ell'}[\exp\{-d(z, z')\}]^2}$.

\section{Consistency of plug-in estimator $E\{S_i(g_n) \mid \hat \theta_n(\mathbf{y}\}$ for $S_i(g^\star_n)$ (Theorem \ref{thm: ARD_consistency})}
\label{sec: proof_plugin}

\begin{proof}[Proof of Theorem \ref{thm: ARD_consistency}]
By the triangle inequality, 
\begin{equation*}
|E\{S_i(g_n) \mid \hat \theta_n(\mathbf{y})\} - S_i(g^\star_n)| \leq |E\{S_i(g_n) \mid \hat \theta_n(\mathbf{y})\} - E\{S_i(g_n) \mid \theta_n\}| + |E\{S_i(g_n) \mid \theta_n\} - S_i(g_n^\star)| \;.
\end{equation*}
By Condition 2 of Theorem \ref{thm: ARD_consistency}, $ |E\{S_i(g_n) \mid \theta_n\} - S_i(g_n^\star)| = o_P(1)$. We now analyze the other term. Under Condition 3, the function $\theta_n \mapsto E\{S_i(g_n) \mid \theta_n\}$ is differentiable, so by the mean value theorem, there exists a sequence of intermediate values $\bar \theta_n$ such that 
\begin{equation*}
E\{S_i(g_n) \mid \hat \theta_n(\mathbf{y})\} = E\{S_i(g_n) \mid \theta_n\} + \nabla E_{\bar \theta_n} \cdot (E_{\hat \theta_n} - E_{\theta_n}) \;.
\end{equation*}
By re-arranging, we see that
\begin{align*}
|E\{S_i(g_n) \mid \hat \theta_n(\mathbf{y})\} - E\{S_i(g_n) \mid \theta_n\}| &= |\sum_{i = 1}^n \partial_i E_{\bar \theta_n} (\hat \theta_{n, i} - \theta_{n, i})| \\ &\leq \sum_{i = 1}^n |\partial_i E_{\bar \theta_n} (\hat \theta_{n, i} - \theta_{n, i})| \\
&\leq  \sup_{\tilde \theta_n} \sum_{i = 1}^n |\partial_i E_{\tilde \theta_n}| \cdot  |(\hat \theta_{n, i} - \theta_{n, i})|
\end{align*}
Under Condition 3, we have that $\sup_{\theta_n} \partial_i E_{\theta_n} \leq C/n$ for some constant $C$, so we can then upper bound 
\begin{equation*}
    |E\{S_i(g_n) \mid \hat \theta_n(\mathbf{y})\} - E\{S_i(g_n) \mid \theta_n\}| \leq \frac{C}{n} \sum_{j} |\hat \theta_{i}(n) - \theta^\star(n)_i| \;,
\end{equation*}
and this last term is $o_P(1)$ by Condition 1 of the theorem. This completes the proof. 
\end{proof}
\section{Proofs of Taxonomy Results (Corollaries \ref{cor:MSE_link} and \ref{cor:MSE_density_diffcent})}
\label{section: Other Proofs}

\begin{proof}[Proof of Corollary \ref{cor:MSE_link}]
This is straightforward to calculate:
\begin{align*}
{E}\left[\Bigg\{E (g_{ij})-g^*_{ij}\Bigg\}^{2}\right] & ={E}\{{E} (g_{ij})^{2}-2E (g_{ij})g^*_{ij}+g_{ij^*}^{2}\} =p^2_{ij}(\theta) - 2p_{ij}(\theta) g^*_{ij} + \left(g^*_{ij}\right)^2
\end{align*}
which completes the proof.
\end{proof}


\begin{proof}[Proof of Corollary \ref{cor:MSE_density_diffcent}]

We begin the proof by verifying that $|E\{S_i(g_n) \mid \theta_n^\star\} - S_i(g_n^\star)| \overset{p}{\rightarrow}$ 0 as $n \rightarrow \infty.$ 

For part 1, density, we have
\begin{align*}
\sum_{j \in \{1,...,n\},j\neq i}\frac{\text{var}(g_{ij})}{(n-1)^2} &=
\sum_{j \in \{1,...,n\},j\neq i}\frac{p_{ij}(\theta)\left(1-p_{ij}(\theta)\right)}{(n-1)^{2}}\\
&\leq\sum_{j \in \{1,...,n\},j\neq i}\frac{1}{(n-1)^{2}}= \frac{1}{n-1}\rightarrow0
\end{align*}
so the Kolmogorov condition is satisfied and
\[
P \Bigg\{ \lim_{n\rightarrow\infty}\frac{d_{i}}{n}=\frac{E\left(d_{i}\right)}{n}\Bigg \}=1
\]
which satisfies the conditions of Proposition 1.

In part 2 we turn to diffusion centrality. Recall that.
\begin{align*}
DC_{i}\left({g};q_{n},K\right) =\sum_{j}\Bigg\{\sum_{t=1}^{K}\left(q_{n}{g}\right)^{t}\Bigg\}_{ij}  =\sum_{j}\sum_{t=1}^{K}\frac{C^{t}}{n^{t}}\sum_{j_{1},...,j_{t-1}}g_{ij_{1}}\cdots g_{j_{t-1}j}.
\end{align*}
For any $t$, we have

\begin{align*}
\text{var}\left(\frac{1}{n^{t}}\sum_{j}\sum_{j_{1},...,j_{t-1}}g_{ij_{1}}\cdots g_{j_{t-1}j}\right) 
&=\frac{1}{n^{2t}} \sum_{j}\sum_{j_{1},...,j_{t-1}} \text{var}(  g_{ij_{1}}\cdots g_{j_{t-1}j} )  \\
& + \frac{1}{n^{2t}} \sum_{j}\sum_{j_{1},...,j_{t-1}}\sum_{k}\sum_{k_{1},...,k_{t-1}} \text{cov}(g_{ij_{1}}\cdots g_{j_{t-1}j}, g_{ik_{1}}\cdots g_{k_{t-1}k}  ) 
\end{align*}
where $j_{0}=k_{0}=i$ and $j_{s}=j, k_{s}=k$. 
$\text{var}(g_{ij_{1}}\cdots g_{j_{t-1}j})$ has variance $\prod_{s=1}^{t}p_{j_{s-1}j_{s}}\left(1-\prod_{s=1}^{t}p_{j_{s-1}j_{s}}\right)\leq1$ and $ \text{cov}(g_{ij_{1}}\cdots g_{j_{t-1}j}, g_{ik_{1}}\cdots g_{k_{t-1}k}  ) \leq 1 $.
In order for $\text{cov}(g_{ij_{1}}\cdots g_{j_{t-1}j}, g_{ik_{1}}\cdots g_{k_{t-1}k}  ) \neq 0$, $g_{ij_{1}}\cdots g_{j_{t-1}j}$ and $g_{ik_{1}}\cdots g_{k_{t-1}k}$ need to have at least one edge in common. Notice that $g_{ij_{1}}\cdots g_{j_{t-1}j}$ has $n^t$ combinations since $i$ is given. Therefore, given a fixed common edge that $g_{ij_{1}}\cdots g_{j_{t-1}j}$ and $g_{ik_{1}}\cdots g_{k_{t-1}k}$ share, $g_{ij_{1}}\cdots g_{j_{t-1}j}$ has $n^{t-2}$ free choices of actors in the path, and $g_{ik_{1}}\cdots g_{k_{t-1}k}$ also has $n^{t-2}$ free choices of actors in the path. Therefore, for a given fixed common edge, there are $n^{2(t-2)}$ non-zero covariance terms. Since there are $n^2$ choices of a common edge, there are a total of $n^{2t-2}$ non-zero covariance terms. Therefore,
\[
\text{var}\left(\frac{1}{n^{t}}\sum_{j}\sum_{j_{1},...,j_{t-1}}g_{ij_{1}}\cdots g_{j_{t-1}j}\right) \leq \frac{n^t + n^{2t-2}}{n^{2t}}.
\]
Let $DC_{i,t}=\frac{1}{n^{t}}\sum_{j}\sum_{j_{1},...,j_{t-1}}g_{ij_{1}}\cdots g_{j_{t-1}j}$,
we have 
\[
\mathbb{P}\Bigg\{ DC_{i,t}-{E}( DC_{i,t}  ) \geq \epsilon\Bigg\} \leq \frac{n^t + n^{2t-2}}{n^{2t} \epsilon^2} \text{ by Chebyshev's inequality}
\] \\ \vspace{-4mm}
\[
\mathbb{P}\Bigg\{ DC_{i,t}-{E}( DC_{i,t} )\mid < \epsilon\Bigg\} \geq 1-\frac{n^t + n^{2t-2}}{n^{2t} \epsilon^2} \rightarrow 1 \text{ as } n \rightarrow \infty
\]
Therefore,
$DC_{i,t}$ goes in probability to $\mathrm{E}( DC_{i,t})$  as $n \rightarrow \infty$ and, 
by continuous mapping theorem,
$$DC_{i}\left({g};q_{n},K\right) = \sum_{t=1}^K C^t \times DC_{i,t}$$ tends to ${E} (DC_{i}\left({g};q_{n},K\right))$ in probability.

For part 3, clustering, the argument is identical to the convergence of clustering in Erdos-Renyi graphs because every link is conditionally edge independent. 
Let $N(i)$ denote the set of neighbors of actor $i$ and $ N(i)$ denote the size of neighbors, then 
\[
{clustering}_i({g}) = \frac{ \sum_{j, k \in N(i)} g_{jk} }  {  N(i)\cdot \{ N(i)\mid-1\}  }
\]
Similar to the proof for density, we have 
\begin{align*}
&\sum_{j, k \in N(i)} \frac{\text{var}( g_{jk})}{[\mid N(i)\mid\times \{\mid N(i)\mid-1\} ]^2}=
\sum_{j, k \in N(i)}\frac{p_{jk}(\theta)\left(1-p_{jk}(\theta)\right)}{[\mid N(i)\mid\times \{\mid N(i)\mid-1\} ]^2} \\
&\leq \sum_{j, k \in N(i)}\frac{1}{[\mid N(i)\mid\times \{\mid N(i)\mid-1\} ]^2}= \frac{1}{\mid N(i)\mid\times \{\mid N(i)\mid-1\}}\rightarrow0
\end{align*}
so the Kolmogorov condition is satisfied and
${clustering}_i({g})$ goes in probability to 
\begin{equation*}
    E_{z_j,\nu_j, z_k, \nu_k  \mid  j,k \in N(i)}\{ \mathbb{P}(g_{jk}=1\mid \nu_j,\nu_k,z_j,z_k) \}
\end{equation*} as $n$ tends to infinity.

Finally, we now verify the conditions of Theorem \ref{thm: ARD_consistency} of the main paper. Specifically, we need to show that the derivative $\partial_i E_{\theta_n}$ is uniformly bounded over $\Theta$.

The degree of a node $i$ is $S_i(g_n) = 1/(n-1)\sum_{j \neq i} p_{ij}(\theta)$. In this case, for any $k$,
\begin{equation*}
\partial_k E\{S_i(g_n) \mid {\theta_n}\} = \frac{1}{n-1} \frac{d}{d\theta_k} p_{ik}(\theta)
\end{equation*}
So, supposing that $\frac{d}{d\theta_k} p_{ik}(\theta)$ is uniformly bounded, we can conclude that $\partial_k E\{S_i(g_n) \mid \theta_n\} \leq C /(n-1)$ for some constant $C$, so Condition 2 holds assuming that $1/n \sum_{j} |\hat \theta_i(n) - \theta_i(n)| = o_P(1)$. A similar argument applies to the clustering coefficient of a node, defined as 
\begin{equation*}
S_i(g_n) = \frac{1}{{N_i \choose 2}}\sum_{j, k \in N_i} g_{ij} g_{jk}
\end{equation*}
where $N_i$ is the set of neighbors of node $i$: $N_i = \{j: g_{ij} = 1\}$.

We finally look at the centrality parameter of a node. We only look at the case of $T = 2$, since the argument for $T > 2$ is similar. We begin by computing $E\{S_i(g_n) \mid {\theta_n}\}$, which is equal to 
\begin{equation*}
    E\{S_i(g_n) \mid {\theta_n}\} = \sum_{j} \frac{C}{n} E[A_{ij}] + \sum_{j} \frac{C^2}{n^2} E\{[A^2]_{ij}\} \;.
\end{equation*}
where $A^2$ is the matrix square of the matrix $A$ and $A$ is the adjacency matrix of the graph $g$. We are interested in the derivative of $E\{S_i(g_n) \mid {\theta_n}\}$. Supposing that $\frac{d}{d\theta_k} p_{ik}(\theta)$ is uniformly bounded, the derivative of the first term satisfies Condition 3.  So we now turn to the second sum and expand
\begin{equation*}
    E\{A_{ij}\}^2 = E\{\sum_{k} A_{ik} A_{kj}\} = \sum_k E\{A_{ik} A_{kj}\} =  \sum_k E\{A_{ik}\} E\{A_{kj}\} = \sum_{k} p_{ik}(\theta) p_{kj}(\theta) \;.
\end{equation*}
Under the same assumption that the derivative $\frac{d}{d\theta_k} p_{ik}(\theta)$ is uniformly bounded, we can conclude that the second sum is also satisfies Condition 2. It follows then that $E\{S_i(g_n) \mid {\theta_n}\}$ satisfies Condition 3. 
\end{proof}

\section{Proof of Consistency of OLS estimators in many networks setting (Theorem \ref{thm: consistenct_OLS_cov})}
\label{sec: consist_OLS_many_network}
\begin{proof}[Proof of Theorem \ref{thm: consistenct_OLS_cov}]
We consider the case where there is no intercept ($\alpha = 0$) to simplify the calculations, but the same argument applies to the case where $\alpha \neq 0$.

We begin by expanding 
\begin{equation*}
    O_r = \beta E\{S_r(g_n) \mid \hat \theta_r(n)\} + \epsilon_r = \beta S_r^\star +  \big(\epsilon_r + \beta E\{S_r \mid \hat \theta_r(n)\} - \beta E\{S_r \mid \theta_r\} + \beta E\{S_r \mid \theta_r\} - \beta S_r^\star\big)
\end{equation*}
Let $\tilde \epsilon_{n, r} =  \big(\epsilon_r + \beta E\{S_r \mid \hat \theta_r(n)\} - \beta E\{S_r \mid \theta_r\} + \beta E\{S_r \mid \theta_r\} - \beta S_r^\star\big)$.  Now, by using the analytic expression for the OLS estimator, we have that
\begin{align*}
|\hat \beta - \beta| &= \frac{1}{\sum_{r = 1}^R E\{S_r \mid \hat \theta_r(n)\}^2} \sum_{r = 1}^R |E\{S_r \mid \hat \theta_r(n)\} \tilde \epsilon_{n, r}| \\
&= \frac{1}{\sum_{r = 1}^R E\{S_r \mid \hat \theta_r(n)\}^2} \sum_{r = 1}^R E\{S_r \mid \hat \theta_r(n)\}\big(|\epsilon_r + \beta E\{S_r \mid \hat \theta_r(n)\} - \beta E\{S_r \mid \theta_r\} + \beta E\{S_r \mid \theta_r\} - \beta S_r^\star|\big) \\
&\leq \frac{1}{\sum_{r = 1}^R E\{S_r \mid \hat \theta_r(n)^2} \sum_{r = 1}^R |E\{S_r \mid \hat \theta_r(n)\} \epsilon_r| +  \frac{\beta}{\sum_{r = 1}^R E\{S_r \mid \hat \theta_r(n)\}^2} \sum_{r = 1}^R |E\{S_r \mid \hat \theta_r(n)\} \left(E\{S_r \mid \hat \theta_r(n)\} - E\{S_r \mid \theta_r\}\right)| + \\& \frac{\beta}{\sum_{r = 1}^R E\{S_r \mid \hat \theta_r(n)\}^2} \sum_{r = 1}^R |E\{S_r \mid \hat \theta_r(n)\} \left(E\{S_r \mid \theta_r\} - S_r^\star\right)| \\
&= I + II + III \;.
\end{align*}
Now, $I$ is $o_P(1)$ assuming that $E(\epsilon_r | E\{S_r \mid \hat \theta_r(n)\}) = 0$. Now, let us look at the second term, 
\begin{equation*}
II = \frac{1}{\sum_{r = 1}^R E\{S_r \mid \hat \theta_r(n)\}^2} \sum_{r = 1}^R E\{S_r \mid \hat \theta_r(n)\} \times |E\{S_r \mid \hat \theta_r(n)\} - E\{S_r \mid \theta_r\}| \;,
\end{equation*}
and the third term is 
\begin{equation*}
III = \frac{1}{\sum_{r = 1}^R E\{S_r \mid \hat \theta_r(n)\}^2} \sum_{r = 1}^R E\{S_r \mid \hat \theta_r(n)\} \times |E\{S_r \mid \theta_r\} - S_r^\star|
\end{equation*}
For the third term, supposing that $E\{S_r \mid \hat \theta_r(n)\} \leq C$, I can upper bound
\begin{equation*}
III \leq \frac{C}{R^{-1}\sum_{r = 1}^R E\{S_r \mid \hat \theta_r(n)\}^2} \frac{1}{R} \sum_{r = 1}^R |E\{S_r \mid \theta_r\} - S_r^\star|
\end{equation*}
Now suppose that that $E\{S_{r}^\star \mid \theta\}$ has finite mean. We then can then conclude that 
\begin{equation*}
III \leq \frac{C}{R^{-1}\sum_{r = 1}^R E\{S_r \mid \hat \theta_r(n)\}^2}\frac{1}{R} \sum_{r = 1}^R |E\{S_r \mid \theta_r\} - S_r^\star| \;.
\end{equation*}
By Hoeffding's inequality, we can conclude that the average $\frac{1}{R} \sum_{r = 1}^R |E\{S_r \mid \theta_r\} - S_r^\star| = o_P(1)$, and so by Slutksy's lemma, we can conclude that $III = o_P(1)$ as $n, R \rightarrow \infty$.  

We now move to the second term $II$. Using a Taylor series expansion, we can write 
\begin{align*}
E\{S_r \mid \hat \theta_r(n)\} - E\{S_r \mid \theta_r(n)\} &= D^T(\bar \theta_n) ||\hat \theta_r(n) - \theta_r(n)|| \\
&= \sum_{i = 1}^n \partial_i E\{S_r \mid \bar \theta_r(n)\} |\hat \theta_r(n) - \theta_r(n)|_i
\end{align*}
for some sequence of intermediate values $\bar \theta_n$. So, 
\begin{align*}
II &\leq \frac{1}{\sum_{r = 1}^R E\{S_r \mid \hat \theta_r(n)\}^2} \sum_{r = 1}^R E\{S_r \mid \hat \theta_r(n)\} \sum_{i = 1}^n \partial_i E\{S_r \mid \bar \theta_r(n)\} \times |\hat \theta_r(n) - \theta_r(n)|_i \\
&\leq \frac{C}{\sum_{r = 1}^R E\{S_r \mid \hat \theta_r(n)\}^2}\sum_{r = 1}^R\frac{1}{n} \sum_{i = 1}^n |\hat \theta_r(n) - \theta_r(n)|_i \\
&= \frac{C}{R^{-1} \sum_{r = 1}^R E\{S_r \mid \hat \theta_r(n)\}^2}\frac{1}{R}\sum_{r = 1}^R \sum_{i = 1}^n |\hat \theta_r(n) - \theta_r(n)|_i
\end{align*}
where the first inequality follows from the Taylor series expansion and the second inequality follows from the assumptions of this theorem.  Supposing that that $E(E\{S_r \mid \hat \theta_r(n)^2) < \infty$, we bound 
\begin{equation*}
II \leq \frac{C}{R^{-1} \sum_{r = 1}^R E\{S_r \mid \hat \theta_r(n)\}^2 }\max_{1 \leq r \leq R} \sum_{i = 1}^n |\hat \theta_r(n) - \theta_r(n)|_i
\end{equation*}
Under the assumptions of the theorem, we have that $\max_{1 \leq r \leq R} \sum_{i = 1}^n |\hat \theta_r(n) - \theta_r(n)|_i = o_P(1)$, so we conclude that $|\hat \beta_{n, R} - \beta| = o_P(1)$, as claimed. 

To prove that the estimator $\hat \gamma_{n, R}$ is consistent, the argument is nearly identical. To see why, we simple re-arrange the supposed data generating model:
\begin{equation}
    S_r^\star = \alpha + \gamma T_r + \epsilon_r - E\{S_i(g_n) \mid \hat \theta_r(n)\} + S_r^\star \;.
\end{equation}
The same argument applies to show that the OLS estimates of $\gamma$ are also consistent under the conditions of the theorem.
\end{proof}

\section{Checking conditions of Theorem \ref{thm: consistenct_OLS_cov} for common network statistics (Theorem \ref{thm: checking_cond_OLS_cov})}
\label{sec: checking_conditions_Thm}
\begin{proof}[Proof of Theorem \ref{thm: checking_cond_OLS_cov}]
We only prove the case for the density. The arguments for the other two statistics are similar.  

From the proof of Theorems \ref{theorem: beta_consistency}, \ref{theorem: SBM_consistency}, \ref{thm: LS_consistency}, we showed that for any network, each estimator $\hat \theta_{i, r}(n)$ satisfies an exponential concentration inequality, and by taking a union bound over all nodes in a network, we see that
\begin{equation*}
\mathbb{P}( \frac{1}{n}\sum_{i = 1}^n  |\hat \theta_{i, r}(n) - \theta_{i, r}^\star| > \epsilon) \leq  \mathbb{P}(\max_{1 \leq i \leq n} |\hat \theta_{i, r}(n) - \theta_{i, r}^\star| > \epsilon) \leq n C \exp(- C'\epsilon^2 n) \;.
\end{equation*}
for some constants $C$ and $C'$. By taking a union bound over all $R$ villages, we conclude that
\begin{equation*}
\mathbb{P}(\max_{1 \leq r \leq R}  \frac{1}{n}\sum_{i = 1}^n  |\hat \theta_{i, r}(n) - \theta_{i, r}^\star| > \epsilon) \leq R n \exp(-\epsilon^2 n) \;.
\end{equation*}
Under the assumptions of the theorem, we have that $R n\exp(-n) \rightarrow 0$, so Condition 2 holds. We now discuss Condition 3 of Theorem \ref{thm: consistenct_OLS_cov}. One way to satisfy this is to require that the network statistic for each network is the same (i.e., we are considering just the centrality of a set of nodes). In this case, since the network statistic $S_{i, r}$ satisfies the required derivative condition, per Theorem \ref{thm: ARD_consistency}, we can then conclude that the maximum also satisfies such a derivative condition. This completes the proof. 
\end{proof}

 \section{Simulations using fully-elicited graphs}
 \label{sec:supsim}
In this section we present additional results using fully-elicited, observed graphs.  We use data from~\cite{banerjeecdj2013}, which consists of completely observed graphs from 75 villages in rural India.  The goal of these results is two-fold.  First, we aim to demonstrate that our results hold in networks that have the level of sparsity and complexity that a user could reasonably find in practice.  Second, we aim to show that the perfomance of our method improves as the graph size increases, as indicated by our results.

\begin{figure*}[]
 \centering
 \includegraphics[width=.45\textwidth]{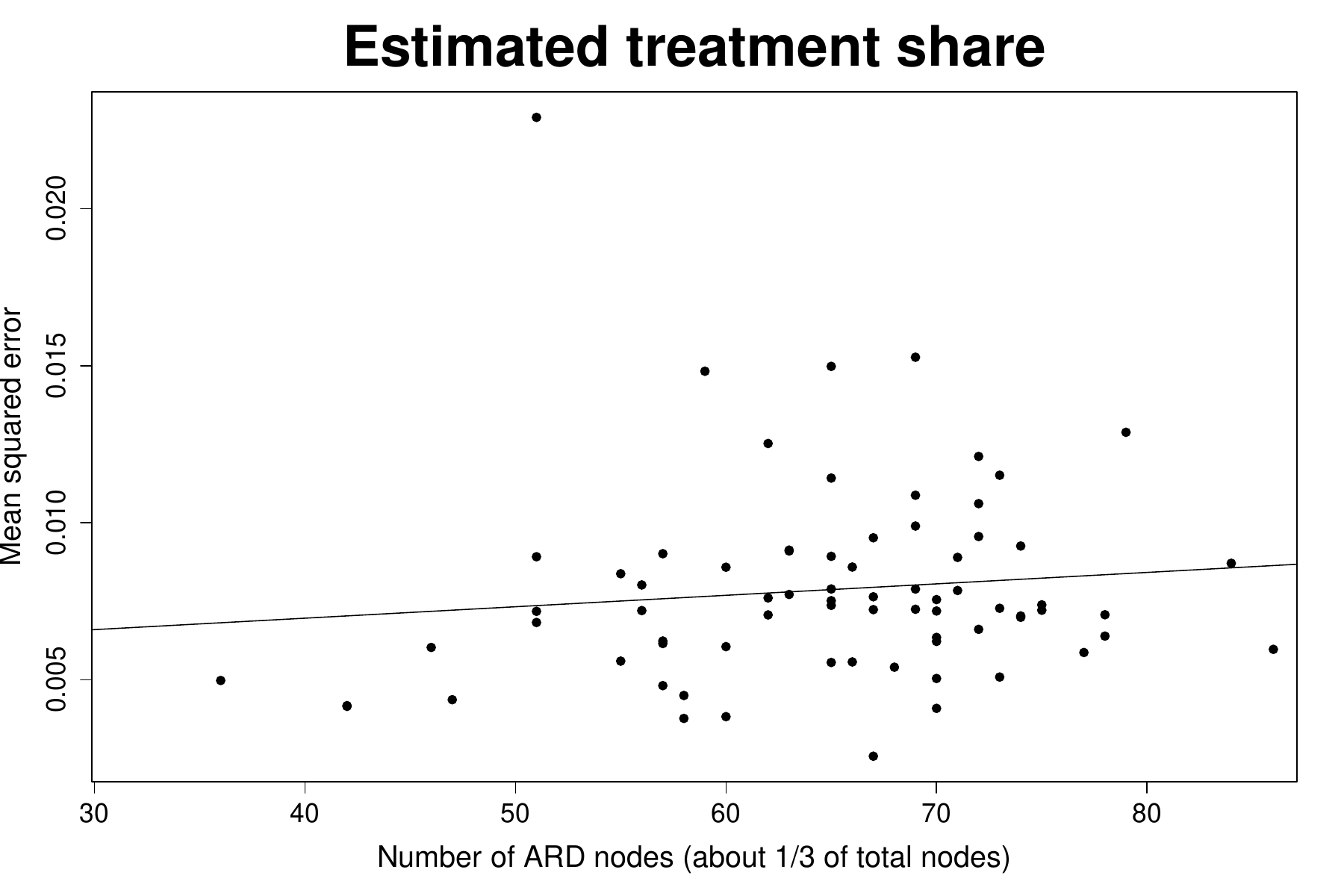}
 \hspace{5pt}
 \includegraphics[width=.45\textwidth]{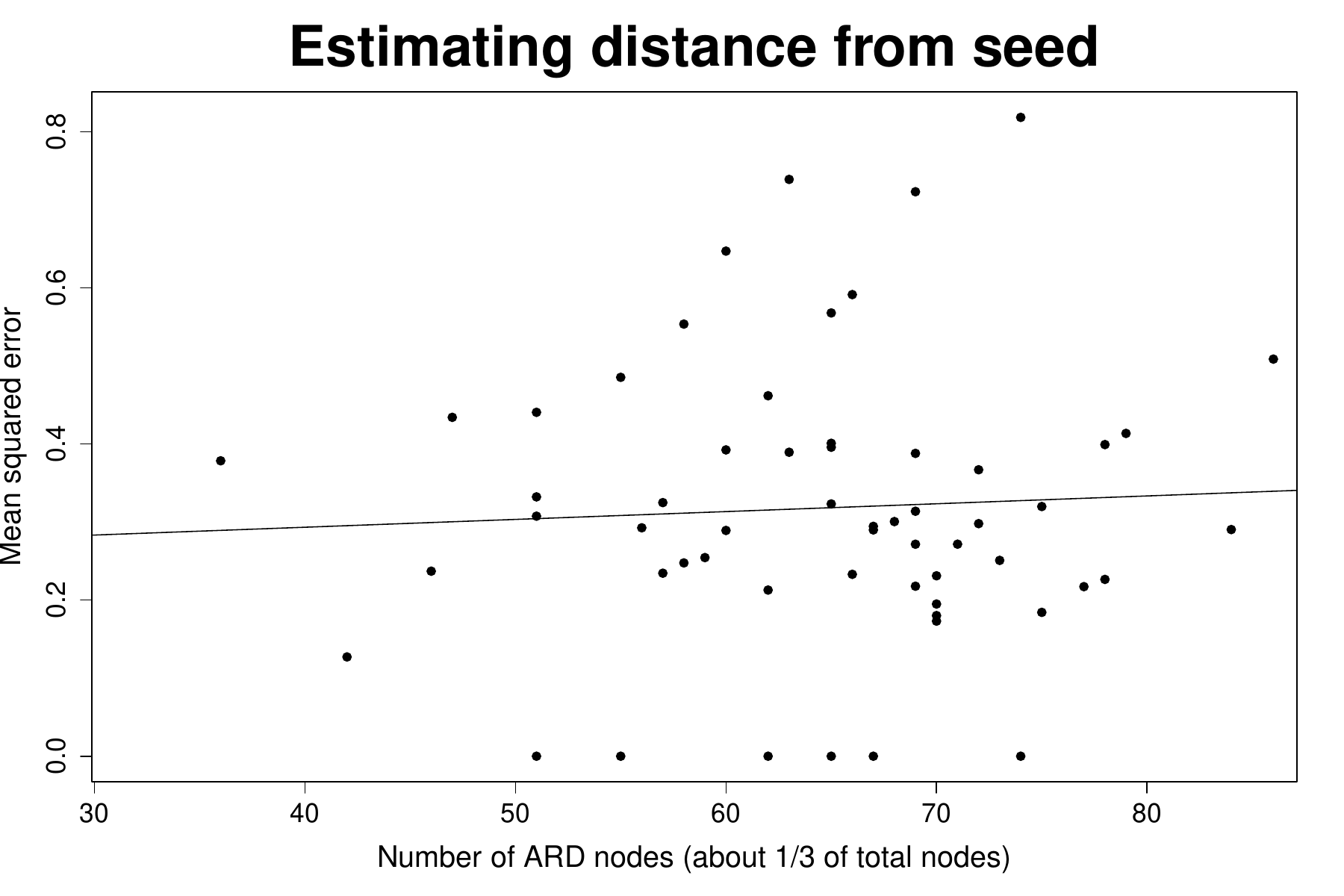}
 \includegraphics[width=.45\textwidth]{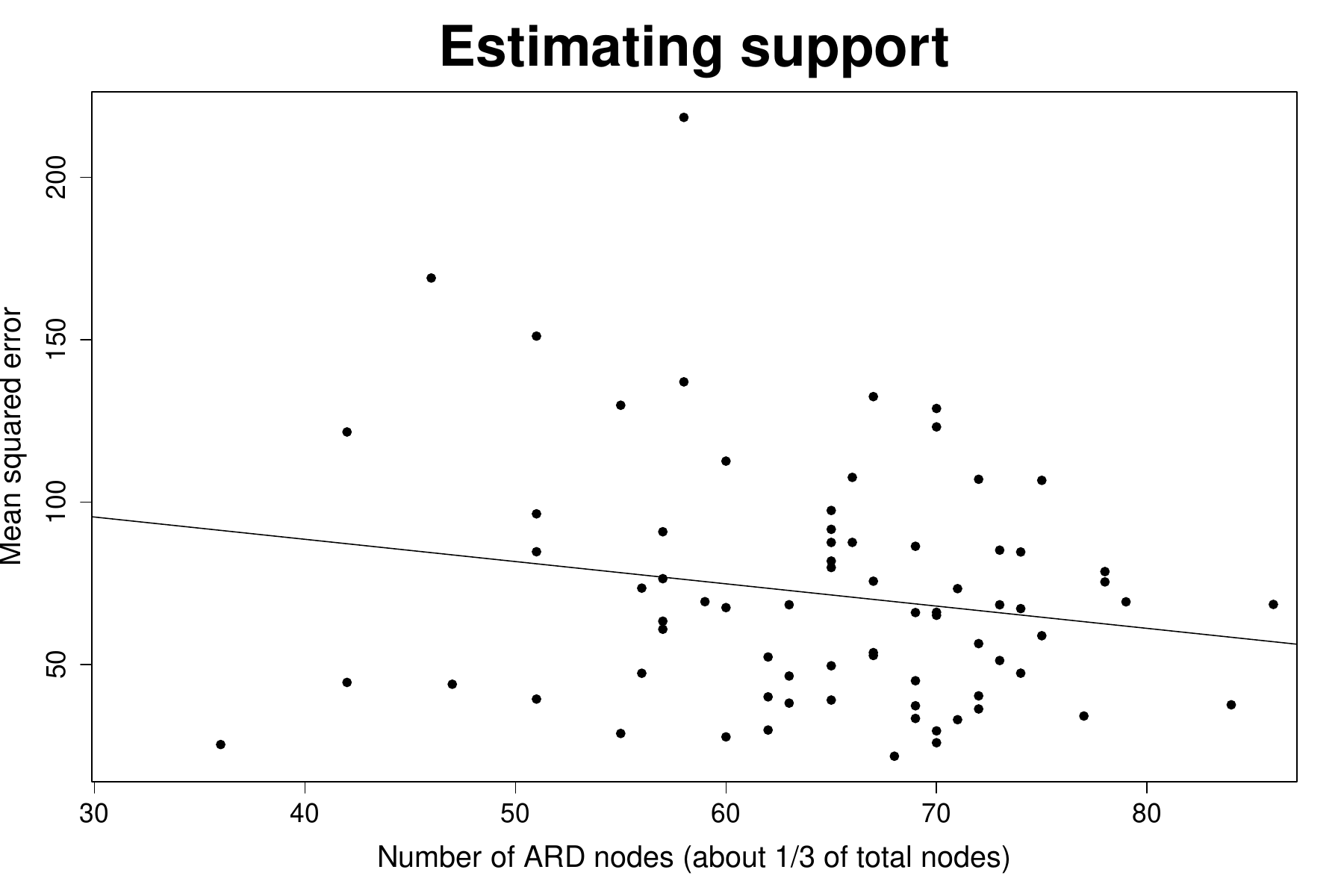}
  \hspace{5pt}
 \includegraphics[width=.45\textwidth]{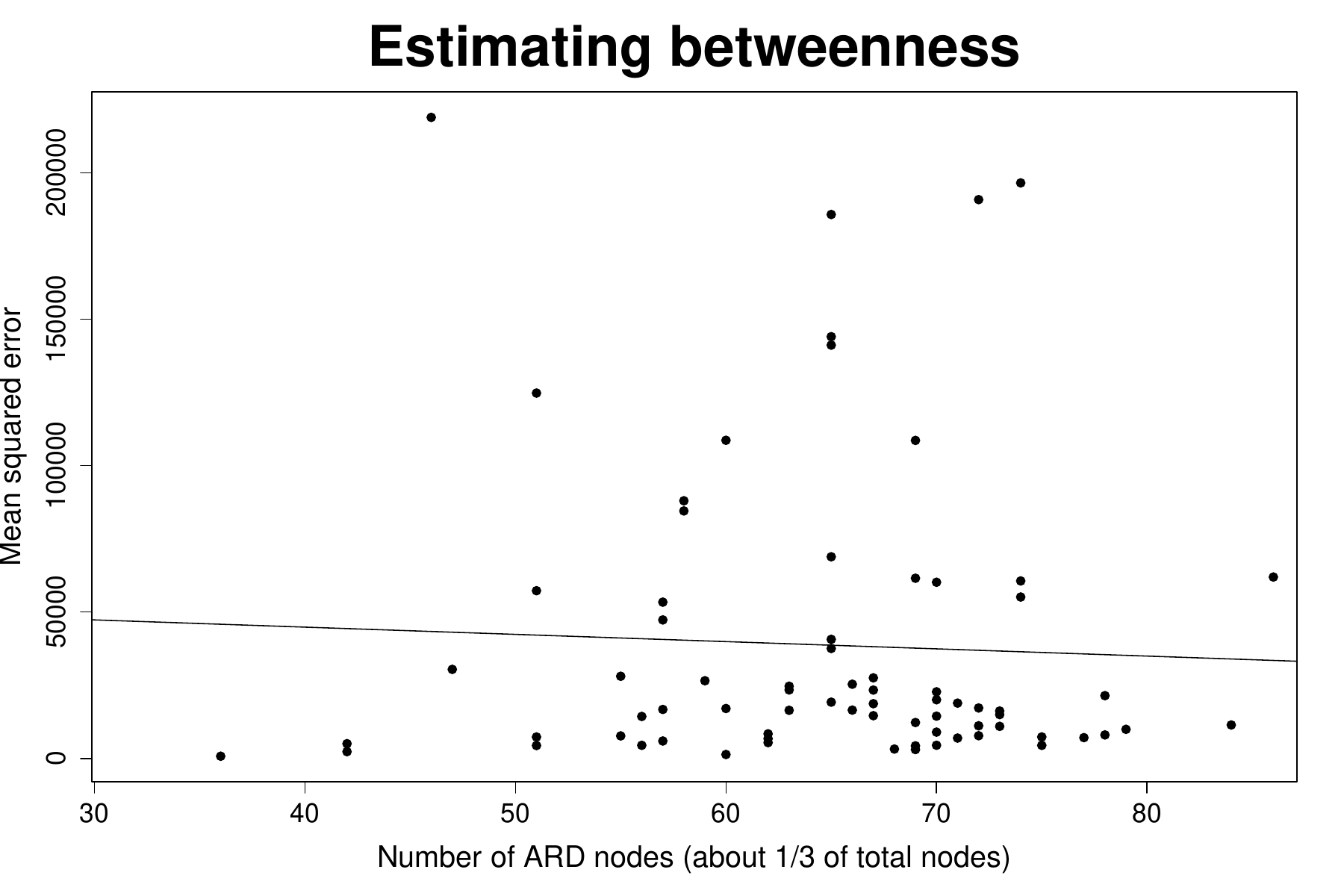}
 \includegraphics[width=.45\textwidth]{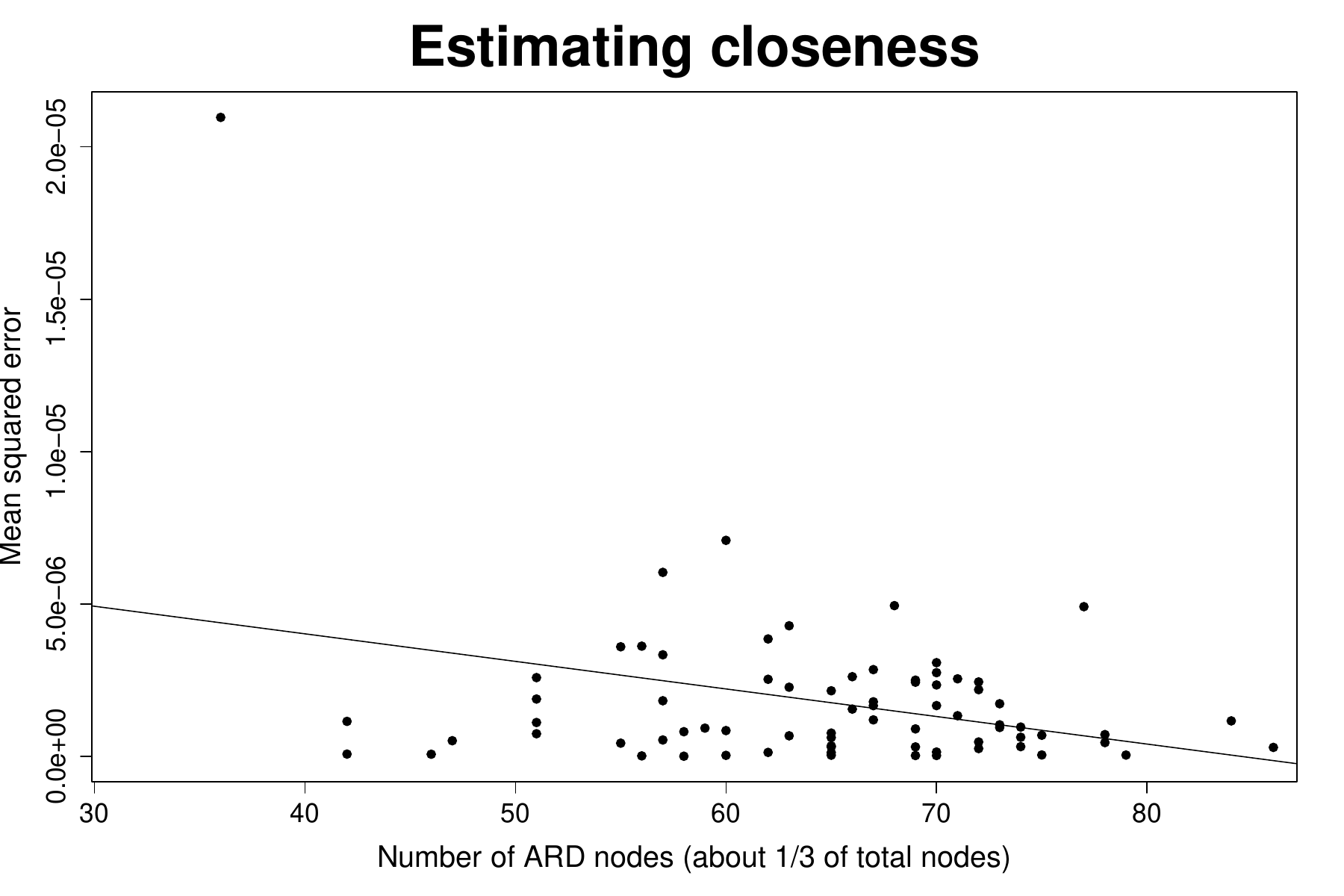}
   \hspace{5pt}
 \includegraphics[width=.45\textwidth]{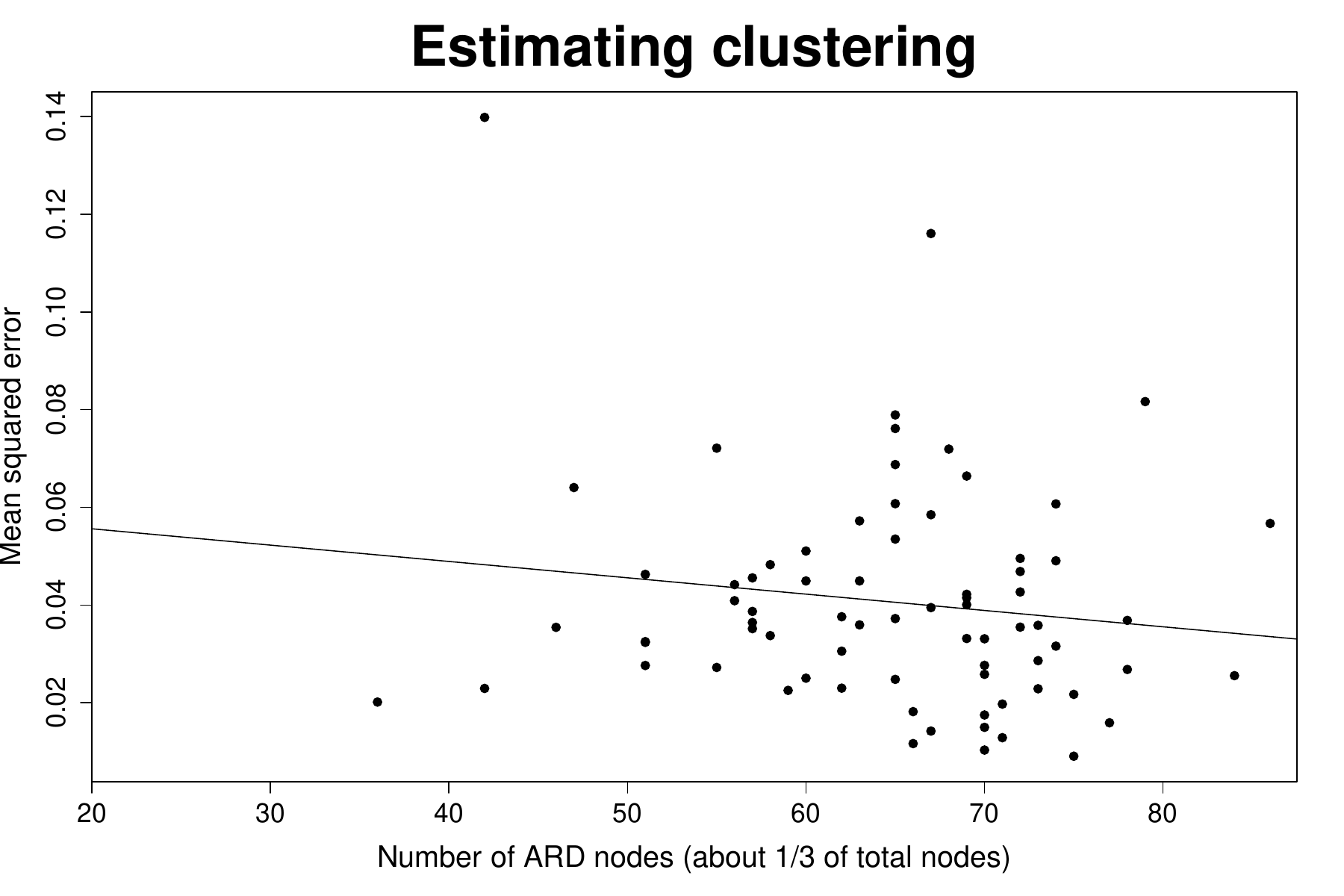}
 \includegraphics[width=.45\textwidth]{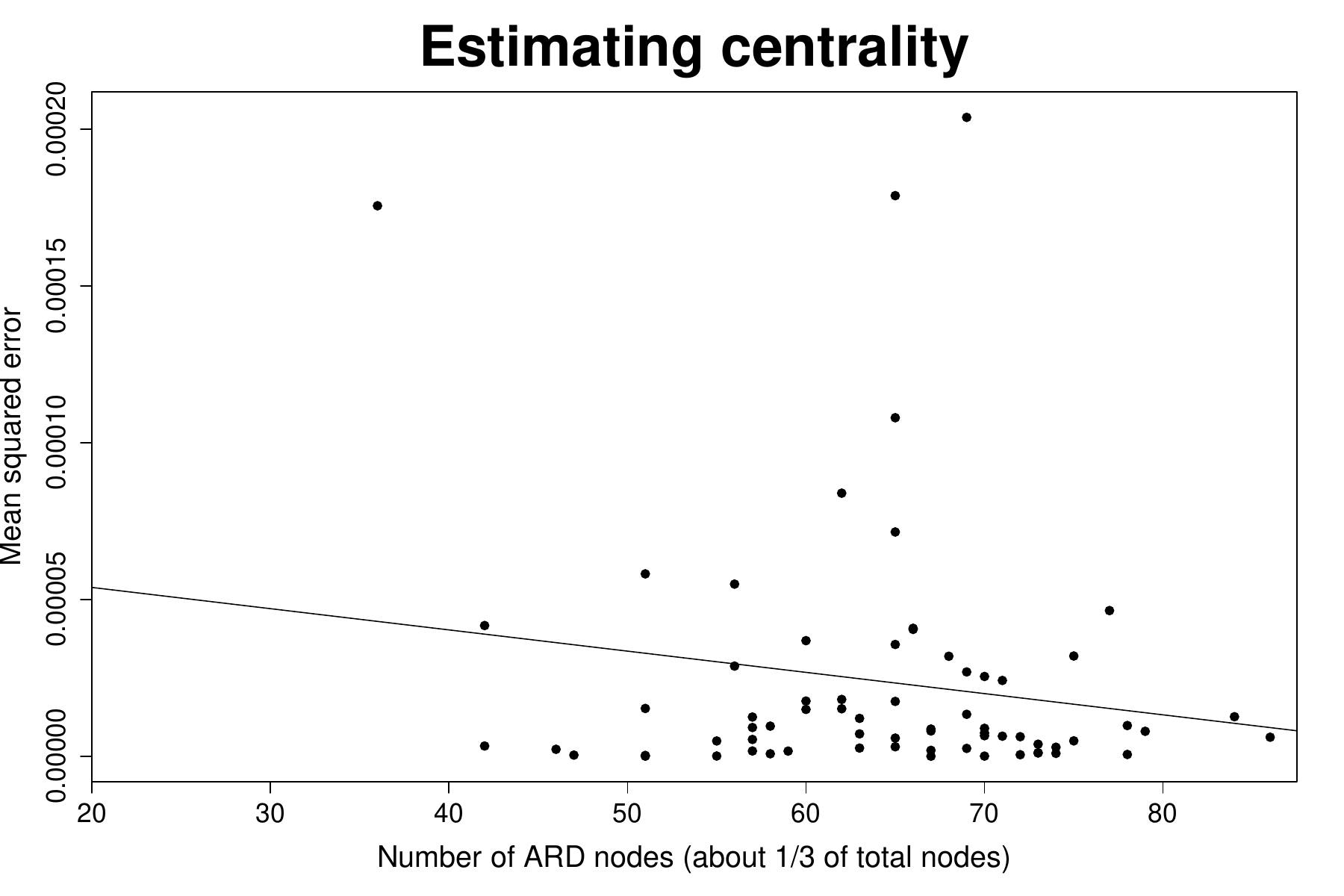}
    \hspace{5pt}
 \includegraphics[width=.45\textwidth]{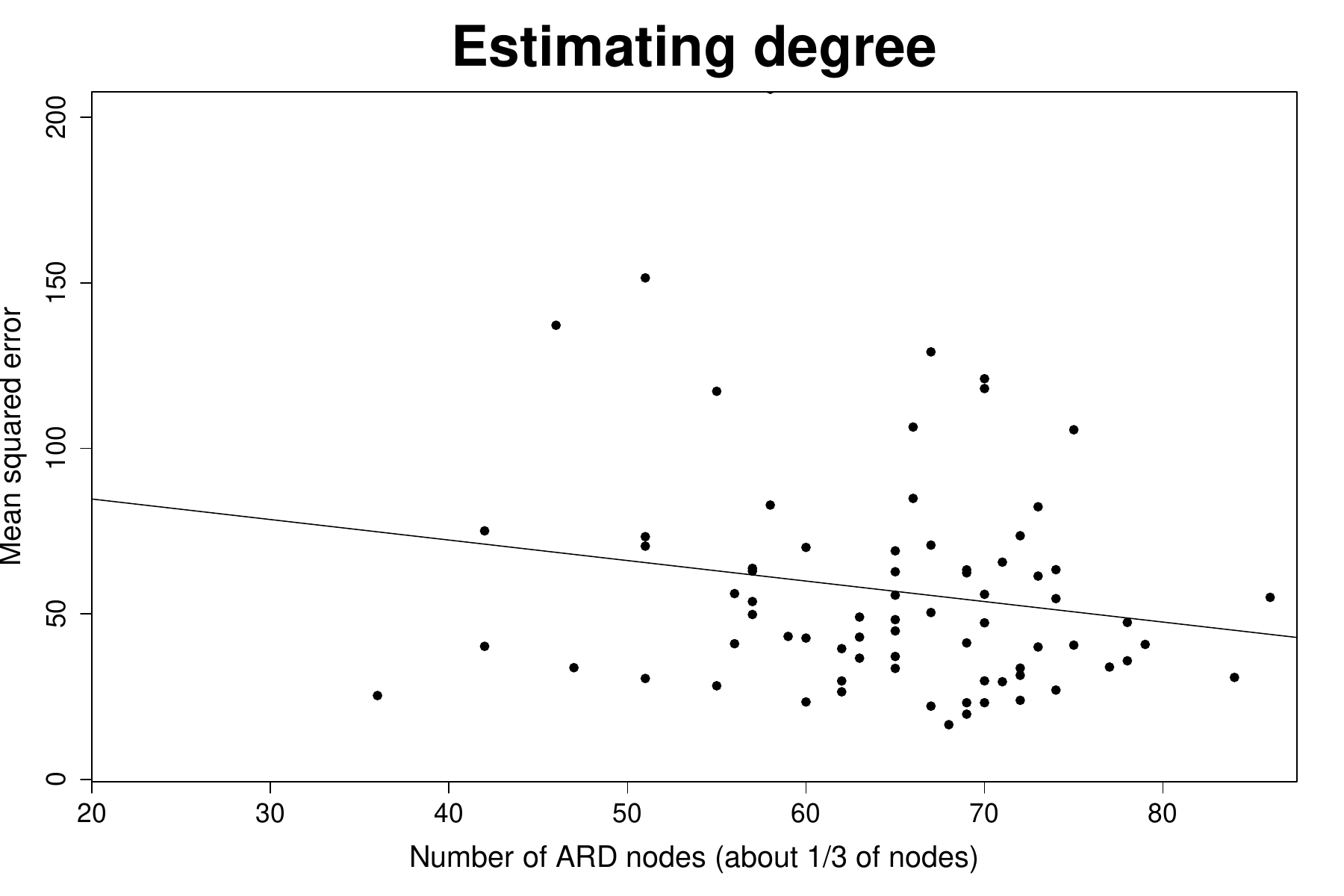}
 \caption{MSE and graph size.  Each plot shows the MSE (computed across nodes) plotted as a function of the number of respondents who received ARD using data from~\cite{banerjeecdj2013}.}
 \label{fig: MSE_graph_size}
 \end{figure*}

In each village, about one-third of respondents were asked ARD questions. ~\cite{breza2017using} compare statistics estimated with ARD from these graphs with the same statistics calculated using the complete graph.  We leverage these results and present a different aspect, how the MSE changes as the size of the graph grows.  We present results for individual-level statistics from these graphs and compute MSE across individuals.  Figure~\ref{fig: MSE_graph_size} presents these results.

\section{Additional simulation results with estimated formation model parameters}
\label{sec: additional_sims}
\begin{figure}
\centering
\includegraphics[width=.9\textwidth]{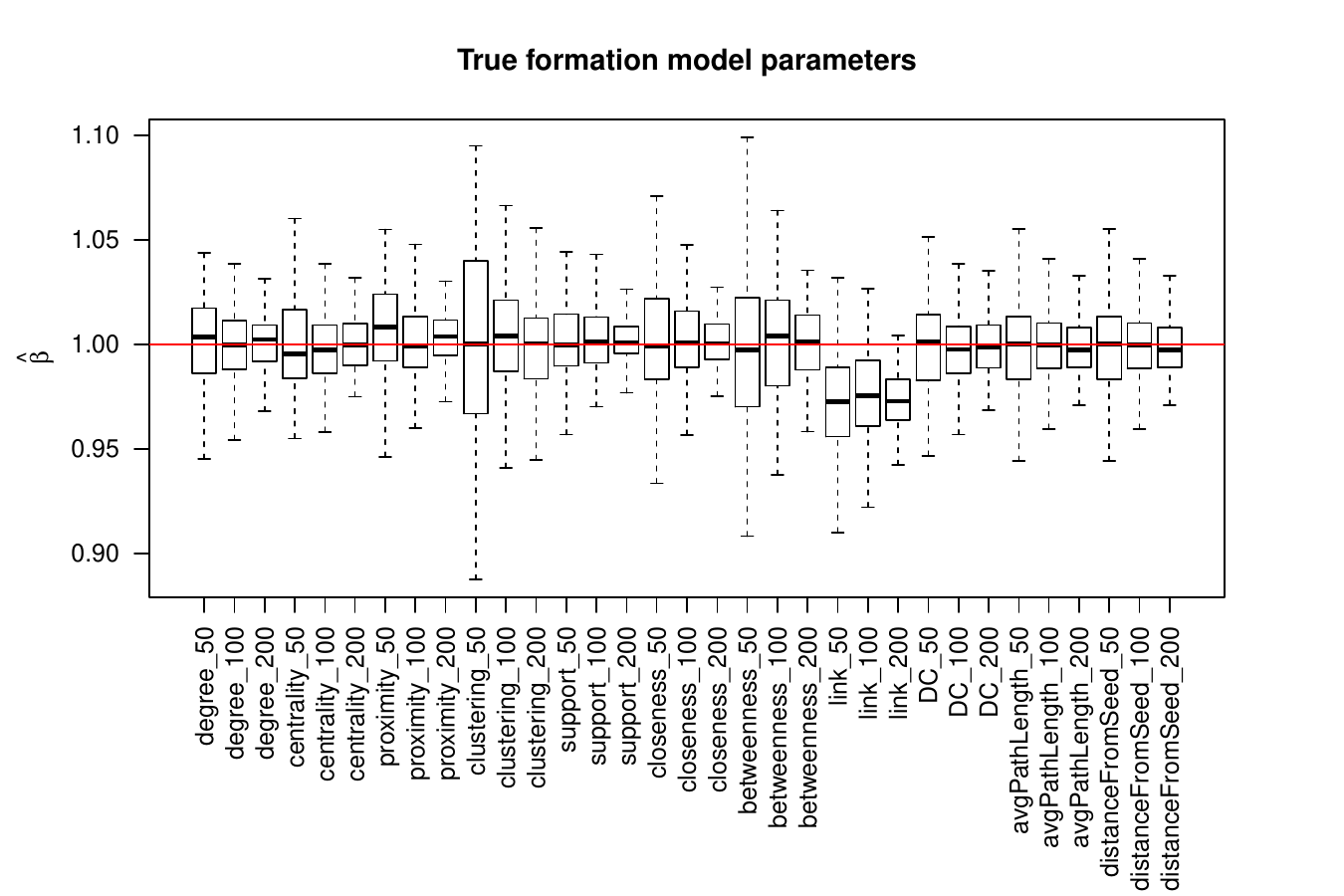}
\smallskip
\vspace{-10mm}
\includegraphics[width=.9\textwidth]{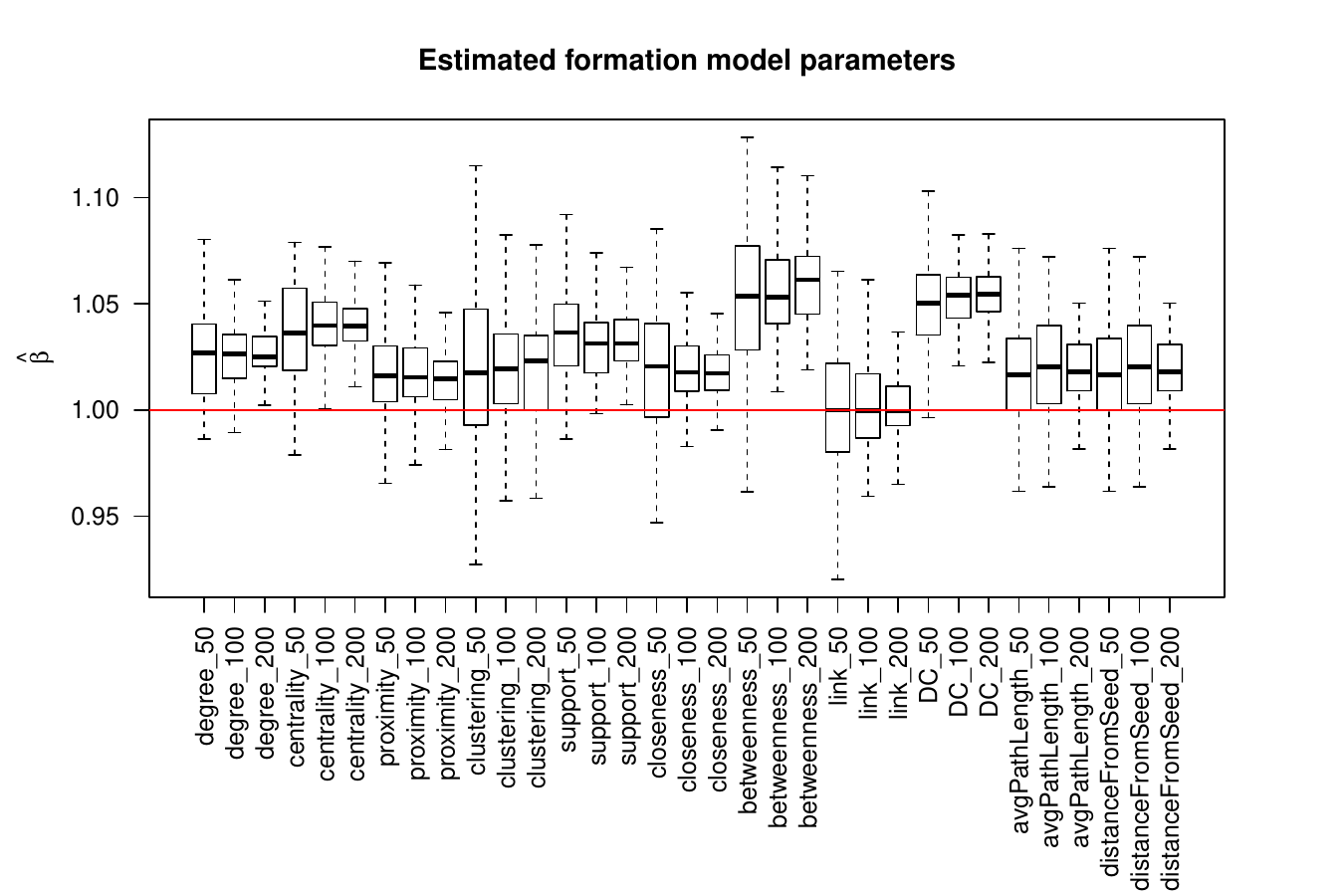}
\caption{Boxplot of $\hat{\beta}$ for $\beta$ in regression $y_{ij,r}=\alpha +\beta \bar{S}_{ij,r} + \epsilon_r$, where ${S}_{ij,r}$ and $\bar{S}_{ij,r}$ represent a true and mean individual-level measure, respectively. Each box represents the distribution of $\hat{\beta}$ for one measure and use of R=50, 100 or 200 networks in regression. 50 actors and 1000 pairs (for link) are randomly selected for each network. The middle line of the boxplot denotes median, and borders of the boxes denote first and third quartile. The red line denotes the true $\beta=1$ used to generate $y_{ij,r}=\alpha +\beta S^*_{ij,r} + \epsilon_r$ in the simulation. These results corroborate the theoretical intuition developed in Theorems \ref{thm: ARD_consistency} and \ref{thm: consistenct_OLS_cov}.}
\label{fig:snode}
\end{figure}

\begin{figure*}[]
\centering
\includegraphics[width=.95\textwidth]{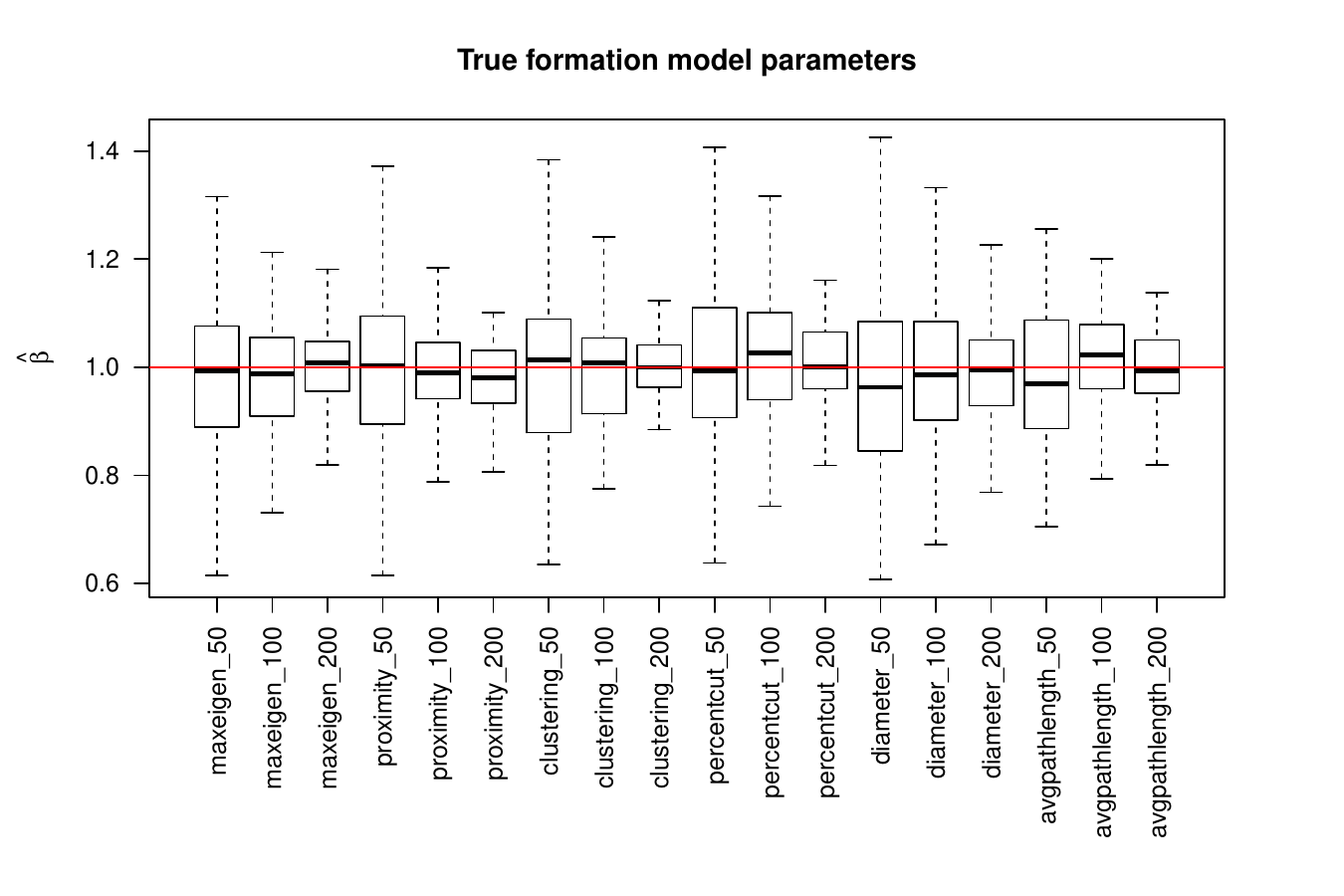}
\smallskip
\vspace{-10mm}
\includegraphics[width=.95\textwidth]{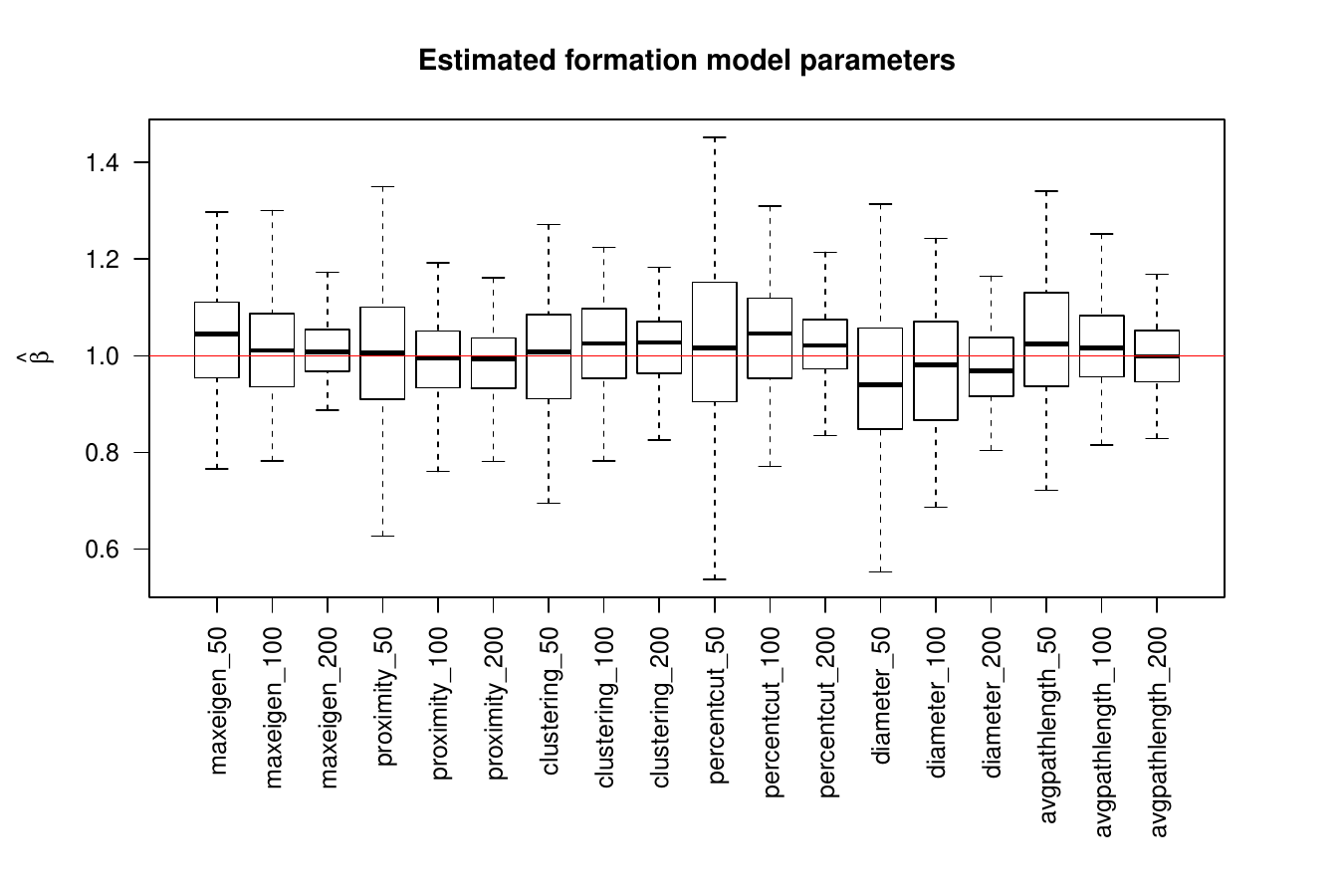}
\vspace{-5mm}
\caption{Boxplot of $\hat{\beta}$ for $\beta$ in regression $y_{r}=\alpha +\beta \bar{S}_{r} + \epsilon_r$, where ${S}_{r}$ and $\bar{S}_r$ represent a true and mean network-level measure, respectively. Each box represents the distribution of $\hat{\beta}$ for one measure and use of R=50, 100 or 200 networks in regression. The middle line of the boxplot denotes median, and borders of the boxes denote first and third quartile. The red line denotes the true $\beta=1$ used to generate $y_{r}=\alpha +\beta S^*_{r} + \epsilon_r$ in the simulation. These results corroborate the theoretical intuition developed in Theorems \ref{thm: ARD_consistency} and \ref{thm: consistenct_OLS_cov}.  }
\label{fig:network}
\end{figure*}

\begin{figure*}[]
\centering
\includegraphics[width=.9\textwidth]{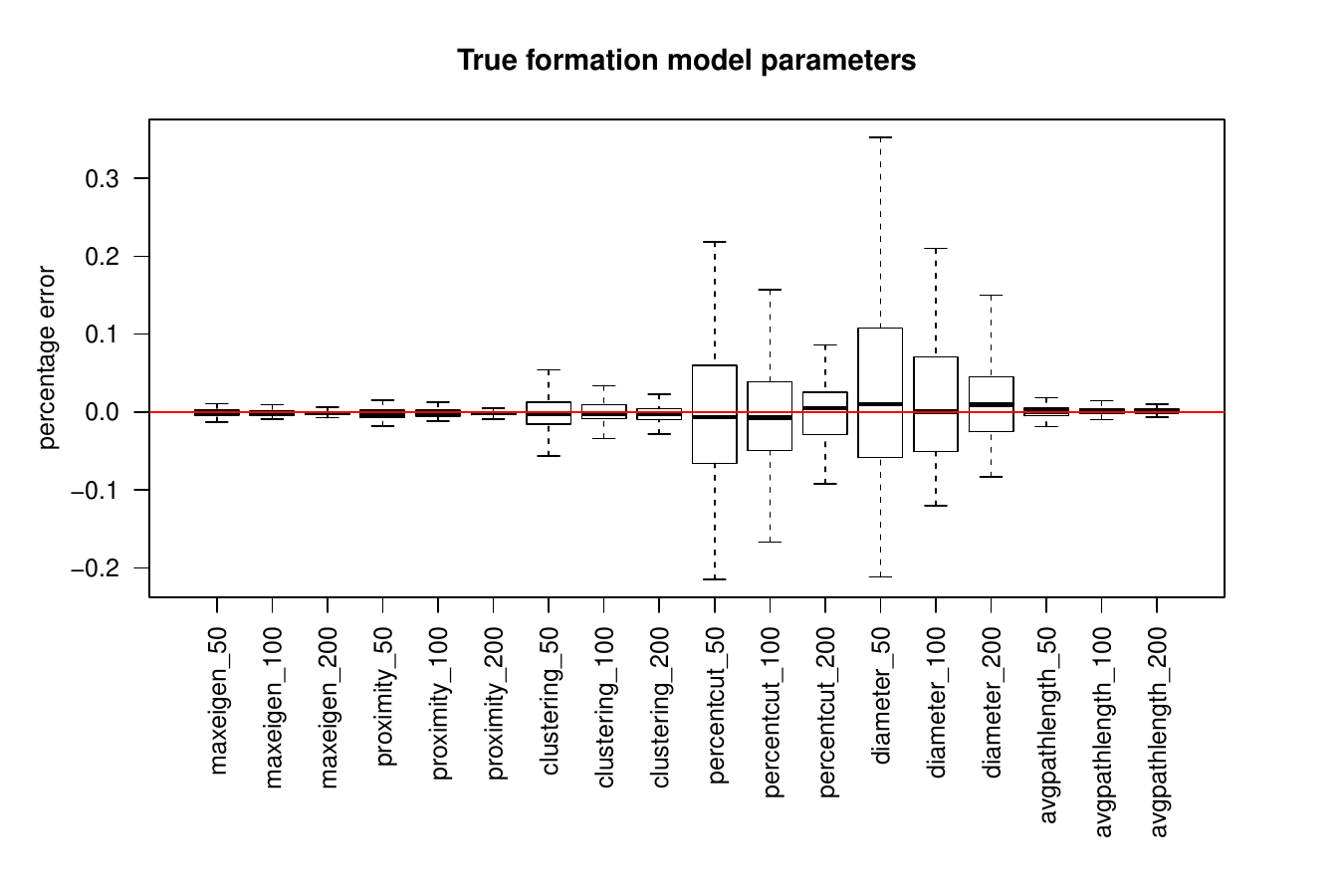}
\smallskip
\vspace{-10mm}
\includegraphics[width=.9\textwidth]{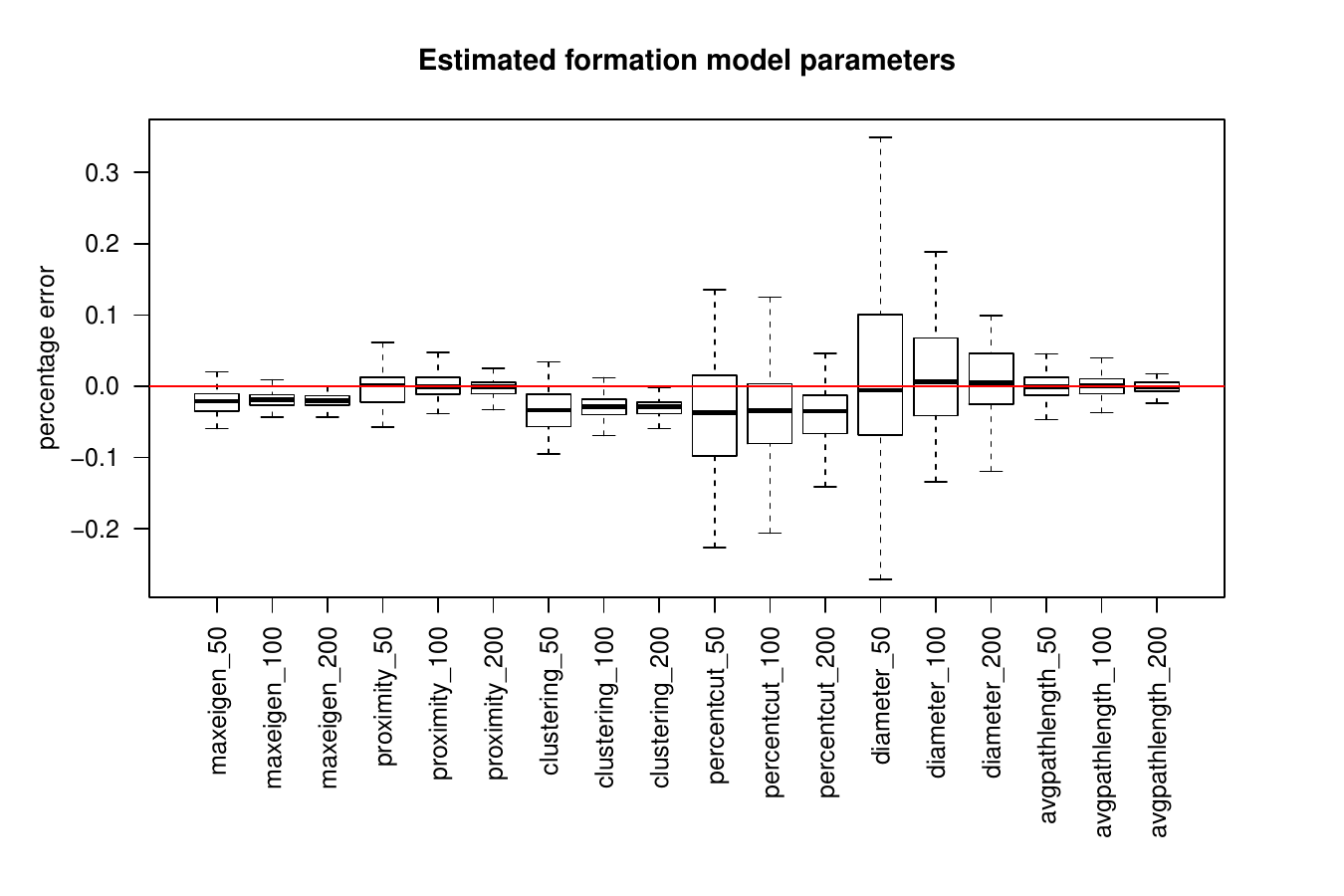}
\vspace{-5mm}
\caption{Boxplot of percentage errors of $\hat{\gamma}$ for $\gamma$ in regression $\bar{S}_{r}=\alpha +\gamma T_r + \epsilon_r$, where ${S}_{r}$ and $\bar{S}_r$ represent a true and mean network-level measure, respectively. Each box represents the distribution of percentage errors for one measure and use of R=50, 100 or 200 networks in regression. The middle line of the boxplot denotes median, and borders of the boxes denote first and third quartile. These results corroborate the theoretical intuition developed in Theorems \ref{thm: ARD_consistency} and \ref{thm: consistenct_OLS_cov}.  }
\label{fig:network_trt}
\end{figure*}

In this section we present additional simulation results to complement the simulations we present in the main text. We present results when the parameters are estimated using the procedure in~\cite{breza2017using}, rather than assumed to be consistently estimated. These simulations are presented in Figures \ref{fig:snode}, \ref{fig:network}, and \ref{fig:network_trt}.  The results we present here use the same simulation setup as Figure~\ref{fig:node} in the main text. 

\section{Simulations to Demonstrate Consistency of Latent Space Model Parameter Estimators}
\label{sec: simulations}
In this section, we study simulation experiments to 
we show that the estimates of $z^\star_i$ and $\nu_i^\star$ are consistent as $n \rightarrow \infty$. 

\begin{figure}
    \centering
    \includegraphics[scale = 0.5]{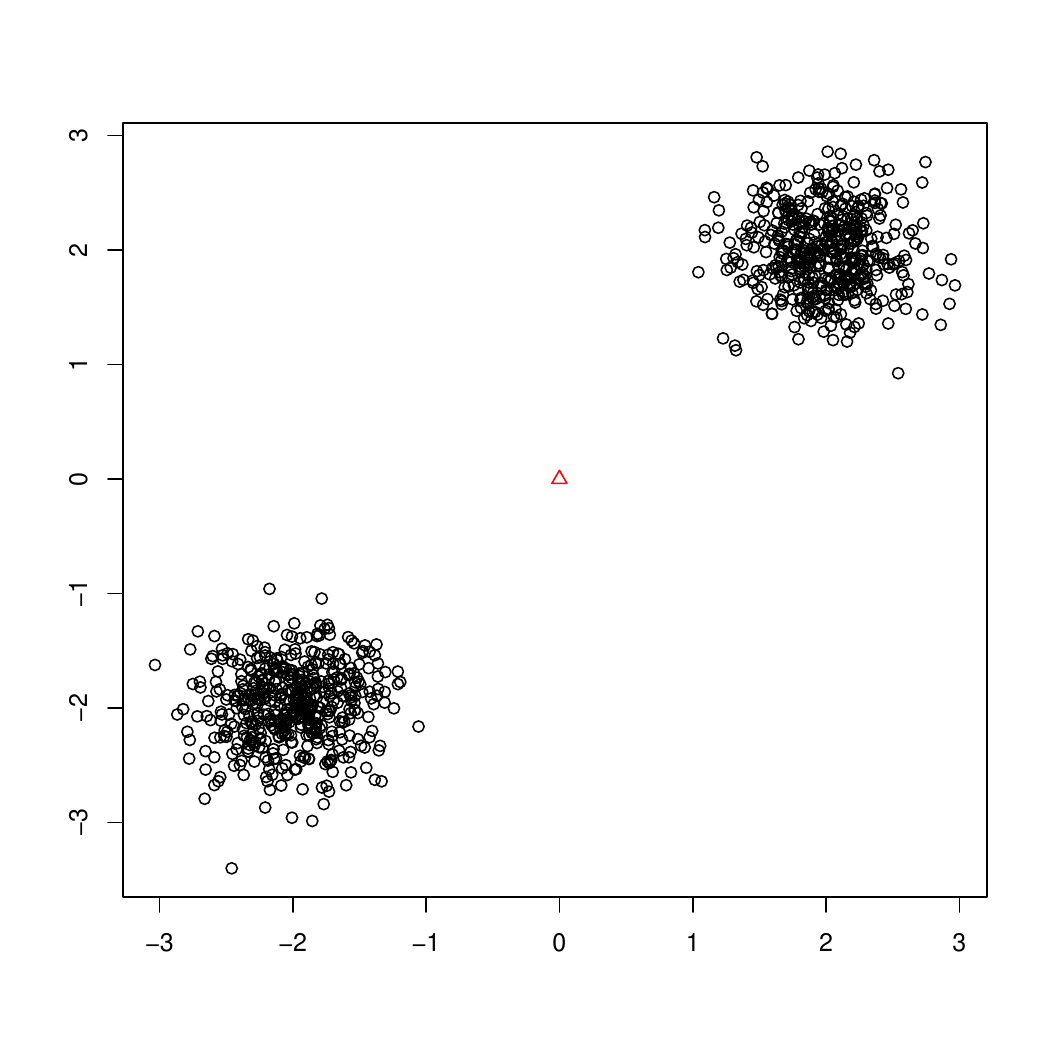}
    \caption{Plot of $n = 500$ locations (black circle) centered at $(2, 2)$ and $(-2, 2)$. The point at $(0, 0)$ (the red triangle) is the location we want to estimate with the ARD.}
    \label{fig: z_i_location}
\end{figure}

We start with the estimates of the node locations. To do this, we create two group centers $\mu_1 = (2, 2)$ and $\mu_2 = (-2, -2)$ and set $z_0 = (0, 0)$. Our goal is to estimate the location of $z_0$. In Figure \ref{fig: z_i_location}, we plot a sample realization of the $z_i$ and $z_0$ for $n = 500.$

\begin{figure}
    \centering
    \includegraphics[scale = 0.5]{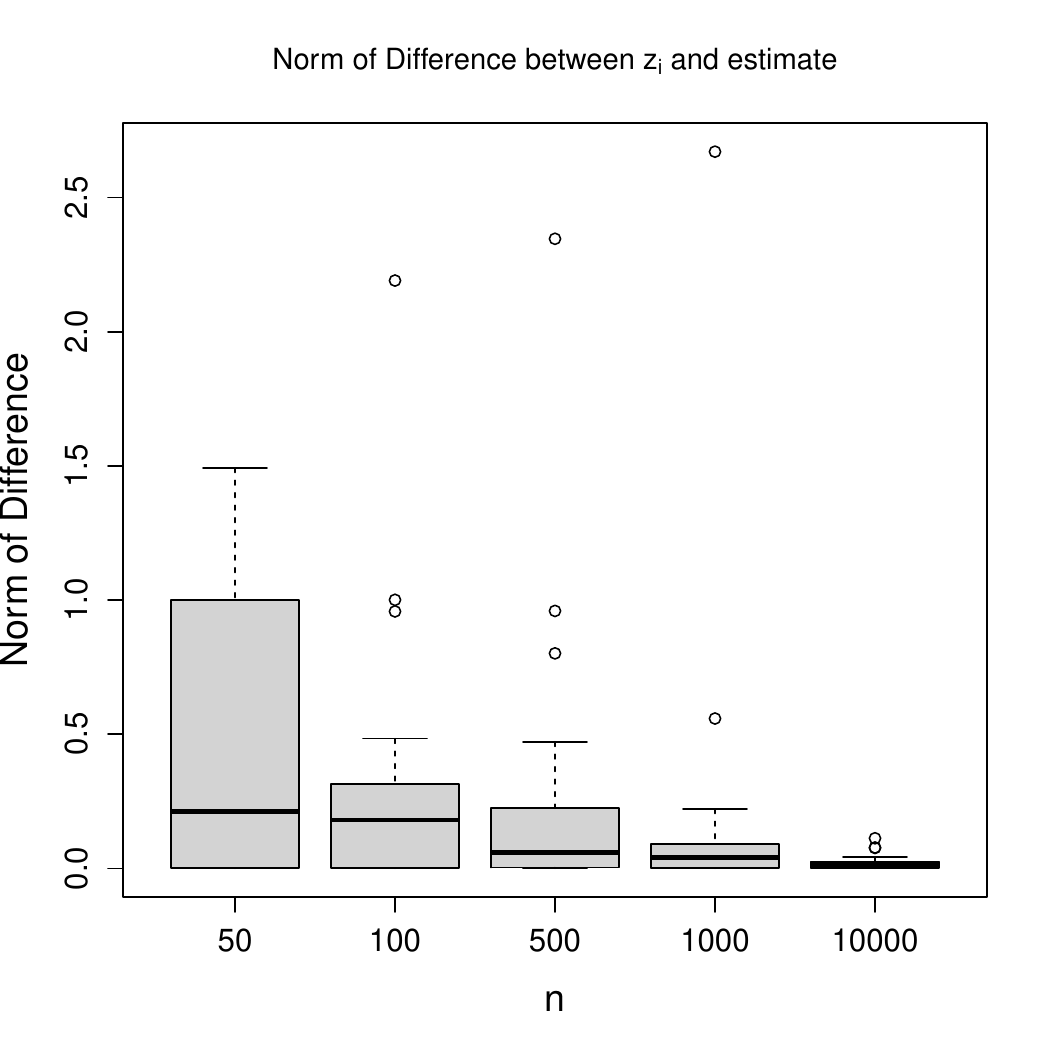}
    \caption{Norm of difference $\hat z_i - z_0$ for various values of $n$ on the $x$-axis.
     }
    \label{fig: error_zi}
\end{figure}

We assign $n$ nodes to be in group 1, and $n$ nodes to be in group 2. Given these group memberships $c_i$, we draw
\begin{equation*}
    z_i\mid\{c_i = j\} \sim N\left(\mu_j, \frac{1}{3}I_2\right) \ \ \ j = 1, 2 \;.
\end{equation*}
where $I_2$ is the $2 \times 2$ identity matrix.
We then create generate edges between the node at location $z_i$ and $z_0$ by 
defining 
\begin{equation*}
P_{i} = \exp(-||z_i - z_0||) = \exp(-||z_i||) \;.
\end{equation*}
where the second equality follows since $z_0 = (0, 0)$. We then generate the edges between nodes in groups 1 and 2 and the node at $z_0$ in this way:
\begin{align*}
    G_{i1} &= \text{Bernoulli}(P_i), \ \ c_i = 1 \\
     G_{i2} &= \text{Bernoulli}(P_i), \ \ c_i = 2 \;.
\end{align*}
The ARD responses are then $y_{i1} = \sum_{i = 1}^n G_{i1}$ and $y_{i2} = \sum_{i = n+1}^{2n} G_{i2}$. We then estimate the node location $z_0$ by the estimation procedure described above. In particular, the estimate $\hat z_i$ solves $\hat z_i = G_1(a)$, where $a = \log(Y_{i1}/n) - \log(Y_{i2}/n)$. We repeat the above process 25 times for each value of $n = 50, 100, 500, 1000, 10^4.$  In Figure \ref{fig: error_zi}, we plot $||\hat z_i - z_i|| = ||\hat z_i||$. We see that the norm is decreasing as $n$ increases.

\begin{figure}
    \centering
    \includegraphics[scale = 0.5]{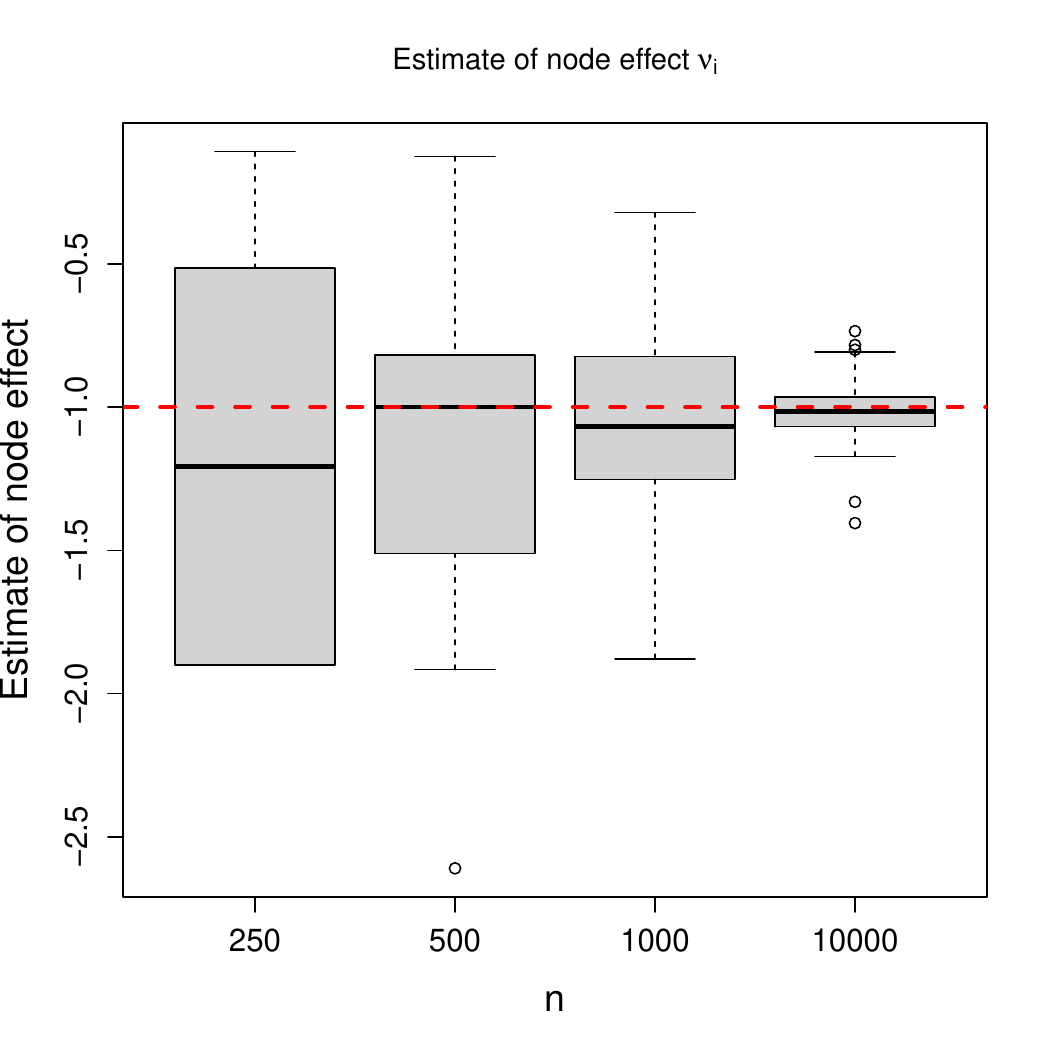}
    \caption{Estimate of the node effect $\nu_i^\star$ using the esimate defined in (\ref{eq: def_hat_nu}). We set $\nu_i^\star = -1$ and generate estimates of this parameter using various values of $n$ on the other $x$-axis. As $n$ increases, we see convergence of the estimate to $\nu_i^\star$. }
    \label{fig: error_nu_i}
\end{figure}

To demonstrate 
 the consistency claim for the node effect estimate $\hat \nu_i$, we simulate $n$ locations $z_i \sim N\left((2,2), \frac{1}{3}I_2\right)$ and $\nu_i^\star \overset{\text{i.i.d.}}{\sim} \text{Unif}(-2, 0)$. We then let $\nu_i^\star = -1.$ Our estimate of the node effects is, recalling (\ref{eq: def_hat_nu}), the $\hat \nu_i$ that solves
\begin{equation*}
   \frac{y_{ik}}{n_k} = E\{\exp(\nu^\star)\} \exp(\hat \nu_i) E[\exp\{-d(z_i, z)\}] \;,
\end{equation*}
where $z \sim F(\mu^\star_k, \sigma^\star_k).$ We suppose that the terms $z_i,$ $E\{\exp(\nu^\star)\}$ and $\mu^\star_k, \sigma^\star_k$ are known, which allows us to solve for the estimate $\hat \nu_i$. We repeat this process 100 times for $n = 250, 500, 1000, 10^4$. In Figure \ref{fig: error_nu_i}, we plot the estimation error and see that as $n$ increases, the error decreases.

\section{Supplemental results used to prove Theorem \ref{thm: LS_consistency}}
\label{sec: add_lemma} 

In this section, we prove Lemma \ref{lemma: consistency_MLE} which is used to prove Theorem \ref{thm: LS_consistency}. To do that, we introduce the pseudo-log likelihood of the ARD. We note here that maximizing the pseudo-likelihood is equivalent to the method-of-moments (or equivalently, Z-estimator) approach taken in Section \ref{sec: define_estimates} but by maximizing the pseudo log likelihood, we are able to use the classical M-estimator results to conclude consistency \citep{Vdv}.

We now discuss the pseudo-likelihood of the ARD.  As described above, the data we observe, when conditioned on the ego's parameters and marginalizing over the alters' parameters, are simply Binomial draws. We can write the log-likelihood for the number of links that $i$ has to a random set of $n_k$ members of group $k$ as
\begin{align}
        \log f(y_{ik}\mid\nu_i, z_i, \eta) &= \log\bigg\{{n_k \choose y_{ik}}\bigg\} + y_{ik} \log(p_{ik}) \\ &+ (n_k - y_{ik}) \log(1 - p_{ik}).
\end{align}
for an arbitrary $\nu_i, z_i, \ \eta.$

We can build our target objective function by summing up over all $k$ traits for each node and then all nodes
\begin{equation*}
   \sum_{i = 1}^m \sum_{k = 1}^K \log f(y_{ik}\mid\nu_i, z_i, \eta).
\end{equation*}
For each $i$, the counts of links across groups are independent conditional on the latent positions.  We describe this as the pseudo-likelihood because the full likelihood also accounts for correlation between $Y_{ik(j)}$ and $Y_{jk(i)}$, where $k(i)$ is person $i$'s group. Nonetheless, this pseudo-likelihood delivers consistent estimates, similar to other recent work in consistent estimators for graph models. See \cite{amini} and its references for a discussion on this point. 
 In practice, we do not know the parameter $\eta^\star$, which contains the means and variances of the distribution of node locations as well as the expxected value of $\exp(\nu_i)$.  Suppose that we have a consistent estimator $\hat \eta \overset{p}{\rightarrow} \eta^\star$. We can then use this plug-in estimator in place of $\eta^\star$, which leads to the final ARD pseudo-likelihood
\begin{equation}
\label{eq: marginal_likelihood}
   \hat \ell_n(\mathbf{y}\mid\theta) = \sum_{i = 1}^m \sum_{k = 1}^K \log f(y_{ik}\mid\nu_i, z_i, \hat \eta).
\end{equation}
We then define the estimates of the node locations and effects as the maximisers of the following pseudo-likelihood:
\begin{equation}
    (\hat \nu_1, \dotsc, \hat \nu_m, \hat z_1, \dotsc, \hat z_m) = \underset{\nu_{[1:m]},z_{[1:m]}}{\arg \max}\hat \ell_n\left( \mathbf{y} \mid  \nu_{[1:m]},z_{[1:m]}, \hat \eta\right).
\end{equation}

We begin by including the following result, Theorem 5.7 of \cite{Vdv}, that allows us to conclude consistency of an M-estimator. This result requires two conditions, which we now state below. 
\begin{condition}
\label{cond: well_sep}
For all $\epsilon > 0$, 
\begin{equation*}
    \sup_{\theta: d(\theta^\star, \theta) \geq \epsilon} Q(\theta) < Q(\theta^\star) \;.
\end{equation*}
\end{condition}

When $\Theta$ is compact, which we assume is true in Condition \ref{cond: compact} below, a sufficient condition for Condition \ref{cond: well_sep} to hold is that $Q$ has a unique maximum at $\theta^\star$. 

\begin{condition}[Uniform law of Large Numbers] We require that
\label{cond: ULLN}
\begin{equation*}
    \sup_{\theta \in \Theta}| \hat Q_n(\theta) - Q(\theta)| \overset{p}{\rightarrow} 0 \;.
\end{equation*}
\end{condition}
Under these two conditions, we can conclude that any M-estimator of the form $\hat \theta_n = \arg \max Q_n(\theta)$ is consistent, in the sense specified below.

\begin{lemma}[Theorem 5.7 of \cite{Vdv}]
Let $\hat Q_n$ be a sequence of random functions indexed by $\theta \in \Theta$, where $(\Theta, d)$ is a metric space. Suppose that Conditions \ref{cond: well_sep} and \ref{cond: ULLN} hold. Then, $d(\hat \theta, \theta^\star) \overset{p}{\rightarrow} 0$ as $n \rightarrow \infty.$
\end{lemma}

There are many ways to verify the uniform law of large numbers result in Condition \ref{cond: ULLN}. See, among others, \cite{Newey_1989, Andrews_Working, PruchaUniform}. In this work, we follow the approach outlined by \cite{Newey_1989}, which requires a compact parameter space, that the functions $\hat Q_n$ converge pointwise to $E(\hat Q_n)$, and that the functions $\hat Q_n$ satisfy a Lipschitz-type condition.

The following two conditions are used in the uniform law of large numbers results from \cite{Newey_1989}.

\begin{condition}[Compact Parameter Space]
\label{cond: compact}
We suppose that $(\Theta, d)$ is a compact metric space.
\end{condition}

\begin{condition}[Pointwise Convergence]
\label{cond: pointwise_conv}
For each $\theta \in \Theta$, $\hat Q_n(\theta) = \bar Q(\theta) + o_P(1)$
\end{condition}

\begin{lemma}[Corollary 2.1 of \cite{Newey_1989}]
\label{lemma: Newey}
Suppose Conditions \ref{cond: compact} and \ref{cond: pointwise_conv} hold and that $\bar Q_n$ is equicontinuous. Also suppose that $\Theta$ is a metric space with metric $d(\theta, \theta')$ and there exists $B_n$ such that for all $\theta, \theta' \in \Theta,\  |\hat Q_n(\theta) - \hat Q_n(\theta')| \leq B_n d(\theta, \theta')$ and $B_n = O_P(1)$. Then $\sup_{\theta \in \Theta} |\hat Q_n(\theta) - \bar Q_n(\theta)| = o_P(1).$
\end{lemma}
As \cite{Newey_1989} points out immediately after Corollary 2.1, if $\bar Q_n = E(\hat Q_n)$ and $E(B_n)$ is bounded, then we can drop the assumption that $\hat Q_n$ is equicontinuous and instead include it as a conclusion to the lemma. In other words, we do not need to check the condition of equicontinuity to use the lemma above.

\begin{lemma}
\label{lemma: unique_max_log_lik}
The likelihood function of the data $y_{ik}$, conditioned on node $i$'s parameters, which we denote by $f(\nu_i, z_i)$, from the proof of Lemma \ref{lemma: consistency_MLE} has a unique maximum at $(\nu_i^\star, z_i^\star, \eta^\star)$ for sufficiently large $K$.
\end{lemma}
\begin{proof}

By the information decomposition, and again using $f$ to denote the likelihood of $y_{ik}$ given node $i$'s parameters, we have that
\begin{equation*}
    E[\log\{f(y_{ik}\mid\nu_i, z_i)\}] = H_{ik}(\theta^\star) - KL_{ik}(\theta\mid\theta^\star) \;.
\end{equation*}
where $H$ is the entropy of $y_{ik} \mid \nu^\star_i, z^\star_i$ and $KL$ is the KL-divergence between $y_{ik} \mid \nu^\star_i, z^\star_i$ and $y_{ik} \mid \nu_i, z_i$. See \cite{cover2012elements} for more information on this decomposition. 

So to maximize the $E[\log\{f(y_{ik}\mid\nu_i, z_i)\}]$, we need to minimize the KL divergence. Hence, by summing over $k = 1,\dotsc, K$, 
\begin{align*}
    \sum_{k = 1}^K KL_k(\theta\mid\theta^\star) &= \sum_{k =1}^K \log\left\{\frac{p_{ik}(\nu_i, z_i)}{p_{ik}(\nu_i^\star, z_i^\star)}\right\} n_k p_{ik}(\nu_i, z_i) + \\
    &\log\left\{\frac{1 - p_{ik}(\nu_i, z_i)}{1 - p_{ik}(\nu_i^\star, z_i^\star)}\right\} n_k \{1 - p_{ik}(\nu_i, z_i)\} \;.
\end{align*}
Now, note first that the KL divergence is always greater than or equal to zero. Second, the KL divergence is zero if and only if $\theta = \theta^\star$. Note that there are just two parameters $\nu_i$ and $z_i$. For any $k = 1, \dotsc, K$, we define the set $A_k$ to be
\begin{equation*}
    A_k = \{(\nu_i, z_i): \exp(\nu_i) H_k(z_i) E\{\exp(\nu)\} = p_{ik}(\nu_i^\star, z_i^\star)\} \;.
\end{equation*}
In words, $A_k$ is the set of parameters $(\nu_i, z_i)$ that lead to the same probability $p_{ik}(\nu_i^\star, z_i^\star).$ Since the KL divergence is always greater than or equal to zero, with equality if and only if the parameters are equal, we see that $\bigcap_{k = 1}^K A_k$ is the set of maximizers of the function $f.$

Clearly, $(\nu_i, z_i) \in A_k$ for each $k$ and thus $(\nu_i^\star, z_i^\star) \in \bigcap_{k = 1}^K A_k$. To argue that $f$ has a unique maximum at $(\nu_i^\star, z_i^\star)$, we now need to argue that $\{(\nu_i^\star, z_i^\star)\} = \bigcap_{k = 1}^K A_k$. Supposing that $p_{ik}(\nu_i^\star, z_i^\star) \neq p_{ik'}(\nu_i^\star, z_i^\star)$ for some $k \neq k'$, meaning we have at least two distinct probabilities, then $f$ has a unique maximum. For $K$ sufficiently large, we will have that $\{(\nu_i^\star, z_i^\star)\} = \bigcap_{k = 1}^K A_k$. Thus, $f$ has a unique maximum.
\end{proof}

\begin{proof}[Proof of Lemma~\ref{lemma: consistency_MLE}]
To show consistency of the estimates based on maximizing the pseudo-likelihood, we first note that each pair $\nu_i, z_i$ appears in exactly $K$ of the terms in the expression from (\ref{eq: marginal_likelihood}). That is, 
\begin{equation*}
    (\hat \nu_i, \hat z_i) = \underset{\nu, z}{\arg \max} \  \sum_{k = 1}^K n_k^{-1} \log f(y_{ik}\mid\nu_i, z_i, \hat \eta) 
\end{equation*}
Thus, we will show that each pair $(\hat z_i, \hat \nu_i)$ converges to the true value.  By recalling that $y_{ik}\mid\nu^\star_i, z^\star_i$ is Binomial, we see that
   \begin{align*}
    \sum_{k = 1}^K n_k^{-1} \log f(y_{ik}\mid\nu_i, z_i) &= \sum_{k = 1}^K \Big\{n_k^{-1} \log {n_k \choose y_{ik}} + \frac{y_{ik}}{n_k}p_{ik} + \\
    &\left(1 - \frac{y_{ik}}{n_k}\right) \log(1 - p_{ik}) \Big\} \;.
\end{align*}
To argue consistency, we will use Theorem 5.7 of \cite{Vdv}. To simplify the analysis, first note that the term $n_k^{-1} \log {n_k \choose y_{ik}}$ does not depend on the parameters, and also $y_{ik}\mid\nu_i, z_i =  \sum_{j \in G_k} g_{ij}$, so the maximum pseudo likelihood estimates $(\hat \nu_i, \hat z_i)$ also satisfy
\begin{align*}
    (\hat \nu_i, \hat z_i) &= \underset{\nu, z}{\arg \max} \  \sum_{k = 1}^K \frac{1}{n_k}\sum_{j \in G_k} \Big\{ g_{ij} \log(\hat p_{ik})  + (1 - g_{ij}) \log(1 -\hat p_{ik}) \Big\} \\
    &= \underset{\nu, z}{\arg \max} \  \hat f_n(\mathbf{y}, \nu_i, z_i, \hat \eta) \;.
\end{align*}
We now define the term $\hat p$ in the expression above.
Given estimates of the structural parameters $E\{\exp(\nu)\}, \mu_k, \sigma_k^2$, we define 
\begin{equation*}
    \hat p_{ik} : =
\exp(\nu_i)\hat E\{\exp(\nu)\} \hat H_{k}(z_i) 
\end{equation*}
where $\hat H_k(z_i) = E[\exp\{-d(z_i, z)\}]$ is computed using $z_j$ drawn iid from $F(\hat \mu_k, \hat \sigma_k^2)$ and $\hat E\{\exp(\nu)\}$ is the estimate of $E\{\exp(\nu)\}$ defined in the previous section.

Define $f_n(\nu_i, z_i) = E\{\hat f_n(\mathbf{y}, \nu_i, z_i, \hat \eta)\}$
and $f(\nu_i, z_i) = \lim_{n \rightarrow \infty} f_n(\nu_i, z_i)$. In the defintion of $f_n$, the expectation is over the distribution of $\mathbf{y}$ (and note that the distribution of $\hat \eta$ is also determined by the distribution of $\mathbf{y})$. To see why, see our discussion where we define particular estimates of $\hat \eta$ and note that these estimates depend on $\mathbf{y}$.
By Lemma \ref{lemma: unique_max_log_lik}, $f$ has a unique maximum at $(\nu_i^\star, z_i^\star, \eta^\star)$. Thus, since $V \times M \times E$ is compact, it follows that Condition \ref{cond: compact} is satisfied.  To verify Condition \ref{cond: ULLN}, we first use the triangle inequality to see that $\sup_{\nu_i, z_i}|\hat f_n(\mathbf{y}, \nu_i, z_i, \hat \eta) - f(\mathbf{y}, \nu_i, z_i, \hat \eta)|$ is upper bounded by
\begin{align*}
     &\sup_{\nu_i, z_i}|\hat f_n(\mathbf{y}, \nu_i, z_i, \hat \eta) - f_n(\mathbf{y}, \nu_i, z_i, \hat \eta)| + \sup_{\nu_i, z_i}| f_n(\mathbf{y}, \nu_i, z_i, \hat \eta) - f(\mathbf{y}, \nu_i, z_i, \hat \eta)| \;.
\end{align*}
The second term, which is deterministic, 
converges to zero uniformly over all $(\nu_i, z_i)$ by the Weierstrassstrass M-test, which we provide for completeness as Lemma \ref{lemma: Weierstrassstrass} and state below:

\begin{lemma}[Weierstrass M-test]
\label{lemma: Weierstrassstrass}
Let $f_n(x) = \sum_{i = 1}^n f_i(x)$ and $f = \lim_n f_n(x)$. Suppose that there exists $M_n$ such that for each $n, \ |f_n(x)| \leq M_n$ for all $x$ and $\sum_{i = 1}^\infty M_i < \infty$. Then $f_n$ converges uniformly to $f$.
\end{lemma}

Hence this second term converges uniformly in probability over all $(\nu_i, z_i)$. We now look at the first term. To show that this converges uniformly in probability to zero, we will use Corollary 2.1 from \cite{Newey_1989} which for completeness we provide in Section \ref{sec: add_lemma}.  In particular, if we can show (1) that $\hat f_n$ converges pointwise to $E(\hat f_n)$ and (2) that $\hat f_n$ satisfies the Lipschitz inequality
\begin{equation}
\label{eq: lip_original}
    |\hat f_n(\mathbf{y}, \nu_i, z_i, \hat \eta) - \hat f_n(\mathbf{y}, \nu_i', z_i', \hat \eta)| \leq B_n d\{(\nu_i, z_i), (\nu_i', z_i')\}  \;,
\end{equation}
where $B_n = O_P(1)$, then Condition 2 holds by Corollary 2.1 of \cite{Newey_1989}. 

We first show the pointwise convergence. By assumption, $\hat p_{ik} = \exp(\nu_i) \hat \tau \hat H(z_i)$ is a continuous function of its arguments, and since $\hat \eta \overset{p}{\rightarrow} \eta^\star$, $\hat p_{ik} \overset{p}{\rightarrow} p_{ik}$ as $n \rightarrow \infty$ by the continuous mapping theorem. Also, conditioned on the ego's parameters, $y_{ik} / n_k \overset{p}{\rightarrow} p_{ik}$ (by Chebyshev's inequality, since $g_{ij}$ are independent and bounded), so we conclude the pointwise convergence. 

To show (\ref{eq: lip_original}), we upper bound the left hand side by $t_1 + t_2$, where 
\begin{align*}
    t_{1k} &= g_{ij}|\log(\hat p_{ik}) - \log(\hat p_{ik})| \leq g_{ij} |\nu_i - \nu_i' + \log \hat H(z_i) - \log \hat H(z_i')|    \\
    t_{2k} &= (1 - g_{ij})|\log(\hat p_{ik}) - \log(\hat p_{ik})| \leq g_{ij} |\nu_i - \nu_i' + \log \hat H(z_i) - \log \hat H(z_i')|    \;. 
\end{align*} 
By assumption, $\hat H$ is Lipschitz in $z$ and so
$|\log\{\hat H(z_i)\} - \log\{\hat H(z_i')\}| \leq C d(z_i, z_i')$ for some constant $C$, so
\begin{equation*}
    t_{1k} \leq g_{ij} \left\{|\nu_i - \nu_i'| + C d( z_i, z_i')\right\} \leq g_{ij} C' d((\nu_i, z_i), (\nu_i', z_i')) \;,
\end{equation*}
and a similar argument holds for $t_{2k}$. Since the left hand side of (\ref{eq: lip_original}) is upper bounded by $\sum_{k  = 1}^K t_{1k} + t_{2k}$, and since $\sum_{j \in G_k} n_k^{-1} g_{ij}$ is $O_P(1)$, we conclude that (\ref{eq: lip_original}) holds and so we conclude by Corollary 2.1 of \cite{Newey_1989} that Condition 2 holds. It follows from Theorem 5.7 of \cite{Vdv} that the maximum pseudo likelihood estimator $(\hat \nu_i, \hat z_i)$ is consistent.
\end{proof}


\bibliographystyle{apalike}

\bibliography{arxiv_main}
 
 \end{document}